\pdfoutput=1 
\documentclass[twocolumn,tighten]{aastex631}

\hypersetup{
    colorlinks=true,
    linkcolor=red,
    filecolor=magenta,      
    urlcolor=cyan,
    citecolor=blue,
    }
    
\usepackage[all]{hypcap}

\shorttitle{MIR/FIR Color-Color Relations}
\shortauthors{Gregg et al.}

\graphicspath{{./}{figures/}}

\begin{document}

\title{Mid and Far-Infrared Color-Color Relations within Local Galaxies}

\author[0000-0003-4910-8939]{Benjamin Gregg}
\affiliation{Department of Astronomy, 
University of Massachusetts, Amherst MA 01003}

\author[0000-0002-5189-8004]{Daniela Calzetti}
\affiliation{Department of Astronomy, 
University of Massachusetts, Amherst MA 01003}

\author[0000-0002-3871-010X]{Mark Heyer}
\affiliation{Department of Astronomy, 
University of Massachusetts, Amherst MA 01003}

\received{Nov. 23 2021}
\revised{Jan. 31 2022 }
\accepted{Feb. 14 2022}
\submitjournal{ApJ}

\begin{abstract}
We present an extensive archival analysis of a sample of local galaxies, combining multi-wavelength data from \textit{GALEX}, \textit{Spitzer} and \textit{Herschel} to investigate ``blue-side" mid-infrared (MIR) and ``red-side" far-infrared (FIR) color-color correlations within the observed infrared spectral energy distributions (IR SEDs). Our sample largely consists of the KINGFISH galaxies, with the important addition of a select few including NGC5236 (M83) and NGC4449. With data from the far-ultraviolet FUV (${\sim}0.15$ $\mu$m) through 500 $\mu$m convolved to common angular resolution, we measure photometry of $kpc$--scale star-forming regions 36$''\times$36$''$ in size. Star formation rates (SFRs), stellar masses and metallicity distributions are derived throughout our sample. Focusing on the $f_{70}/f_{500}$ ``FIR" and $f_{8}/f_{24}$ ``MIR" flux density ratios (colors), we find that a sub-sample of galaxies demonstrate a strong IR color-color correlation within their star-forming regions, while others demonstrate uncorrelated colors. This division is driven by two  main  effects:  1)  the  local strength of star formation (SF) and 2) the metal content of the interstellar medium (ISM). Galaxies uniformly dominated by high surface densities of SF (e.g. NGC5236) demonstrate strong IR color-color correlations, while galaxies that exhibit lower levels of SF and mixed environments (e.g. NGC5457) demonstrate weaker or no correlation---explained by the increasing effect of varying ISM heating and metal content on the IR colors, specifically in the MIR.  We find large dispersion in the SFR--$L_{8}$ (8 $\mu$m luminosity) relation that is traced by the metallicity distributions, consistent with extant studies, highlighting its problematic use as a SFR indicator across diverse systems/samples.
\end{abstract}


\hypertarget{1}{\section{Introduction}}

Discovering how galaxies evolve through time is an essential stepping-stone in our fundamental understanding of the Universe. The temporal evolution of galaxies is tracked by the change of important physical parameters such as their star formation rate (SFR), mass, and composition (metals, gas, dust) and the change in their dynamical state (mergers, interactions, inflows, outflows). 

Typically, the important physical parameters in galaxies are measured by the strength of light they emit at a variety of wavelengths. For example, dust and gas masses are usually derived from the far-infrared (FIR) regime, where the strength of optically thin dust continuum emission is directly proportional to the total amount of gas and dust. SFRs are usually derived directly from the ultraviolet (UV) regime, where the thermal blackbody emission from massive, newly formed stars peaks, and indirectly from the mid-infrared (MIR), as UV and optical light from embedded young stars is absorbed by foreground dust and thermally re-radiated into the infrared. A key result from recent years is that in the past 11 billion years, about 75\% of the cosmic SFR density has been processed by dust into the infrared \citep{2014ARA&A..52..415M}, highlighting the importance of dust correction in understanding the evolution of galaxies.

There exists an important physical connection between two galactic parameters, the SFR and gas content---as the dense gas within galaxies gravitationally coalesces to form new stars. Measuring and understanding the physical relation between the SFR and gas content, and thus the process by which galaxies regulate their star formation (SF), has been the subject of vast effort in recent years. This connection between the SFR and gas reservoirs in galaxies is described by the well known Schmidt–Kennicutt Law \citep{1998ApJ...498..541K,2012ARA&A..50..531K}. This tight relation has been shown to hold at both low and high redshift, as long as the SFR and gas contents are globally integrated over whole galaxies \citep{2010ApJ...714L.118D,2010MNRAS.407.2091G}. On the contrary, this relation has been shown to weaken at smaller scales until breaking down on sub-galactic scales of ${\sim}200 \, pc$, where the molecular gas is shown to be spatially uncorrelated with current SF \citep{2010ApJ...721..383M,2010ApJ...722L.127O,2010ApJ...722.1699S}. The breakdown of this strong relation at sub-galactic scales has been typically explained in two ways: the finite timescale of the physical association between young stars and their dense birth clouds \citep{2009ApJS..184....1K,2014ApJ...795..156W}; and the increasing dispersion in tracers of SFR and gas caused by small number statistics at small scales \citep{2012ApJ...752...98C,2014MNRAS.439.3239K}. On scales of  ${\sim}1 \, kpc$, there does appear to be a physical relation between the SFR and gas reservoirs, but it is comparatively weak and the functional form remains debated \citep{2007ApJ...671..333K,2008AJ....136.2846B,2013ApJ...778..133L,2013AJ....146...19L}. 

Based on these previous results, we expect the physical link between SF and gas contents to be generally weak at sub-galactic scales. Yet, the results of \cite{2018ApJ...852..106C} suggested a strong correlation between the ``red" (FIR) and ``blue" (MIR) sides of the infrared spectral energy distribution (IR SED) in sub-galactic (${\sim}500 \, pc$) star-forming regions within the galaxy NGC4449. Thus, the results of \cite{2018ApJ...852..106C} indicate the need to investigate whether the observed correlation indicates a physical or trivial connection between the derived SFRs and gas contents at these spatial scales and whether these correlations hold in general. A trivial correlation can arise within dusty star-forming galaxies from the fact that luminosities on both sides of the IR SED can receive contributions from the underlying older, cool stellar population. Either result is interesting as the SFR and gas masses derived on these scales are typically treated as independent. 

Currently, there exists a number of studies in the literature that investigate the SFR--gas mass relation as a function of sub-galactic scale (e.g. \citealt{2007ApJ...671..333K,2008AJ....136.2782L,2008AJ....136.2846B,2009AJ....137.4670L,2010ApJ...722.1699S,2011ApJ...730...72R,2011ApJ...735...63L,2012ApJ...745..183R,2015MNRAS.448.1847H,2016ApJ...825...12J,2016MNRAS.459.1440W}) in addition to the recent results of \cite{2018ApJ...852..106C}. With the exception of \cite{2018ApJ...852..106C}, none of these works explore the use of the IR SED to measure SFRs and gas content from the ``blue" and ``red" sides, respectively---an application that will  be common at high redshift for dusty galaxies probed by the \textit{James Web Space Telescope} (\textit{JWST}) and the \textit{Atacama Large Millimeter/submillimeter Array} (\textit{ALMA}). Thus, presently it is important that a large variety of galactic environments and differing star formation activities are investigated on a range of spatial scales in order to study whether the use of the IR SED to simultaneously measure both SFRs and gas content can be extended to galaxies in general at all redshifts.

In this study, we investigate the relation between derived star formation and dust/gas contents on $kpc$--scales in a large sample of 60 local galaxies. Using archival data, the FIR and MIR colors are derived from the IR SED in numerous sub-galactic regions. Correlations between these colors are explored across our sample, as well as the underlying processes that may be driving them. We pay special attention to three interesting test cases that span some of the diversity in our large sample---NGC5236, NGC5457, and NGC4449.

NGC5236 (M83) is a local, grand-design spiral galaxy, located at a distance of $4.5 \, Mpc$ \citep{2003ApJ...590..256T}. It has a measured radius in B band equal to 25 magnitude ($r_{25}$) of 6.44 arcminute ($'$) or $8.43 \, kpc$ \citep{1991rc3..book.....D}. It is classified as type Sc, viewed relatively face-on, and is metal rich, with metallicity greater than solar \citep{1999ApJ...523..136S}. The mean SFR surface density of NGC5236 is 0.017 $M_{\odot}\,yr^{-1}\,kpc^{-2}$ and total SFR is 2.3  $M_{\odot}\,yr^{-1}$ \citep{2010ApJ...714.1256C}---putting NGC5236 well within the regime of actively star-forming galaxies in the local Universe, often considered a starburst. NGC5236 hosts a prominent central bar and dynamical studies have shown that gas is funneled along the bar,  producing very high rates of star formation at the galactic center \citep{2010MNRAS.408..797K}. NGC5236 is also rich in molecular gas, fueling the star formation, with 13 percent of the total galactic disk mass in the form of molecular gas \citep{2004A&A...413..505L}.

NGC5457 (M101) is a large face-on spiral galaxy, with a measured size of $r_{25}=12'$ or $23.4 \, kpc$ at a distance of $6.7 \, Mpc$ \citep{Freedman_2001,2014A&A...570A..13M}. This makes its disk nearly 3 times larger in physical size compared to NGC5236. Because of its large size and steady radial metallicity gradient,  NGC5457 has the largest overall metallicity change between center and outskirts among galaxies where direct electron temperature based measurements are available, ranging from about 7.5$\leq$ 12$+log(O/H)$ $\leq$ 8.8 \citep{2003PASP..115..928K,Croxall_2016}. NGC5457 contains a diverse range of star-forming and passive environments \citep{Watkins_2017}, with a mean SFR surface density of 0.002 $M_{\odot}\,yr^{-1}\,kpc^{-2}$ \citep{2013AJ....146...19L} and total SFR of 2.3 $M_{\odot}\,yr^{-1}$ \citep{2011PASP..123.1347K}, similar to the SFR and surface density values of the Milky Way \citep{2012ARA&A..50..531K}. NGC5457 an ideal example of a complex galaxy with diverse environments and highly varying interstellar medium (ISM) composition. 

NGC4449 is a local (${\sim}4.2 \, Mpc$; \citealt{2003A&A...398..467K}) Magellanic irregular dwarf galaxy hosting a central starburst, with a measured size of $r_{25}=3.1'$ or $3.79 \, kpc$. The specific SFR (SFR/stellar mass) of NGC4449 puts it on average well above the main sequence of local star-forming galaxies \citep{2014MNRAS.445..899C}. The mean SFR surface density is 0.015 $M_{\odot}\,yr^{-1}\,kpc^{-2}$ \citep{2010ApJ...714.1256C}, comparable to NGC5236. It has sub-solar metallicity, about 1/3 solar,  with a relatively modest metallicity gradient \citep{2015MNRAS.450.3254P}. 

One major goal of this project is to inform future studies of high redshift galaxies with \textit{ALMA} and the upcoming \textit{JWST}. The unprecedented synergy in angular resolution and depth between \textit{JWST} and \textit{ALMA} will revolutionize the study of the SF and gas contents of distant systems. These two facilities will obtain complementary spatially resolved data on the SFR from the MIR (\textit{JWST}) and the dust and gas contents from the FIR (\textit{ALMA}) out to high redshift. Specifically, \textit{JWST} will probe SF in galaxies in the restframe 8 $\mu$m out to redshift $z{\sim}2$, where \textit{ALMA} can detect the restframe $\sim$300~$\mu$m dust emission. Understanding these resolved systems in detail at high redshift will first require an in-depth understanding in local galaxies. We will need a solid understanding of the two sides of the IR SED---whether they are independent of one another and provide unbiased measures, or if instead they are correlated, in which case we need to identify precisely what drives those correlations. 

As part of this study, we investigate the 8 $\mu$m emission within our sample of galaxies using archival data from IRAC on the \textit{Spitzer Space Telescope} (\textit{SST}). The restframe 8 $\mu$m emission from galaxies has historically been used as a monochromatic SFR indicator. This is justified by the fact that  polycyclic aromatic hydrocarbons (PAHs) and small dust grains can be heated by the photodissociation regions (PDRs) that surround actively star-forming regions and the subsequent emission from PAHs is brightest at about 8 $\mu$m \citep{2007ApJ...657..810D,2007ApJ...656..770S,2009ApJ...703.1672K,2011A&A...533A.119E}. However, there is an important caveat; PAHs are destroyed by ionizing radiation from newly formed stars \citep{2004ApJS..154..253H,2007ApJ...660..346P,2008MNRAS.389..629B,Rela_o_2009}, leading to a deficit in the 8 $\mu$m luminosity in galaxies with low metallicity, where PAHs are less well shielded by metals \citep{2005ApJ...628L..29E,2007ApJ...666..870C,2007ApJ...656..770S,2014MNRAS.445..899C,Shivaei_2017}. It has also been argued that metals can act as catalysts for the formation and growth of PAHs, leading to smaller average sizes in low metallicity environments \citep{2012ApJ...744...20S}. More recently, \cite{Lin_2020} find an anti-correlation between the dust-only 8 $\mu$m luminosity and the age of young stellar clusters, suggesting the 8 $\mu$m luminosity decreases with increasing age of the stellar population. The existence of a strong interstellar radiation field is also found to suppress the emission from PAHs, independent of metallicity \citep{2006A&A...446..877M,Gordon_2008,2011ApJ...728...45L,Shivaei_2017,2018ApJ...864..136B}. Adding to the complexity, spatially resolved studies have shown that a significant amount of 8 $\mu$m emission is associated with the cold, diffuse ISM, which suggests an important heating source other than recent ($<100 \, Myr$) star formation \citep{2008MNRAS.389..629B,2014ApJ...784..130C,2014ApJ...797..129L}. These results suggest that the luminosity at 8 $\mu$m is not a straightforward indicator of the SFR. In this work, we further investigate these issues by exploring how metallicity gradients within our sample affect the observed ``red" side versus ``blue" side IR color-color correlations---where we derive the ``blue" side color from the flux density ratios at 8 and 24 $\mu$m. 

This paper is organized as follows: In Section \hyperlink{2}{2}, our galaxy sample is presented. In Section \hyperlink{3}{3}, we discuss the archival multi-wavelength images utilized in this work. In Section \hyperlink{4}{4}, the analysis of our data is described, including measuring photometry and deriving SFRs and stellar masses. We present the results of our work in Section \hyperlink{5}{5}. In Section \hyperlink{6}{6}, we discuss our results in the context of how they compare with other previous studies and outline directions for the future. In Section \hyperlink{7}{7}, we summarize the important takeaways.

\hypertarget{2}{\section{Sample}}
For our sample of galaxies, we include the majority of the Key Insights on Nearby Galaxies: a Far Infrared Survey with \textit{Hershel} (KINGFISH) sample \citep{2011PASP..123.1347K}. These are 61 local galaxies (D$<30\, Mpc$) observed by the \textit{Herschel Space Observatory} (\textit{HSO}) with PACS 70, 100, 160, SPIRE 250, 350, and 500 $\mu$m. The KINGFISH galaxies were chosen to cover a large range of galaxy properties and ISM environments found in the nearby universe \citep{2011PASP..123.1347K}. NGC0584, NGC1377, IC342, NGC2146, NGC3077, NGC3184, NGC5398, NGC5408, and NGC6946 are not included in our sample as not all relevant data is publicly available on the NASA/IPAC Extragalactic Database (NED).

 We also add three galaxies to our sample from the Very Nearby Galaxy Survey (VNGS) \citep{2012MNRAS.419.1833B}; NGC5236, NGC2403, NGC5194. This survey observed 13 very nearby galaxies with \textit{HSO}, chosen to probe as wide a region in galaxy parameter space as possible, while maximizing the achievable spatial resolution. Additionally, we include NGC4449 in our sample, observed by \textit{HSO} as part of the Dwarf Galaxy Survey (DGS) \citep{2013PASP..125..600M}. This is in order to compare to \cite{2018ApJ...852..106C}, who first identified the strong color-color correlation between the ``red" and ``blue" sides of the IR SED for NGC4449. This brings our total sample of galaxies to 56, which is presented in Table \hyperlink{t1}{1}, along with relevant galactic properties.
 
\begin{deluxetable*}{lrrrrccllcc}
\tabletypesize{\footnotesize}
\tablenum{1}
\capstart
\tablecaption{\hypertarget{t1}{Galaxy Sample }}
\tablehead{
\colhead{Galaxy} & \colhead{$RA$}\tablenotemark{a} & \colhead{$DEC$} \tablenotemark{a} & \colhead{$D$} \tablenotemark{b} & \colhead{$i$} \tablenotemark{c} & \colhead{Morph.} \tablenotemark{d} & \colhead{12$+log(O/H)$} $^{\mbox{\textit{\small e}}}$ & \colhead{$SFR$ $\,\,\,^{\mbox{\textit{\small f}}}$ }  & \colhead{$log(M_{\bigstar})$ $^{\mbox{\textit{\small g}}}$}  & \colhead{Scale [36$''$] $^{\mbox{\textit{\small h}}}$} & \colhead{$N_{SNR}$ $^{\mbox{\textit{\small i}}}$} \\[-2mm]
\colhead{} & \colhead{($^{\circ}$)} & \colhead{($^{\circ}$)} & \colhead{($Mpc$)} & \colhead{($^{\circ}$)} & \colhead{} & \colhead{} & \colhead{($M_{\odot} \, yr^{-1}$)} & \colhead{($M_{\odot}$)} & \colhead{($kpc$)} & \colhead{}
}
\startdata
\rule{0pt}{3ex}
NGC0337 & 14.958708 & -7.577972 & 18.5 & 56.0 & SBd & 8.18 & 1.194 & 9.283 & 3.2 & 15\\
NGC0628 & 24.173946 & 15.783662 & 9.2 & 31.0 & SAc & 8.35 & 1.11 & 9.773 & 1.6 & 140\\
NGC0855 & 33.514555 & 27.877341 & 8.83 & 60.0 & E & 8.29 & ... & 8.586 & 1.5 & 3\\
NGC0925 & 36.820333 & 33.579167 & 8.9 & 64.0 & SABd & 8.25 & 0.514 & 9.469 & 1.6 & 85\\
NGC1097 & 41.579375 & -30.274889 & 25.0 & 57.0 & SBb & 8.47 & 12.925 & 10.971 & 4.4 & 100\\
NGC1266 & 49.003125 & -2.427361 & 32.2 & 32.0 & SB0 & ... & ... & 10.184 & 5.6 & 6\\
NGC1291 & 49.327445 & -41.108065 & 7.5 & 45.0 & SBa & ... & 0.182 & 10.506 & 1.3 & 24\\
NGC1316 & 50.673825 & -37.208227 & 17.2 & 44.0 & SAB0 & ... & ... & 11.287 & 3.0 & 13\\
NGC1404 & 54.716333 & -35.594389 & 19.0 & 26.0 & E1 & ... & ... & 10.827 & 3.3 & 1\\
NGC1482 & 58.662353 & -20.502680 & 19.6 & 55.0 & SA0 & 8.11 & 2.685 & 9.866 & 3.4 & 11\\
NGC1512 & 60.976167 & -43.348861 & 11.6 & 62.0 & SBab & 8.56 & 0.36 & 9.920 & 2.0 & 37\\
HoII & 124.770750 & 70.720028 & 3.05 & 0.0 & Im & 7.72 & 0.036 & 7.590 & 0.5 & 0\\
DDO053 & 128.530000 & 66.181667 & 3.6 & 29.0 & Im & 7.6 & 0.006 & 6.348 & 0.6 & 0\\
NGC2403 & 114.214167 & 65.602556 & 3.1 & 52.0 & SABcd & 8.46 & 0.389(C) & 9.827(C) & 0.5 & 134\\
NGC2798 & 139.344970 & 41.999729 & 23.7 & 55.0 & SBa & 8.34 & 2.852 & 9.966 & 4.1 & 9\\
NGC2841 & 140.510975 & 50.976519 & 14.1 & 63.0 & SAb & 8.54 & 2.45 & 10.170 & 2.5 & 53\\
NGC2915 & 141.548042 & -76.626333 & 4.1 & 55.0 & I0 & 7.94 & 0.024 & 7.641 & 0.7 & 2\\
HoI & 145.146278 & 71.179555 & 3.9 & 36.0 & IABm & 7.61 & 0.004 & 6.870 & 0.7 & 0\\
NGC2976 & 146.814417 & 67.916389 & 3.6 & 68.0 & SAc & 8.36 & 0.084 & 8.972 & 0.6 & 34\\
NGC3049 & 148.706517 & 9.271094 & 18.5 & 64.0 & SBab & 8.53 & 0.566 & 8.548 & 3.2 & 12\\
M81DwB & 151.377500 & 70.364444 & 8.8 & 49.0 & Im & 7.84 & 0.006 & 7.136 & 1.5 & 0\\
NGC3190 & 154.523468 & 21.832295 & 24.0 & 74.0 & SAap & ... & 0.588 & 10.219 & 4.2 & 12\\
NGC3198 & 154.978966 & 45.549623 & 14.0 & 70.0 & SBc & 8.34 & 0.996 & 9.824 & 2.4 & 53\\
IC2574 & 157.097833 & 68.412139 & 3.9 & 0.0 & SABm & 7.85 & 0.06 & 8.225 & 0.7 & 24\\
NGC3265 & 157.778220 & 28.796670 & 18.5 & 46.0 & E & 8.27 & 0.339 & 8.650 & 3.2 & 4\\
NGC3351 & 160.990417 & 11.703806 & 10.0 & 28.0 & SBb & 8.6 & 0.666 & 10.300 & 1.7 & 60\\
NGC3521 & 166.452421 & -0.035864 & 13.5 & 59.0 & SABbc & 8.39 & 2.833 & 10.852 & 2.4 & 80\\
NGC3621 & 169.568792 & -32.814056 & 7.2 & 61.0 & SAd & 8.27 & 0.616 & 9.462 & 1.3 & 91\\
NGC3627 & 170.062351 & 12.991538 & 10.6 & 57.0 & SABb & 8.34 & 2.171 & 10.596 & 1.9 & 62\\
NGC3773 & 174.553648 & 12.112048 & 17.0 & 0.0 & SA0 & 8.43 & 0.301 & 8.584 & 3.0 & 2\\
NGC3938 & 178.206042 & 44.120722 & 21.0 & 18.0 & SAc & ... & 2.436 & 9.599 & 3.7 & 47\\
NGC4236 & 184.175500 & 69.462583 & 4.5 & 0.0 & SBdm & 8.17 & 0.133 & 8.370 & 0.8 & 64\\
NGC4254 & 184.706682 & 14.416509 & 15.0 & 18.0 & SAc & 8.45 & 4.253 & 9.595 & 2.6 & 62\\
NGC4321 & 185.728463 & 15.821818 & 15.9 & 39.0 & SABbc & 8.5 & 3.227 & 10.392 & 2.8 & 79\\
NGC4449 & 187.046261 & 44.093630 & 4.02 & 49.0 & IBm & 8.26(P) & 0.322(C) & 9.44(C) & 0.7 & 55\\
NGC4536 & 188.612707 & 2.188137 & 15.3 & 59.0 & SABbc & 8.21 & 2.416 & 9.487 & 2.7 & 51\\
NGC4559 & 188.990195 & 27.959992 & 8.9 & 62.0 & SABcd & 8.29 & 0.602 & 8.971 & 1.6 & 64\\
NGC4569 & 189.207470 & 13.162940 & 17.0 & 66.0 & SABab & ... & 0.862 & 10.473 & 3.0 & 24\\
NGC4579 & 189.431342 & 11.818194 & 21.0 & 46.0 & SABb & ... & 1.804 & 10.235 & 3.7 & 38\\
NGC4594 & 189.997633 & -11.623054 & 9.55 & 57.0 & SAa & ... & 0.199 & 11.074 & 1.7 & 41\\
NGC4625 & 190.469671 & 41.273965 & 11.8 & 26.0 & SABmp & 8.35 & 0.084 & 8.927 & 2.1 & 8 \\
NGC4631 & 190.533375 & 32.541500 & 7.5 & 73.0 & SBd & 8.12 & 1.647 & 9.746 & 1.3 & 110\\
NGC4725 & 192.610755 & 25.500805 & 12.7 & 59.0 & SABab & 8.35 & 0.501 & 10.577 & 2.2 & 81\\
NGC4736 & 192.721088 & 41.120458 & 4.6 & 38.0 & SAab & 8.31 & 0.37 & 10.329 & 0.8 & 53\\
DDO154 & 193.521875 & 27.149639 & 4.04 & 0.0 & IBm & 7.54 & 0.002 & 6.576 & 0.7 & 0\\
NGC4826 & 194.181837 & 21.682970 & 4.8 & 55.0 & SAab & 8.54 & 0.216 & 9.859 & 0.8 & 31\\
DDO165 & 196.603542 & 67.706944 & 4.6 & 0.0 & Im & 7.63 & 0.002 & 7.046 & 0.8 & 1\\
NGC5055 & 198.955542 & 42.029278 & 9.0 & 55.0 & SAbc & 8.4 & 1.336 & 10.659 & 1.6 & 128\\
NGC5194 & 202.469629 & 47.195172 & 8.58 & 47.0 & SAbc & 8.55 & 2.718(C) & 10.841(C) & 1.5 & 112\\
NGC5236 & 204.253958 & -29.865417 & 4.55 & 40.0 & SABc & 8.75(H) & 2.392(C) & 10.665(C) & 0.8 & 179\\
NGC5457 & 210.802267 & 54.348950 & 7.24 & 8.0 & SABcd & 8.68(B) & 2.721(K) & 10.717(C) & 1.3 & 473\\
NGC5474 & 211.256708 & 53.662222 & 6.82 & 31.0 & SAcd & 8.31 & 0.092 & 8.703 & 1.2 & 32\\
NGC5713 & 220.047938 & -0.288975 & 18.4 & 8.0 & SABbcp & 8.24 & 1.863 & 9.939 & 3.2 & 17\\
NGC5866 & 226.622912 & 55.763213 & 14.0 & 65.0 & S0 & ... & 0.218 & 9.943 & 2.4 & 10\\
NGC7331 & 339.266724 & 34.415519 & 15.0 & 61.0 & SAb & 8.34 & 2.932 & 10.589 & 2.6 & 86\\
NGC7793 & 359.457625 & -32.591028 & 3.4 & 35.0 & SAd & 8.31 & 0.197 & 8.879 & 0.6 & 108\\
\enddata
\tablecomments{
\tablenotetext{a}{Location of the galactic center in decimal degrees obtained from NED. RA = Right Ascension and DEC = Declination.}
\tablenotetext{b}{The distance to each galaxy obtained as the average of various Cephid/TRGB measurements listed in NED.}
\tablenotetext{c}{Inclination Angle obtained from the Two Micron All-Sky Survey (2MASS) Large Galaxy Atlas (LGA) \citep{2003AJ....125..525J} $K_{s}$ band $cos^{-1}(a/b)$ axis ratio, listed in NED.} 
\tablenotetext{d}{Morphological type as listed in NED}
\tablenotetext{e}{Characteristic oxygen abundance or ``metallicity". Obtained from \cite{Moustakas_2010} using \cite{2005ApJ...631..231P} empirically calibrated values. (P) obtained from \cite{2015MNRAS.450.3254P}, (H) \cite{Hernandez_2017}, (B) \cite{2004ApJ...615..228B}.}
\tablenotetext{f}{Star Formation Rates as listed in \cite{2011PASP..123.1347K} obtained from \cite{2010ApJ...714.1256C}. Derived from the combination of H$\alpha$ and 24 $\mu$m luminosity described in \cite{2009ApJ...703.1672K}. (C) obtained directly from \cite{2010ApJ...714.1256C} and (K) from \cite{2011PASP..123.1347K}. The SFRs are
rescaled to our distance values.}
\tablenotetext{g}{Stellar masses listed in \cite{2011PASP..123.1347K}, obtained from \cite{2011ApJ...738...89S} by the multi-color method described in \cite{2009MNRAS.400.1181Z}. (C) obtained from \cite{2014MNRAS.445..899C} using 3.6 $\mu$m photometry. The masses are rescaled to our distance values.}
\tablenotetext{h}{The physical scale in $kpc$ corresponding to 36$''$ at the adopted distance to each galaxy.} 
\tablenotetext{i}{The total number of 36$''\times$36$''$ regions within each galaxy found to meet the applied SNR cutoffs (see Section \hyperlink{4.1}{4.1}); SNR $\geq8$ for SPIRE 500 $\mu$m and SNR $\geq10$ for IRAC 8, MIPS 24, and PACS 70 $\mu$m. }
}
\end{deluxetable*}

\vspace{-8mm}
This sample of galaxies fits well with the goals of this paper. Using a diverse sample of local galaxies that span a wide range of morphologies and environments and are observed by \textit{SST} and \textit{HSO}, we can effectively investigate IR  color-color relations, drawn from the  ``red" (FIR) and ``blue" (MIR) sides of the dust emission SED. We can determine whether they hold for galaxies in general or when/why they begin to break down. We aim to characterize these correlations and what is driving them.

\hypertarget{3}{\section{Data}}

For our analysis, we use a variety of multi-wavelength archival images covering from the Far-UltraViolet (FUV), ${\sim}$0.15 $\mu$m, to the FIR, ${\sim}$500 $\mu$m. These images are from the Galaxy Evolution Explorer (\textit{GALEX}) FUV and Near-UltraViolet (NUV), \textit{SST} 3.6–24 $\mu$m
and \textit{HSO} 70–500 $\mu$m. We retrieve all of our data from the NED database. 

The \textit{GALEX} FUV and NUV images for our sample are obtained from the \textit{GALEX} Ultraviolet Atlas of Nearby Galaxies \citep{2007ApJS..173..185G}. This is a \textit{GALEX} survey that observed 1034 local galaxies in both the FUV and NUV. All \textit{GALEX} images are in units of $CPS$ or counts per second, which are readily converted to physical units using the photometric keywords published in \cite{2007ApJS..173..682M}.

For the majority of our galaxies, archival \textit{SST} IRAC 3.6, 4.5, 5.8, 8.0 and MIPS 24 $\mu$m images are obtained from the \textit{Spitzer} Infrared Nearby Galaxy Survey (SINGS), which observed 75 galaxies in the nearby ($<30 \, Mpc$) universe \citep{2003PASP..115..928K}. For the few galaxies that were not part of SINGS, we use the \textit{SST} images from the Local Volume Legacy (LVL) project \citep{2009ApJ...703..517D}, a survey that has observed near 260 galaxies in the local ($<11\, Mpc$) universe.  All \textit{SST} images are in units of surface brightness $MJy/sr$.

The \textit{HSO} PACS 70, 160 and SPIRE 250, 350, 500 $\mu$m images are obtained from KINGFISH, the VNGS and the DGS, discussed in Section \hyperlink{2}{2}. For NGC4449, the \textit{HSO} images from the DGS are not publicly available on NED. We obtain these images from the \textit{Herschel} archive\footnote{see \url{http://archives.esac.esa.int/hsa/whsa/}} and reprocess the data using Scanamorphos v24.0 \citep{2013PASP..125.1126R}. All PACS images are in $Jy/pixel$, while SPIRE images are in $MJy/sr$ for the KINGFISH sample and $Jy/beam$ for the VNGS and DGS. For the SPIRE beam areas, we assume 469, 831 and 1804 $\, ('')^{2}/beam$ for $250$, $350$, and 500 $\mu$m respectively. Table \hyperlink{t2}{2} lists the central wavelength, full-width-at-half-maximum (FWHM) of the point spread function (PSF), and representative image sensitivities for each photometric band.

\setcounter{table}{1}
\begin{table*}
\centering
\begin{center}
\caption{\hypertarget{t2}{ Measured Flux Densities/Sensitivities}} 
\hspace{-16mm}
\begin{tabular*}{\textwidth}{c @{\extracolsep{\fill}} ccccc}
\hline
\hline
\rule{0pt}{3ex}
Wavelength & Facility/Instrument$^{\mbox{\textit{a}}}$ & PSF$^{\mbox{\textit{b}}}$ & $f_{\nu}$ \textbf{:} NGC5236$^{\mbox{\textit{c}}}$ & $f_{\nu}$ \textbf{:} NGC5457$^{\mbox{\textit{c}}}$ & $f_{\nu}$ \textbf{:} NGC4449$^{\mbox{\textit{c}}}$ \\
( $\mu$m) & & ($''$) & (mJy) & (mJy) & (mJy) \\
\hline
\rule{0pt}{3ex}
0.15 (FUV) & \textit{GALEX} & 4.48 & 11.0066 $\pm$ 0.0010 & 1.0447 $\pm$ 0.0001 & 14.6893 $\pm$ 0.0001\\
\rule{0pt}{3ex}
0.23 (NUV) & \textit{GALEX} & 5.05 & 14.0074 $\pm$ 0.0021 & 1.6318 $\pm$ 0.0002 & 16.7769 $\pm$ 0.0002 \\
\rule{0pt}{3ex}
3.6 & \textit{SST}/IRAC & 1.90 & 404.8584 $\pm$ 0.0119 & 40.2126 $\pm$ 0.0038 & 54.8638 $\pm$ 0.0075 \\
\rule{0pt}{3ex}
4.5 & \textit{SST}/IRAC & 1.80 & 277.3978 $\pm$ 0.0065 & 25.5724 $\pm$ 0.0046 & 36.8415 $\pm$ 0.0095 \\
\rule{0pt}{3ex}
5.8 & \textit{SST}/IRAC & 2.10 & 1017.7421 $\pm$ 0.0067 & 50.9747 $\pm$ 0.0077 & 90.2820 $\pm$ 0.0046\\
\rule{0pt}{3ex}
8.0 & \textit{SST}/IRAC & 2.80 & 2444.4489 $\pm$ 0.0180 & 100.2364 $\pm$ 0.0104 & 165.6912 $\pm$ 0.0161\\
\rule{0pt}{3ex}
24.0 & \textit{SST}/MIPS & 6.40 & 6897.5677 $\pm$ 0.0178 & 96.5818 $\pm$ 0.0086 & 442.5403 $\pm$ 0.0160\\
\rule{0pt}{3ex}
70.0 & \textit{HSO}/PACS & 5.70 & 68430.1527 $\pm$ 1.3096 & 1352.1522 $\pm$ 1.2244 & 6051.8881 $\pm$ 1.2130\\
\rule{0pt}{3ex}
160.0 & \textit{HSO}/PACS & 11.20 & 66253.6734 $\pm$ 2.6256 & 3630.4586 $\pm$ 1.8835 & 5025.3862 $\pm$ 1.6208\\
\rule{0pt}{3ex}
250.0 & \textit{HSO}/SPIRE & 18.20 & 20539.8548 $\pm$ 7.1825 & 1705.4042 $\pm$ 1.2904 & 1717.6835 $\pm$ 2.4529\\
\rule{0pt}{3ex}
350.0 & \textit{HSO}/SPIRE & 24.90 & 7368.0864 $\pm$ 6.6831 & 744.8033 $\pm$ 1.5510 & 672.2131 $\pm$ 2.7887\\
\rule{0pt}{3ex}
500.0 & \textit{HSO}/SPIRE & 36.10 & 2207.9534 $\pm$ 7.2721 & 240.4996 $\pm$ 1.8204 & 226.4679 $\pm$ 3.3260\\
\hline
\end{tabular*}
\begin{flushleft} 
\currtabletypesize{\sc Note}--- \\
\rule{0pt}{3ex}
$^{\mbox{\textit{a}}}$ Facility/Instrument: \textit{GALEX} = \textit{Galaxy Evolution Explorer}; \textit{SST} = \textit{Spitzer Space Telescope}, IRAC and
MIPS Cameras; \textit{HSO} = \textit{Herschel Space Observatory}, PACS and SPIRE
instruments. \\
\rule{0pt}{3ex}
$^{\mbox{\textit{b}}}$ The FWHM of the PSF for each band, obtained from \cite{2011PASP..123.1218A}.\\
\rule{0pt}{3ex}
$^{\mbox{\textit{c}}}$ Measured flux density ($f_{\nu}$) for a central 36$''\times$36$''$ aperture within each galaxy. The sensitivities are measured on the background surrounding each galaxy and correspond to the standard deviation of the iteratively $3\sigma$ clipped images, multiplied by the square root of the number of pixels in the aperture over the relevant beam area. All images are convolved to SPIRE 500 $\mu$m resolution. \\
\rule{0pt}{3ex}
\rule{0pt}{3ex}
\end{flushleft}
\end{center}
\end{table*}

\vspace{-0.5mm}
\hypertarget{4}{\section{Analysis}}
\hypertarget{4.1}{\subsection{Photometry}}

 After obtaining and organizing all of our multi-wavelength images for the entire sample of galaxies, we process the data for further analysis. We initially perform a global background subtraction on all images. An iterative sigma clipping is used to remove all pixels more than 3$\sigma$ above the median. The background is estimated as the mode of the clipped images and is subtracted from the original data. 
 
 We convolve all of the images to common resolution using the kernels from \cite{Aniano_2011}. We chose the high resolution kernels, which are sampled into 0.25 arcsecond ($''$) pixels. All kernels are re-sampled to the appropriate pixels sizes for each of our images using the Image Reduction and Analysis Facility (IRAF) \textit{magnify} function, which performs a linear interpolation across the kernels and re-samples \citep{1986SPIE..627..733T,1993ASPC...52..173T}. The kernels are re-normalized to an enclosed volume of unity (by dividing by the sum of all pixels) to ensure the total flux is conserved. The convolution is then performed using these kernels. From the PSF FWHM's provided in Table \hyperlink{t2}{2} for each image, it is clear that SPIRE 500 $\mu$m has the lowest angular resolution and so we convolve all images to the resolution of SPIRE 500 $\mu$m, using the \textit{convolve\_fft} function from the \textit{astropy.convolution} module within Python \citep{astropy:2013,astropy:2018}. 
 
Since our data come from various projects, the images were reduced differently with varying image pixel sizes and units. For example, the SPIRE 500 $\mu$m image of NGC5457, a galaxy part of the KINGFISH project, has units of $MJy/sr$ and a pixel size of $14''$. On the other hand, this image of NGC5236, part of the VNGS, has units of $Jy/beam$ and a pixel size of $12''$. These and all such differences are accounted for in the convolution, kernels, and unit conversions. We convert the units for all images to $Jy/pixel$, using the relevant pixel sizes/beam areas for each image. For the \textit{GALEX} FUV and NUV, we use the following formulas from \cite{2007ApJS..173..682M} to convert from CPS to AB magnitudes ($m_{AB}$) and flux density ($f_{\nu}$): $m_{AB}(FUV) = -2.5 \, log(CPS) + 18.82$, $m_{AB}(NUV) = -2.5 \, log(CPS) + 20.08$ and $f_{\nu} \, [Jy]=3631 * 10^{(m_{AB}/-2.5)}$. 
 
Multi-wavelength aperture photometry is then measured on all the convolved images. We construct photometric grids across each galaxy, with individual square grid boxes/apertures 36$''\times$36$''$ in size. This aperture size is selected to match the common angular resolution of the images (SPIRE 500 $\mu$m $\sim$ 36.1$''$), which ensures that measurements from adjacent regions are mostly independent, while also maximizing the total number of regions. For each aperture in the grid, we measure the photometry in all images by summing all interior pixels. To do this, we use the aperture photometry provided by the Astropy \textit{photutils} package within Python \citep{larry_bradley_2019_3568287}. A ``sub-pixel" summation method is used, by which all pixels are subdivided into smaller 32$\times$32 sub-pixels in order to more accurately estimate the total flux within each individual aperture. 

Image uncertainties are estimated by first performing an iterative 3$\sigma$ clipping on all of the convolved images. The image uncertainty $\sigma_{im}$ in an individual pixel is calculated as the standard deviation of the clipped, flattened images. The total flux uncertainty within each grid box is estimated as  $\sqrt{N_{pix,ap}/BA}*\sigma_{im}$, where $N_{pix,ap}$ is the number of pixels within each aperture box and $BA$ is the relevant beam area. Since all aperture grid boxes are the same size, all flux measurements in each individual image have the same estimated uncertainties.  

We apply Signal-to-Noise Ratio (SNR) cutoffs to the photometric grid measurements for each galaxy, focusing on four images: IRAC 8, MIPS 24, PACS 70, and SPIRE 500 $\mu$m. This is in order to investigate the correlations in the IR SED between the ``blue side" $f_{8}/f_{24}$ and the ``red side" $f_{70}/f_{500}$ flux density ratios (colors). For our main SNR cutoff, we require that all measurements/regions have a minimum SNR $\geq8.0$ for SPIRE 500 $\mu$m and SNR $\geq10$ for IRAC 8, MIPS 24, and PACS 70 $\mu$m. For NGC5457, where there is an abundance of bright regions, we apply a slightly stricter SNR cutoff, requiring SNR $\geq10$ for SPIRE 500 $\mu$m as well. 
 
 For the IRAC 8 $\mu$m and MIPS 24 $\mu$m images, the known stellar contribution is removed. We use the formulae from \cite{2004ApJS..154..253H} and \cite{2007ApJ...666..870C} given by: $f_{\nu,D} (8) = f_{\nu}(8) - 0.25\, f_{\nu}(3.6)$ and $f_{\nu,D} (24) = f_{\nu}(24) - 0.035\, f_{\nu}(3.6)$, where the
flux densities are in $Jy$ and ``D" refers to 
the dust-only emission component. These formulations use the 3.6 $\mu$m emission as a tracer of the old stellar populations to remove their relative contribution to the 8 and 24 $\mu$m bands, giving the dust-only emission components. This has typically been justified by the fact that the IRAC 3.6 $\mu$m flux is mostly comprised of the photospheric emission from old stars; however, the contamination by the 3.3 $\mu$m PAH emission feature is ${\sim}5 –15\%$
\citep{2012ApJ...744...17M}. For the purposes of the subtraction of the 8 and 24 $\mu$m, this contamination has a relatively minor effect. Even in the presence of significant contribution to the 3.6 $\mu$m  emission from hot dust \citep{2015ApJS..219....5Q}, its impact on our subtracted 8 and 24 $\mu$m images is less than $10\%$. 
 
\vspace{-1mm}
\hypertarget{4.2}{\subsection{Star Formation Rates and Stellar Masses}}

Using a dust-corrected hybrid indicator \citep{Kennicutt_2012}, we derive SFRs by a linear combination of the \textit{GALEX} FUV and MIPS 24 $\mu$m emission. We use the calibration from \cite{2011ApJ...735...63L}, given by 
\begin{equation}
\label{eq:1}
    SFR(M_\odot \, yr^{-1}) = 4.6\times 10^{-44} \big[L(FUV)+\, 6.0 \cdot L(24) \big]
\end{equation}
where the luminosities are in $erg \, s^{-1}$. The luminosities are calculated using $L_{\nu}(erg \, s^{-1}) = \nu \, f_{\nu} \, (10^{-23}) \, 4 \pi d^{2}$, where $\nu$ are the band center frequencies, $f_{\nu}$ are the photometric flux densities in $Jy$, and $d$ is the distance to each galaxy in $cm$. The calibration assumes a \cite{2001MNRAS.322..231K} stellar initial mass function (IMF). 

This SFR indicator combines the direct stellar light from massive, young stars (FUV) and the thermal emission from warm dust heated by newly formed stars (24 $\mu$m). The 24 $\mu$m emission dominates the contribution to the SFR in the densest, most actively star-forming regions, where there tends to be abundant nearby absorbing dust. The FUV emission may dominate in the less actively star-forming regions, where there may be  significantly less dust obscuration. However, an older stellar population (${>100}\,  Myr$) that does not track recent star formation also contributes UV photons and to dust heating, affecting the measured FUV and 24 $\mu$m luminosities. In particular, this could be a problem for the central regions of galaxies, where the older stellar populations may significantly contribute to the emission. 

 \cite{2005ApJ...633..871C} found that the 24 $\mu$m emission in NGC5194 is almost entirely concentrated in HII knots. They determine 14\% of the integrated 24 $\mu$m emission across the galaxy is diffuse, compared to 21\% in the central region. For the UV, \cite{1995AJ....110.2665M} found that UV-bright clusters are not the dominant contributors to the UV luminosity in starburst galaxies; typically 50\% to 80\% of the UV light from starburst galaxies is diffuse. \cite{2001ApJ...555..322T} and \cite{2005ApJ...628..210C} explain this large fraction of diffuse UV light in starburst galaxies as due to either disrupted star clusters that are old enough that only B-stars and older survive or possibly a different IMF in the diffuse field. \cite{2015ApJ...808...76C} investigate the nature of the diffuse FUV light in a more typical star-forming galaxy, NGC5457, finding that the diffuse fraction of the FUV depends on location in the galaxy; ranging from 14\% to 61\% for different interarm regions, generally decreasing with distance from spiral arm structure. It remains unclear what fraction of this diffuse FUV emission is associated with older stellar populations and what fraction is scattered light from young stars.  Although there is much left to be uncovered, the 24 $\mu$m emission seems to be tracing current star formation well, within a correction of at most $\sim$20\% for the central zone of galaxies. The FUV is more uncertain and strongly depends on assumptions for the IMF, likely suffering from a higher contribution from stellar populations older than 100 $Myr$, perhaps on the order of $\sim$30-40\% on average.  Removing the contributions to the SFR from older stellar populations is a complex process, little understood and rarely done, sometimes simplified by first modeling the smooth component of the old stellar light distribution and subsequently normalizing and subtracting it from the SFR tracers. In this study, we ignore this contribution and assume that the FUV and 24 $\mu$m luminosities are exclusively tracing recent (${<100}\,  Myr$) star formation, which leads to derived SFRs that may be biased high, particularly in regions of higher stellar mass surface density (e.g. the central bulge and spiral arms). 

Using the inclination and distances given in Table \hyperlink{t1}{1}, we calculate the SFR surface densities ($\Sigma_{SFR}$) from the SFRs given by Equation \ref{eq:1}. The deprojected area (in $kpc^2$) of a single 36$''\times$36$''$ grid box is calculated for each galaxy as the physical area divided by the cosine of the inclination angle. $\Sigma_{SFR}$ is found as the SFRs divided by these deprojected areas. 

In a small number of cases, the FUV measurements are of low significance. Deeply embedded regions with high extinction will exhibit very little FUV flux, as foreground dust absorbs the high energy radiation from young stars; yet these regions may be actively forming stars and shine brightly in the IR. To account for this, we require that the FUV SNR $\geq5.0$. If this condition is not met, then we set the FUV luminosity equal to zero and calculate the SFR using only the 24 $\mu$m luminosity. For the vast majority of the star-forming regions within our galaxies ($>$ 99\%), regions meeting the initial SNR cutoffs also have FUV SNR $\geq5.0$. In all cases, these regions have 24 $\mu$m SNR $\geq10.0$.

We derive the stellar mass of our regions from the IRAC 3.6 $\mu$m photometry, assuming a constant 3.6 $\mu$m mass-to-light-ratio $(M/L)_{3.6} = \, 0.5 \, M_{\odot}/L_{\odot,3.6}$. A number of previous studies have shown reasonable consistency with this ``flat" value for the $(M/L)_{3.6}$ \citep{Oh_2008,2012AJ....143..139E,McGaugh_2014,2014ApJ...789..126B,Meidt_2014}. This has been justified by the fact that the relation between $(M/L)_{3.6}$ and metallicity is found to be almost orthogonal to the relation between $(M/L)_{3.6}$ and age, such that old, metal-rich populations have comparable $(M/L)_{3.6}$ to younger, metal-poor populations \citep{Meidt_2014}. These studies suggest that old stellar populations
across different galaxy morphologies are typically well-described by a constant $(M/L)_{3.6}$. The stellar mass is determined as the product of the 3.6 $\mu$m luminosity and the $(M/L)_{3.6}$, where we calculate the 3.6 $\mu$m luminosity in units of ``in--band" solar luminosities, using $L_{\odot,3.6}= 1.4\times 10^{32} \, erg\, s^{-1}$ \citep{Oh_2008}. For this calculation, an additional SNR cutoff is applied to the 3.6 $\mu$m photometry, requiring SNR $\geq10.0$. 

To estimate the uncertainties in the SFR, $\Sigma_{SFR}$, and stellar mass measurements, we perform a Monte Carlo calculation with the assumption that the photometric errors are normally distributed with a standard deviation given by their uncertainties. Assuming a 10\% error on all unknown sources of uncertainty (distances from Table \hyperlink{t1}{1}, deprojected grid box areas, coefficients of Equation \ref{eq:1}), 100 random draws are simulated and the resulting distribution in each parameter is calculated. The uncertainities in the parameters are derived as the standard deviation of the distributions. 

\begin{figure*}
\centering
\includegraphics[width=0.75\textwidth]{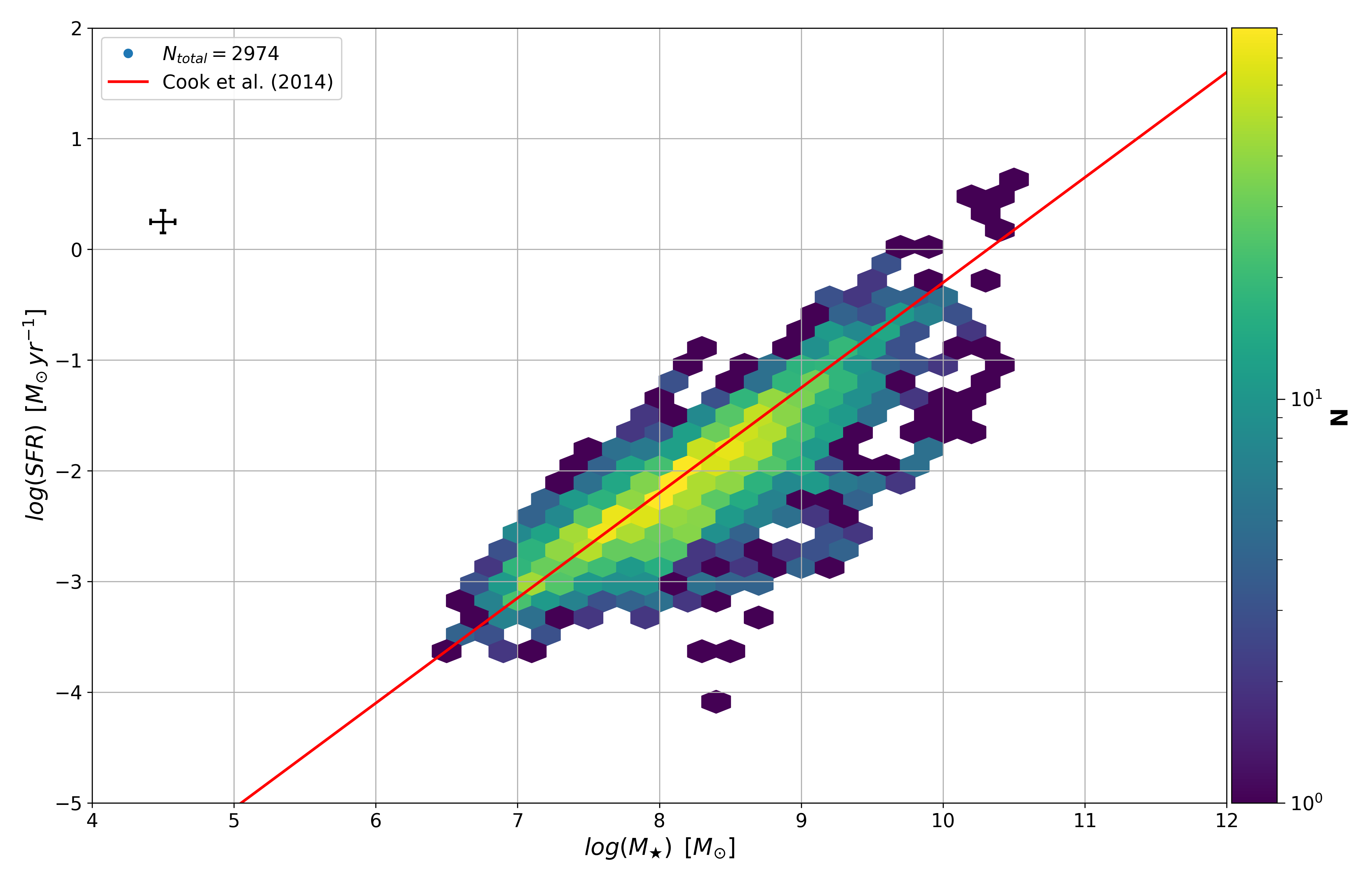}
\caption{The galactic ``main sequence" or SFR--stellar mass relation for all 36$''\times$36$''$ star-forming regions within our ``classified" (Regimes 1 and 2; Section \hyperlink{5.2}{5.2}) sample of galaxies that meet the SNR cutoffs, at the angular resolution of SPIRE 500 $\mu$m. The image shows a 2D logarithmic binning of the data, depicting the number density of our sub-galactic regions along the well-known main sequence relation. Shown in red is the \cite{2014MNRAS.445..899C} relation for the LVL survey. The mean errors are shown by the black bars. The total number of regions is shown in the legend. Stellar masses are derived assuming a constant $(M/L)_{3.6}$ $=0.5 \, M_{\odot}/L_{\odot,3.6}$, similar to \cite{2014MNRAS.445..899C}.  }
\label{fig:f1}
\end{figure*}

For a sanity check, we plot the galactic ``main sequence" or  SFR--stellar mass relation for all 36$''\times$36$''$ star-forming regions within our sample that satisfy the SNR cutoffs, shown in Figure \ref{fig:f1}. The well-known strong relation between the SFR and stellar mass can clearly be seen for our sample. Figure \ref{fig:f1} shows that our star-forming regions follow a SFR-stellar mass trend analogous to that found in the LVL galaxies by \cite{2014MNRAS.445..899C}, where they similarly assume a constant $(M/L)_{3.6} = \, 0.5 \, M_{\odot}/L_{\odot,3.6}$. Clearly, the slope of the trends are consistent and our measurements align on top of the relation from \cite{2014MNRAS.445..899C}.  

There is likely some contamination in the IRAC 3.6 $\mu$m luminosity from PAH emission features and dust heated by young stars. \cite{2012ApJ...744...17M} find that the non-stellar contribution to the galaxy integrated 3.6 $\mu$m is on the order of 10\%; while locally it can reach much higher values, contributing an average of ${\sim}22\%$ in star-forming regions. As a result, our derived stellar masses will tend to overestimate the true stellar masses, specifically in dusty star-forming regions. Yet, they are useful to compare our sample to \cite{2014MNRAS.445..899C} who also ignore the non-stellar component of the 3.6 $\mu$m. We exclude these stellar mass estimates from subsequent analysis.

\hypertarget{5}{\section{Results}}
\hypertarget{5.1}{\subsection{IR Color-Color relations}}

\begin{figure*}
\centering
\includegraphics[width=14.8cm,height=6.4cm]{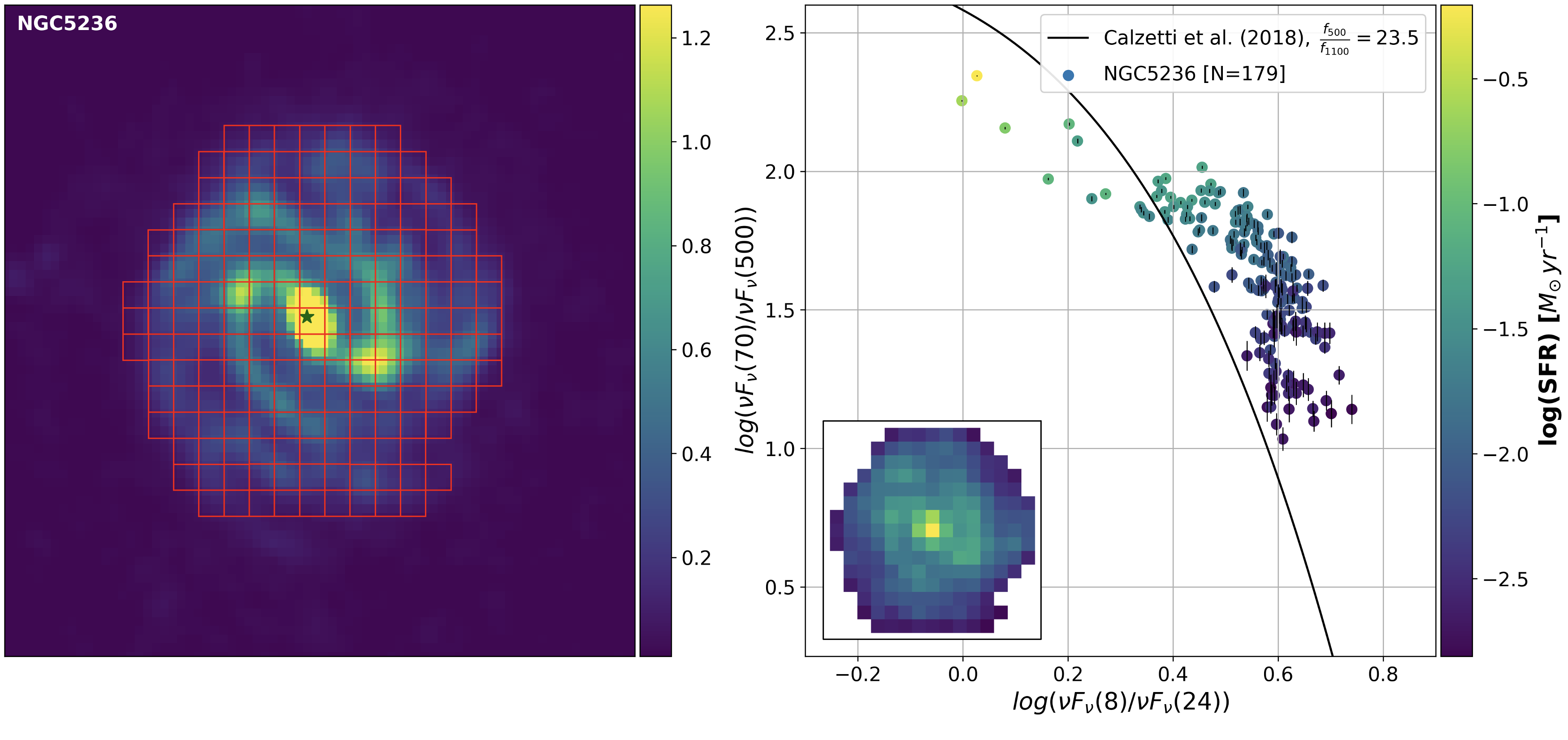}
\includegraphics[width=14.8cm,height=6.4cm]{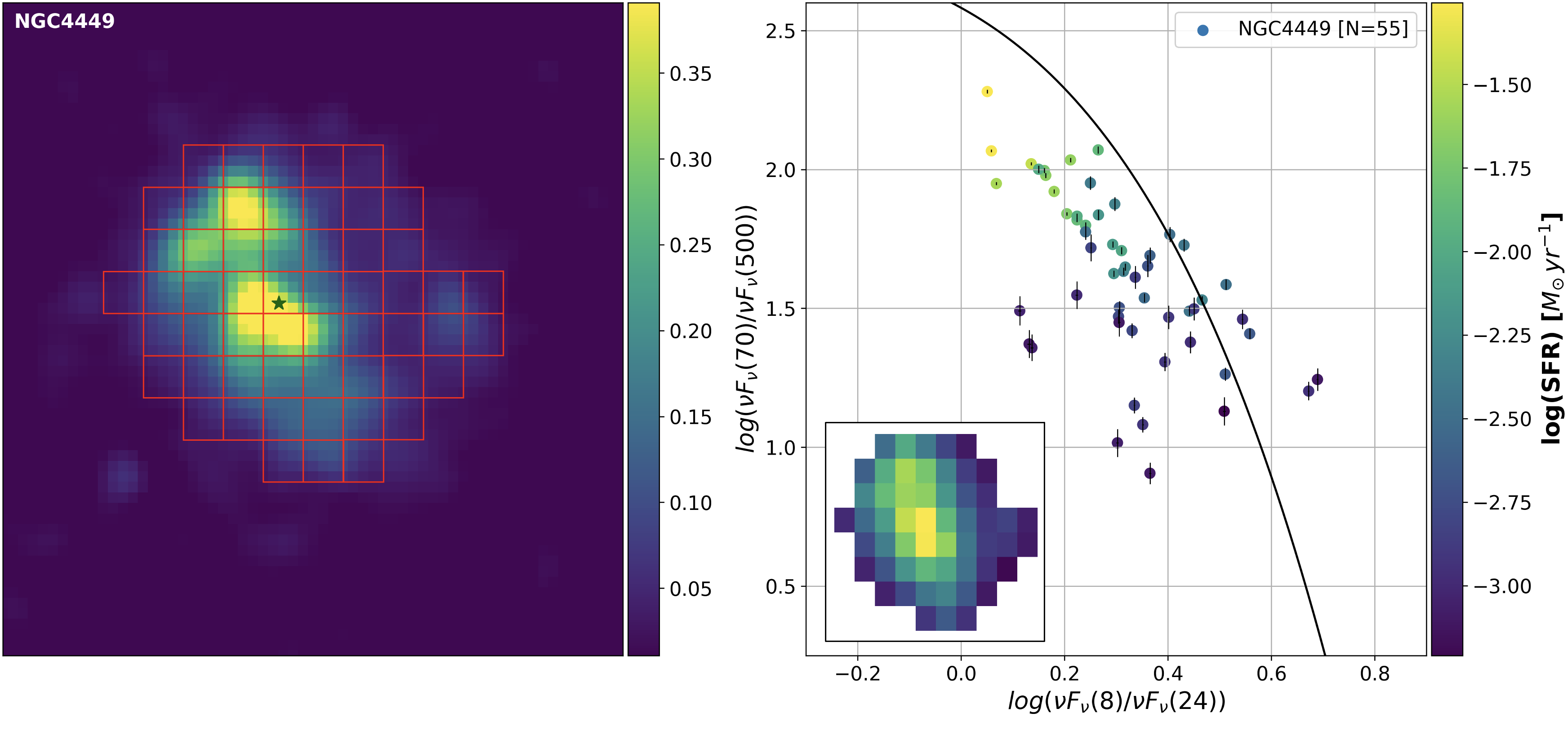}
\includegraphics[width=14.8cm,height=6.4cm]{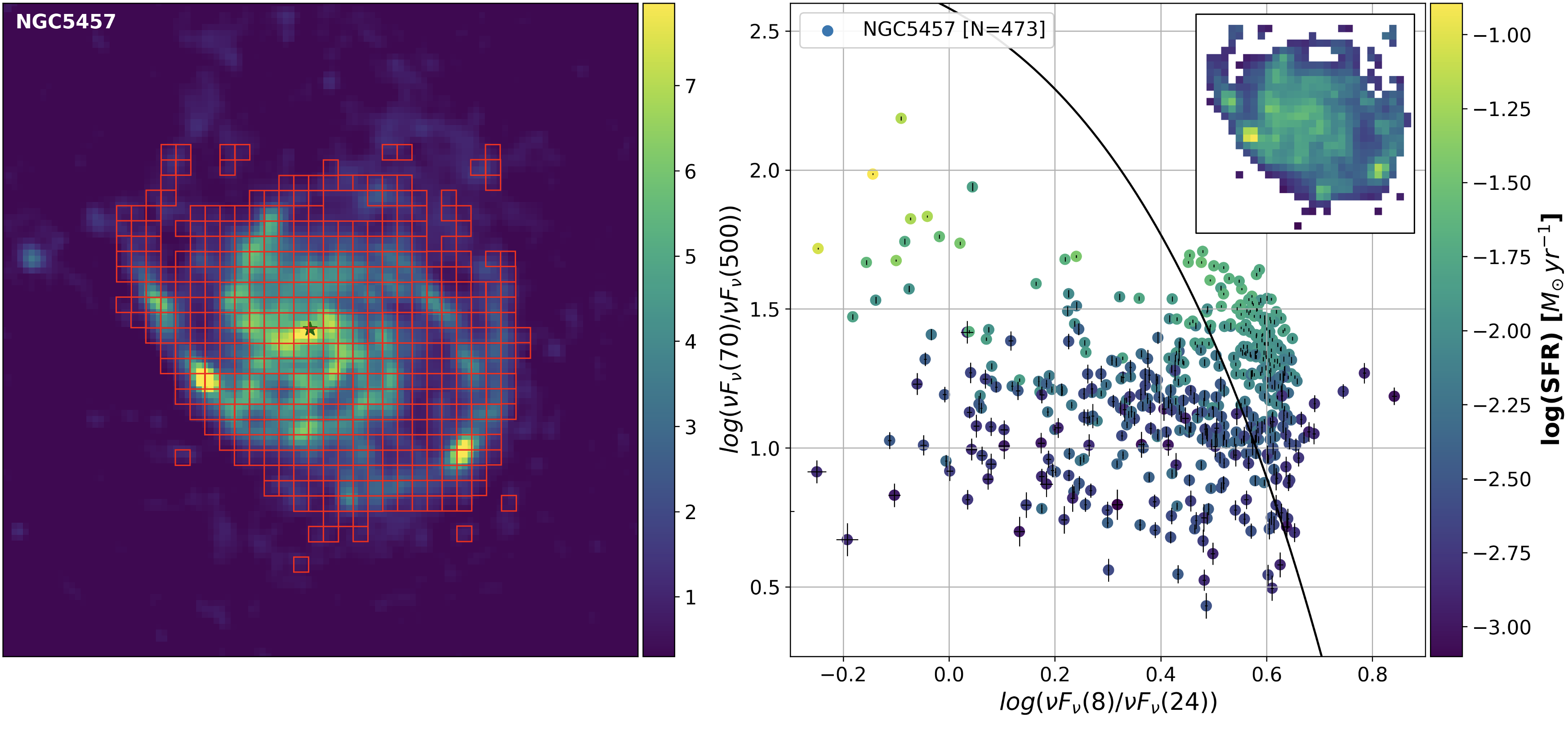}
\caption{ The MIR/FIR color-color relations for the galaxies NGC5236, NGC4449, and NGC5457 (top, middle, bottom). Left panels: the SPIRE 500 $\mu$m image in arbitrary units. The photometric grid with the applied SNR cutoff is over-plotted in red, with 36$''\times$36$''$ individual grid boxes. The star shows the location of the center of the galaxy given by NED. Right panels: The $f_{70}/f_{500}$ versus $f_{8}/f_{24}$ color-color relation for each galaxy at the angular resolution of SPIRE 500 $\mu$m, with regions corresponding to the spatial scales listed in Table \hyperlink{t1}{1}. Plotted are the ratios of the flux densities multiplied by their respective frequencies $\nu$. The data points represent the red regions from the left panel and are color-coded by log($SFR$). The inset images show the spatial distribution of log($SFR$) for each galaxy, color-coded on the same scale as the data points. The black curve shows the relation for NGC4449 found by \cite{2018ApJ...852..106C}, with an applied correction factor from the 1100 $\mu$m flux to the 500 $\mu$m flux given in the legend. The axis ranges between the three scatter plots are matched and centered around the color distribution of NGC5236; this excludes 6 data points at low $f_{8}/f_{24}$ color in NGC5457.} 
\label{fig:f3}
\end{figure*}

With the photometry derived for our entire sample, we investigate the IR color-color relations and the resulting implications. We first focus on three test cases that span some of the diversity of our galaxy sample: NGC5236, NGC4449, and NGC5457. For these three galaxies, the $f_{70}/f_{500}$ versus $f_{8}/f_{24}$ flux density ratios or ``color-color" scatter plots are shown in Figure \ref{fig:f3}. There is a very diverse spread in the correlation between the ``blue" $f_{8}/f_{24}$ and ``red" $f_{70}/f_{500}$ colors within our sample of galaxies. A subset of our galaxies (such as NGC5236) exhibit a very strong anti-correlation between the ``red" and ``blue" IR colors, while other galaxies (NGC4449) exhibit a much weaker color-color trend or no trend at all (NGC5457). 

\begin{figure*}
\centering
\includegraphics[width=.494\textwidth]{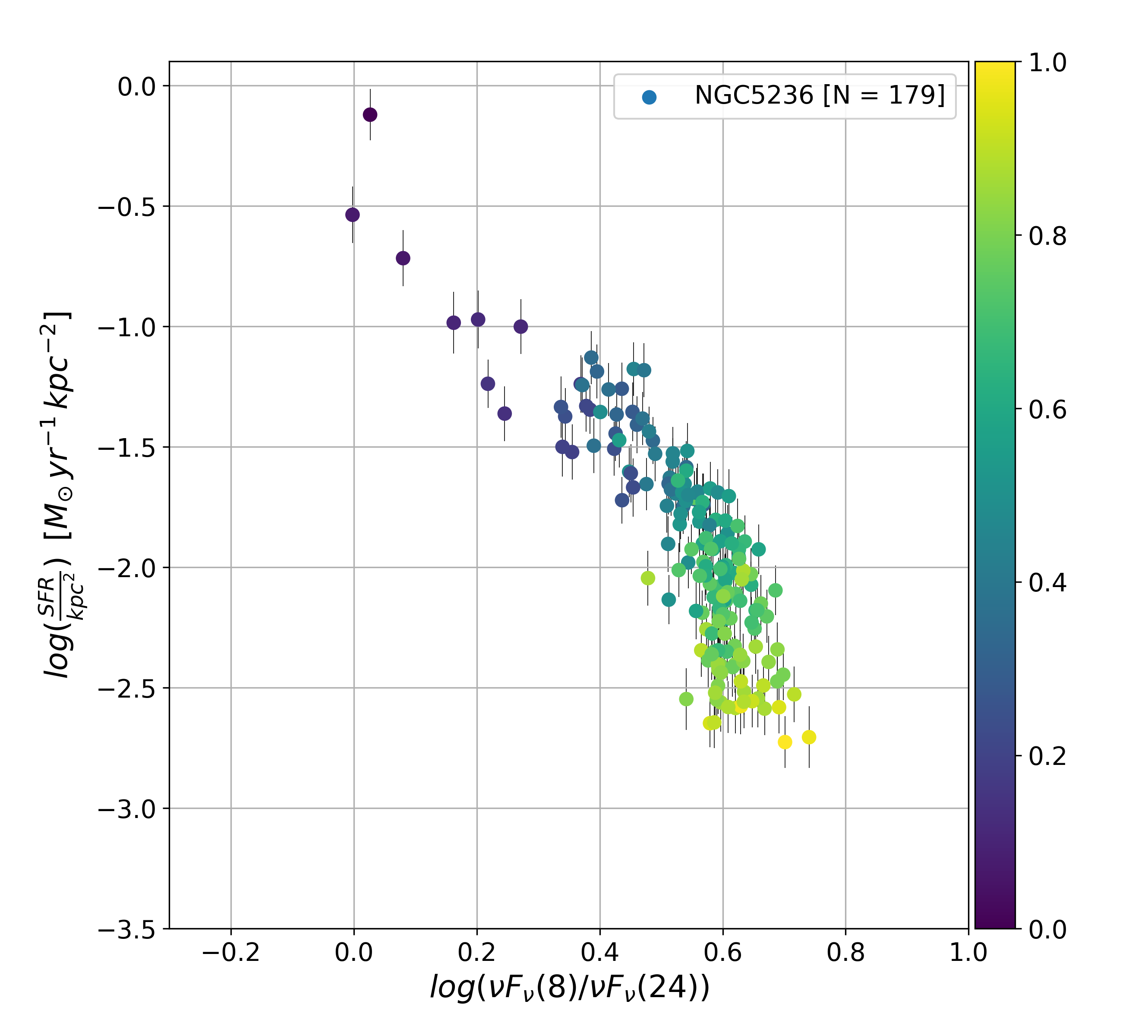}
\includegraphics[width=.494\textwidth]{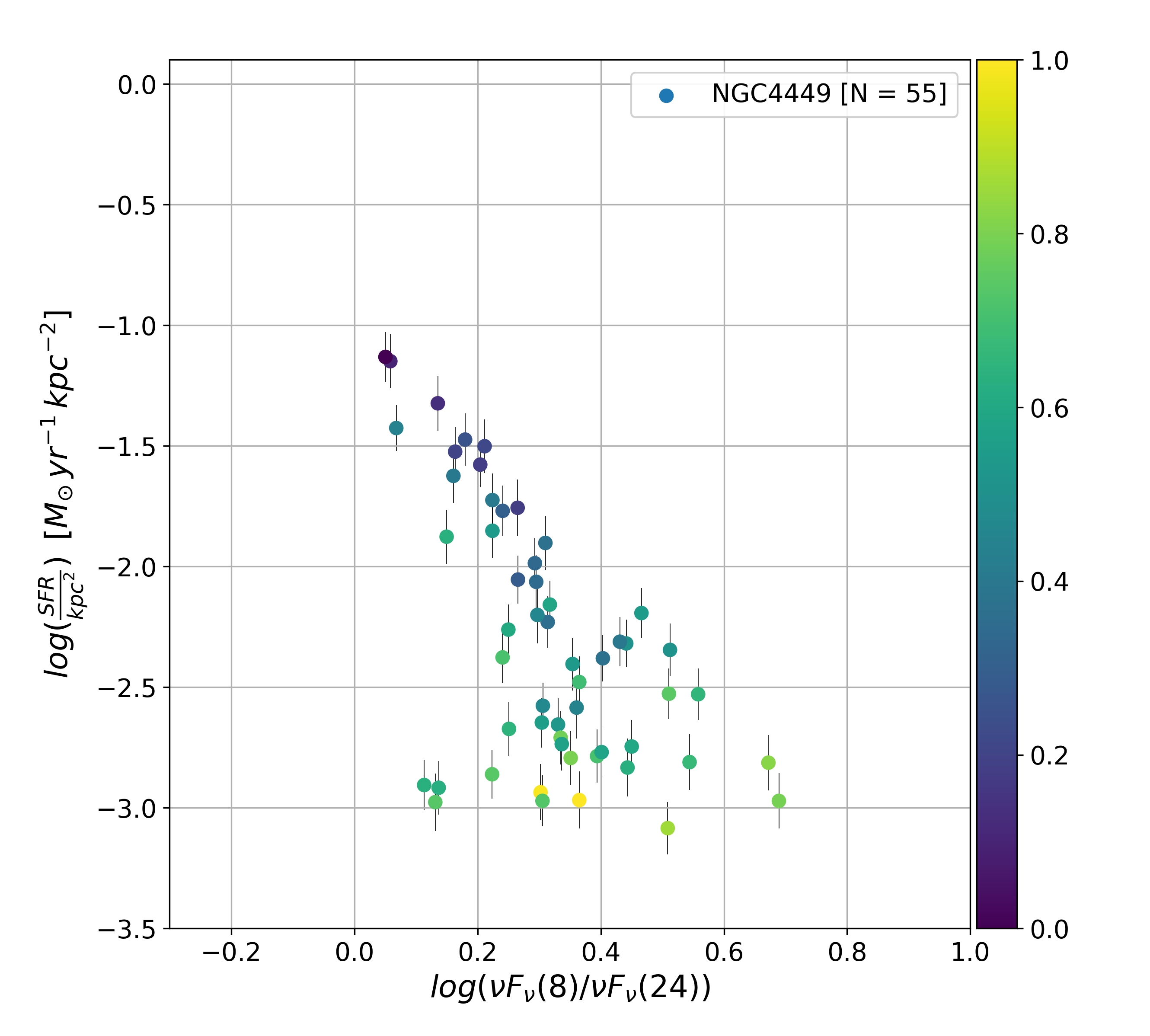}
\includegraphics[width=.494\textwidth]{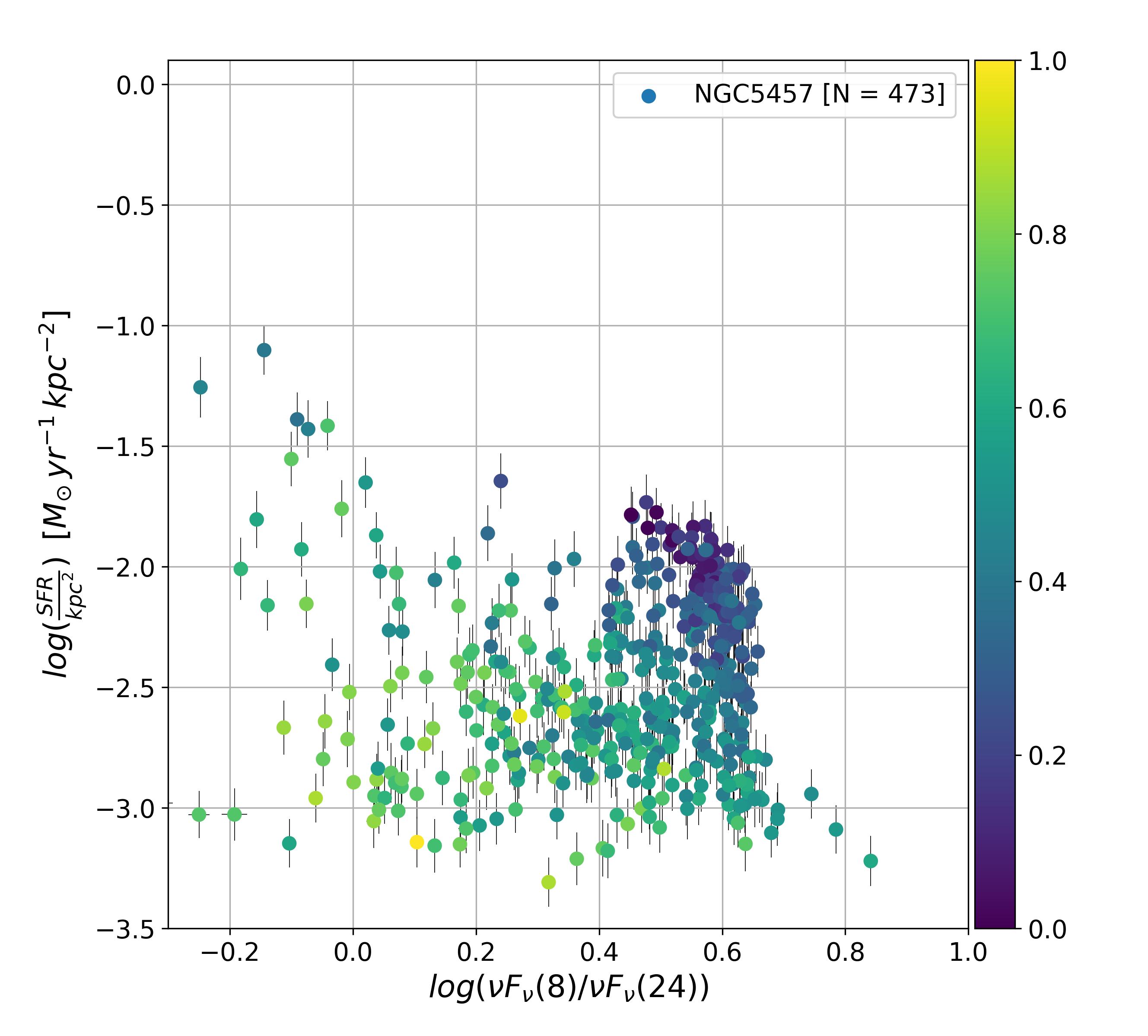}
\caption{The $\Sigma_{SFR}$ (SFR surface density) versus $f_{8}/f_{24}$ color for the three test galaxies: NGC5236, NGC4449, and NGC5457 (left, right, bottom). The points represent the 36$''\times$36$''$ star-forming regions within each galaxy, color-coded by the distance from the galactic center, normalized between 0 and 1. Blue corresponds to central regions while yellow corresponds to the outer edge of the galaxies. Surface densities are corrected for each galaxy's inclination angle. The total number of sub-galactic regions is given in each legend. The axis ranges between the plots are matched and centered around the distribution of NGC5236, excluding 6 data points at low $f_{8}/f_{24}$ color in NGC5457.} 
\label{fig:f4}
\end{figure*}

The individual points in the right panels of Figure \ref{fig:f3} are color-coded by the log($SFR$). It is clear that for NGC5236, both the $f_{8}/f_{24}$ color and the $f_{70}/f_{500}$ color are good tracers of the SFR. For example, regions within NGC5236 with high SFRs trace high $f_{70}/f_{500}$ and low $f_{8}/f_{24}$ color, and are found in the top left of the plot. Conversely, low SFRs trace low $f_{70}/f_{500}$ and high $f_{8}/f_{24}$ color and are found in the bottom right of the plot. This can be understood based on the fact that the 24 and 70 $\mu$m flux are mainly comprised of thermal dust emission heated by massive, recently formed stars, while the 8 and 500 $\mu$m flux receive significant contribution from dust heated by older stellar populations. Yet, this does not appear to be so simple for the other two galaxies.  As the SFR decreases in these two (NGC4449 and NGC5457), the $f_{70}/f_{500}$ color clearly decreases, but the $f_{8}/f_{24}$ color does not increase; only weakly in NGC4449. The $f_{70}/f_{500}$ color consistently appears to be a strong indicator of the SFR in these galaxies, but the $f_{8}/f_{24}$ color does not. These discrepancies are not due to the different spatial scales sampled by our 36$''\times$36$''$ regions across the three galaxies. NGC5236 and NGC4449 are virtually at the same distance, yet they display different trends between their MIR/FIR colors. NGC5457 is about 60-80\% more distant than either galaxy, however the lack of correlation between the two colors is far more extreme than can be accounted for by this difference in distance. See Section \hyperlink{5.3}{5.3} for further discussion.

In Figure \ref{fig:f4}, we plot the $\Sigma_{SFR}$ (SFR surface density or $SFR/kpc^{2}$) versus the ``MIR" $f_{8}/f_{24}$ color. We see that in the case of NGC5236, the $f_{8}/f_{24}$ color closely traces the $\Sigma_{SFR}$. On the contrary, for NGC5457 there is clearly no relation. For NGC4449, it appears that as $\Sigma_{SFR}$ decreases, the $f_{8}/f_{24}$ color ``spreads out" at lower SFRs. NGC4449 appears to be an interesting ``in-between" case, where at higher $\Sigma_{SFR}$ the IR color-color trend appears to hold strong, but not in the lower SFR, outer regions of the galaxy. Keeping in mind that we use the MIPS 24 $\mu$m flux in combination with the FUV to derive the SFR (Section \hyperlink{4.2}{4.2}), $\Sigma_{SFR}$ and the 24 $\mu$m flux must be correlated. As a result, Figure \ref{fig:f4} suggests that the IRAC 8 $\mu$m flux is not a good tracer of the SFR throughout NGC4449 and NGC5457, which contributes to the differences in the observed IR color-color relations.

Tangentially, we also derive the IR color-color relations for these test galaxies by replacing the FIR $f_{70}/f_{500}$ color with $f_{160}/f_{500}$.  Similar to the results of \cite{2018ApJ...852..106C}, we find that galaxies that demonstrate a $f_{70}/f_{500}$ versus $f_{8}/f_{24}$ trend also demonstrate a $f_{160}/f_{500}$ versus $f_{8}/f_{24}$ trend, but that it is generally weaker. This is likely explained by the fact that compared to the 70 $\mu$m, the PACS 160 $\mu$m flux can be more significantly influenced by the underlying stellar population \citep{2010ApJ...714.1256C,2013ApJ...768..180L} and thus is a less direct indicator of the SFR.

\vspace{1mm}
\hypertarget{5.2}{\subsection{Regimes}}

As can be seen from the three example galaxies in Figure \ref{fig:f3}, there appears to be two regimes within our sample of galaxies. There are galaxies in Regime 1 (like NGC5236), where there is an observed anti-correlation between the $f_{70}/f_{500}$ and $f_{8}/f_{24}$ colors. And there are galaxies in Regime 2 (like NGC5457), where there is clearly no trend between their IR colors. We divide our full sample of galaxies into these two regimes and investigate the important differences between them. 

Initially, we visually categorize each galaxy into the regimes. Taking a more rigorous approach, the regimes are also divided based on correlation coefficients. For each galaxy in our sample, the Spearman correlation coefficient ($\rho$) is calculated for both the $f_{70}/f_{500}$ versus $f_{8}/f_{24}$  color-color relation and the $\Sigma_{SFR}$ versus $f_{8}/f_{24}$ color relation, listed in Table \hyperlink{t3}{3}. The Spearman correlation coefficient is a non-parametric measure of the rank correlation of two variables. It is a measure of how well the relationship between two variables can be described by a single monotonic function. We find good agreement between our visual classifications and the classifications based on the correlation coefficients if the regimes are divided at $|\rho|$ = 0.5. For Regime 1, we require that $|\rho| \geq$ 0.5 for both the $f_{70}/f_{500}$ versus $f_{8}/f_{24}$ relation and the $\Sigma_{SFR}$ versus $f_{8}/f_{24}$ relation. Conversely for Regime 2, we require that $|\rho| < $ 0.5 for either of the two relations. See Section \hyperlink{5.2.1}{5.2.1} for an explanation of this cutoff between the regimes. Table \hyperlink{t3}{3} lists the results of the regime classifications for our entire sample of galaxies. A sub-sample of galaxies could not be placed into either regime due to an insufficient number of regions (N $\leq$ 10) meeting the SNR cutoffs and the resulting poorly constrained correlation coefficients. These sources tend to be either Dwarf irregulars, physically small, and/or far away. They are listed under Unclassified in Table \hyperlink{t3}{3}. 

\begin{figure*}
\centering
\includegraphics[width=15.5cm]{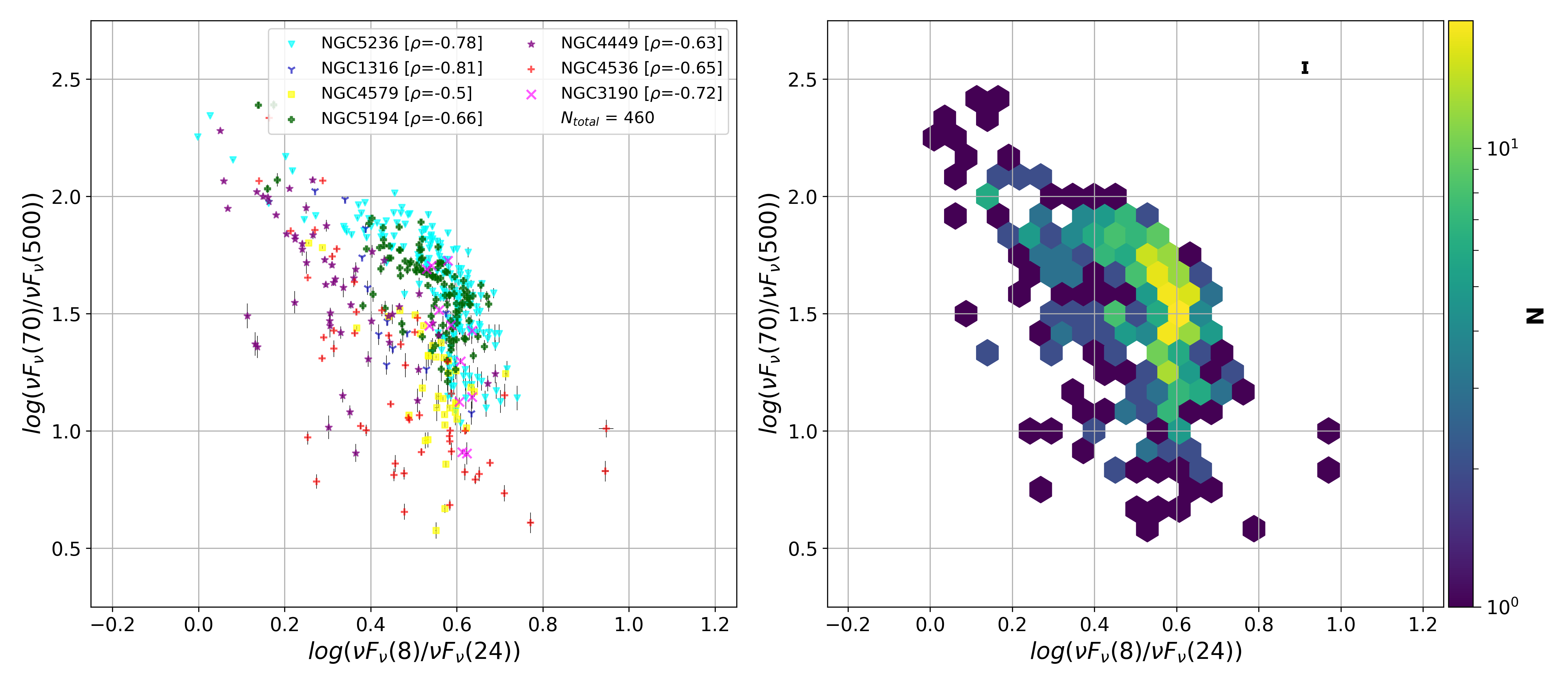}
\includegraphics[width=15.5cm]{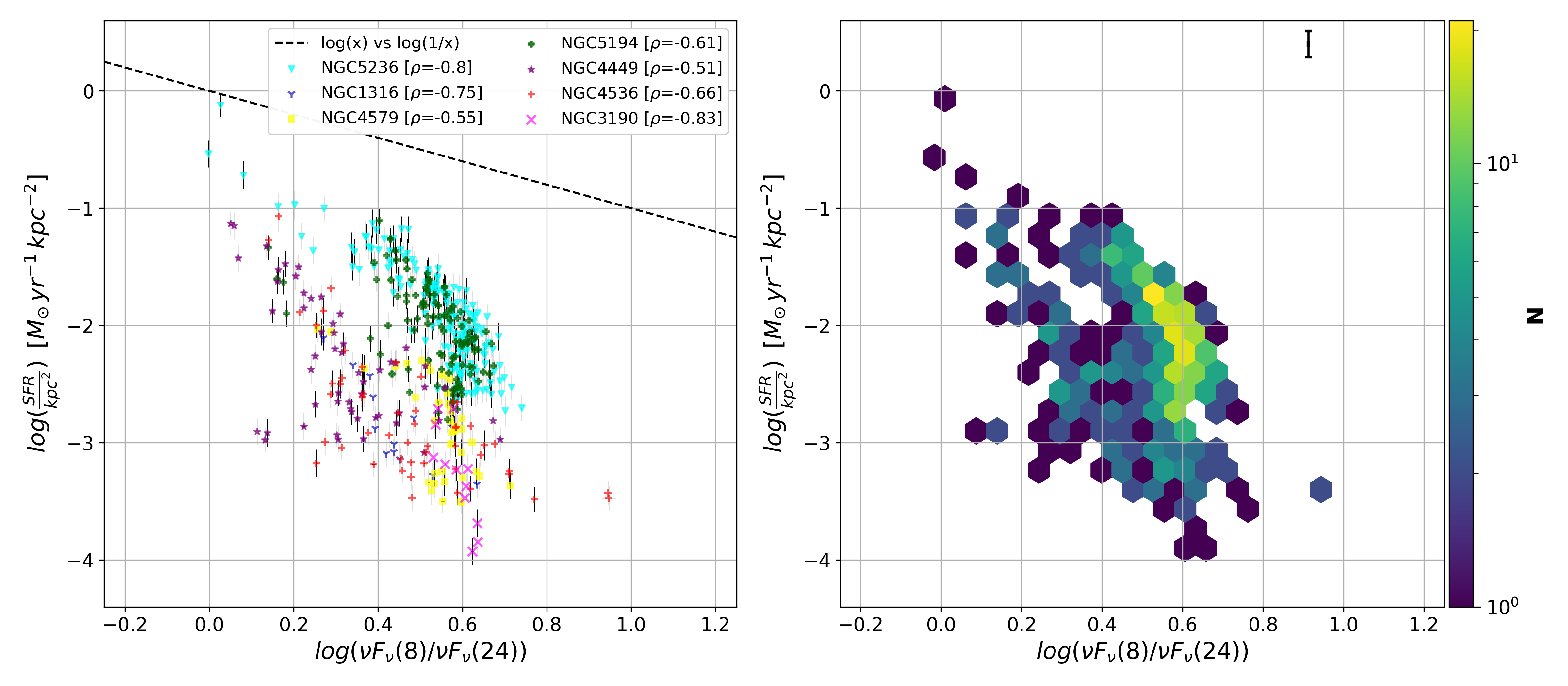}
\includegraphics[width=15.5cm]{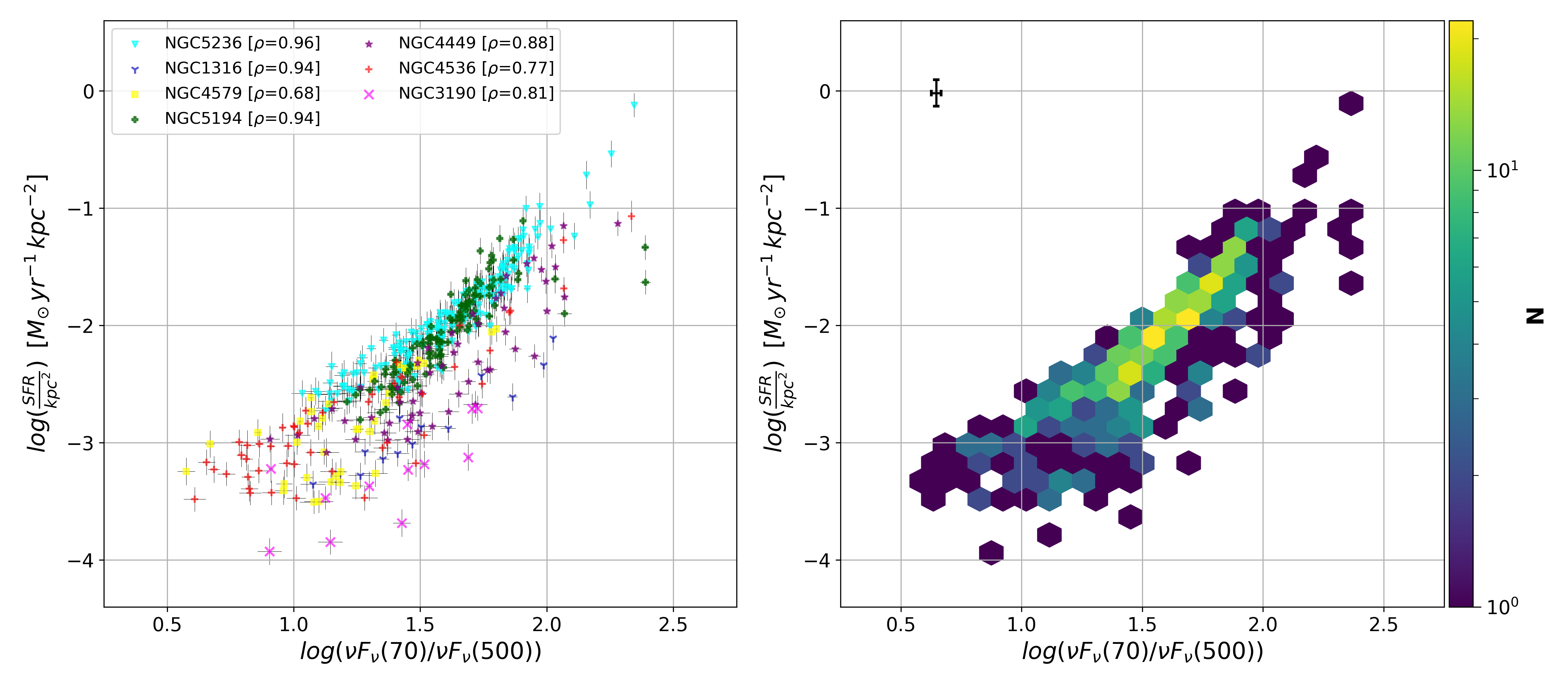}
\caption{ \textit{Regime 1}. Shows all galaxies placed in Regime 1. Left: The points with various colors and shapes correspond to the 36$''\times$36$''$ regions within the different galaxies at the angular resolution of SPIRE 500 $\mu$m. Right: 2D hexagonal binning showing the number density of the points in the left plots. The black brackets show the mean error in all regions. Top panel: the $f_{70}/f_{500}$ versus $f_{8}/f_{24}$ color-color relation. Middle panel: $\Sigma_{SFR}$ versus $f_{8}/f_{24}$ color. Plotted in black dashes is the log($x$) versus log($1/x$) relation, showing that the trends are more than just the trivial relation between the 24 $\mu$m and the SFR. Bottom panel: $\Sigma_{SFR}$ versus $f_{70}/f_{500}$ color. In the legend, $\rho$ refers to the value of the Spearman correlation coefficient calculated for each galaxy/relation and $N_{total}$ is the total number of regions. }
\label{fig:f5}
\end{figure*}

\begin{figure*}
\centering
\includegraphics[width=15.5cm]{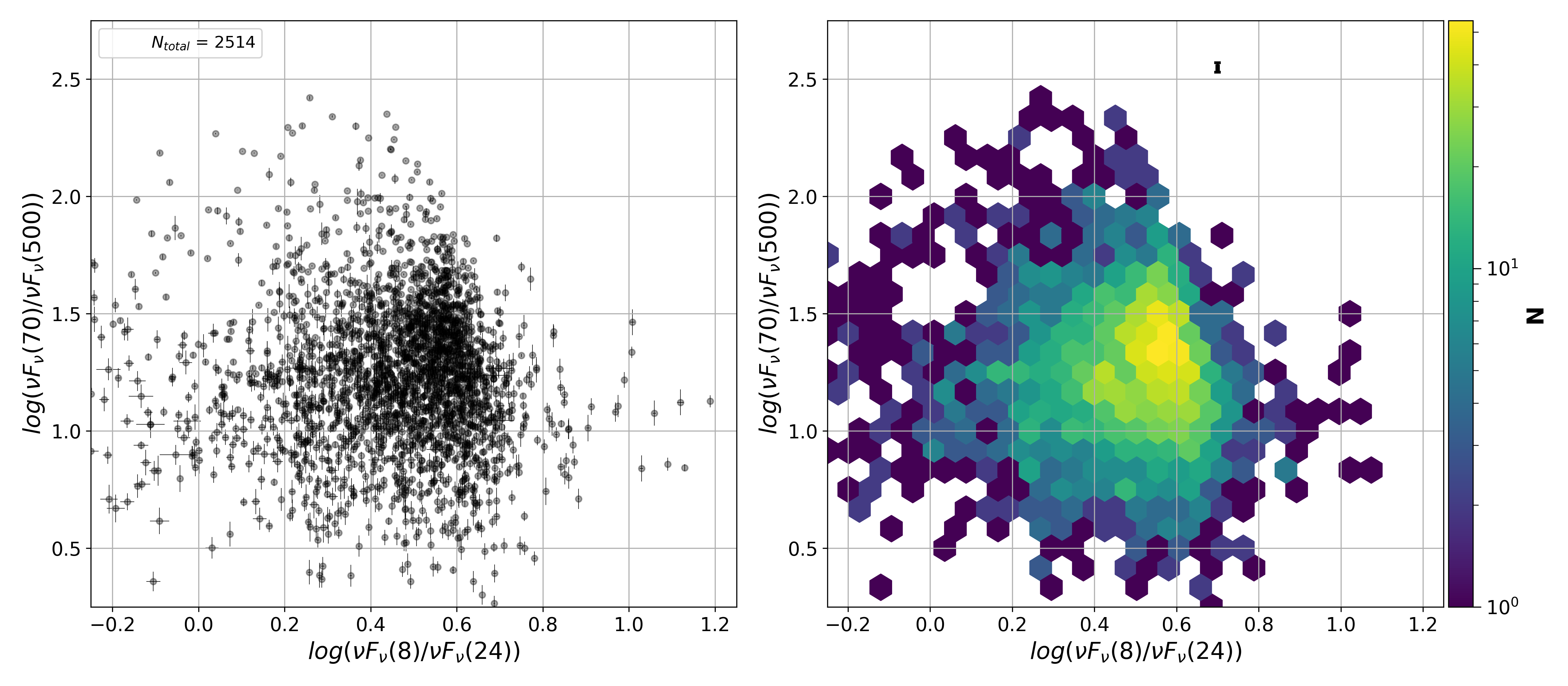}
\includegraphics[width=15.5cm]{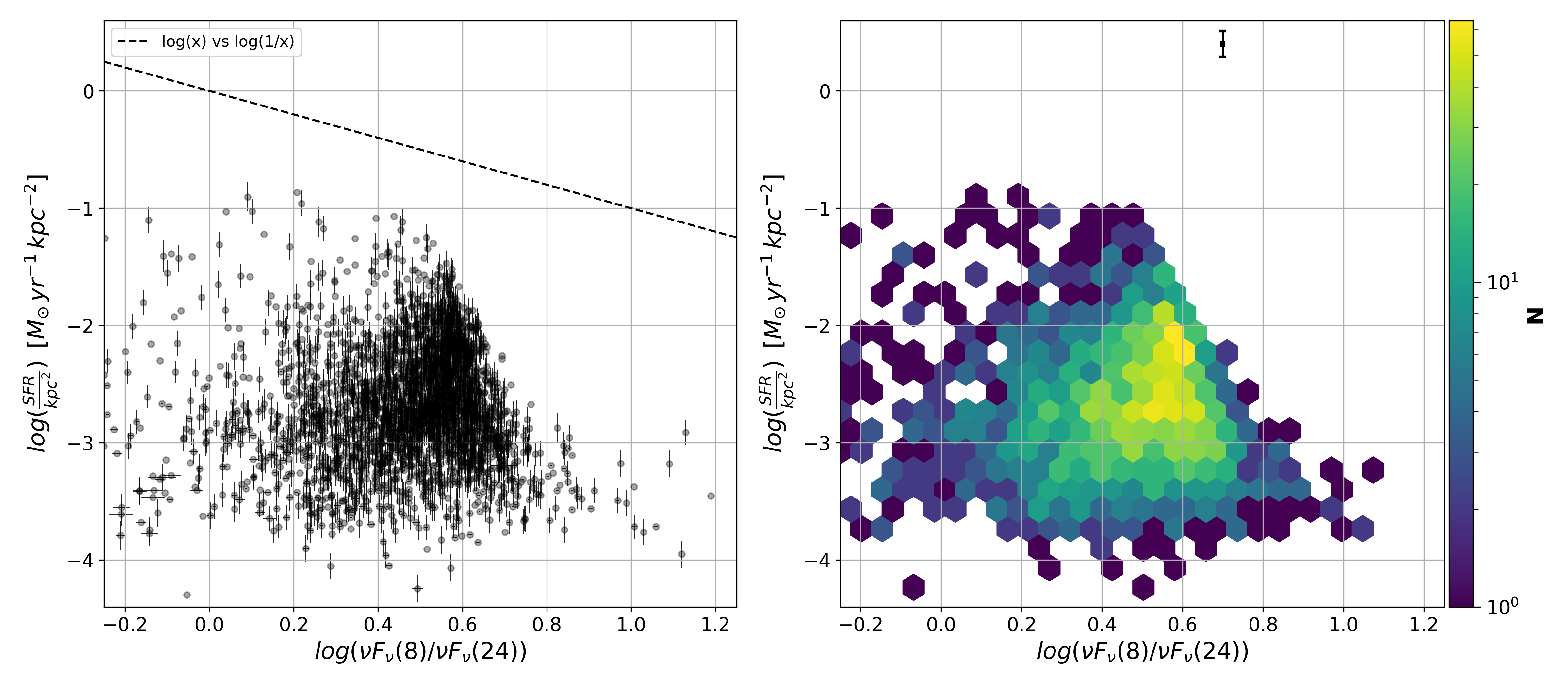}
\includegraphics[width=15.5cm]{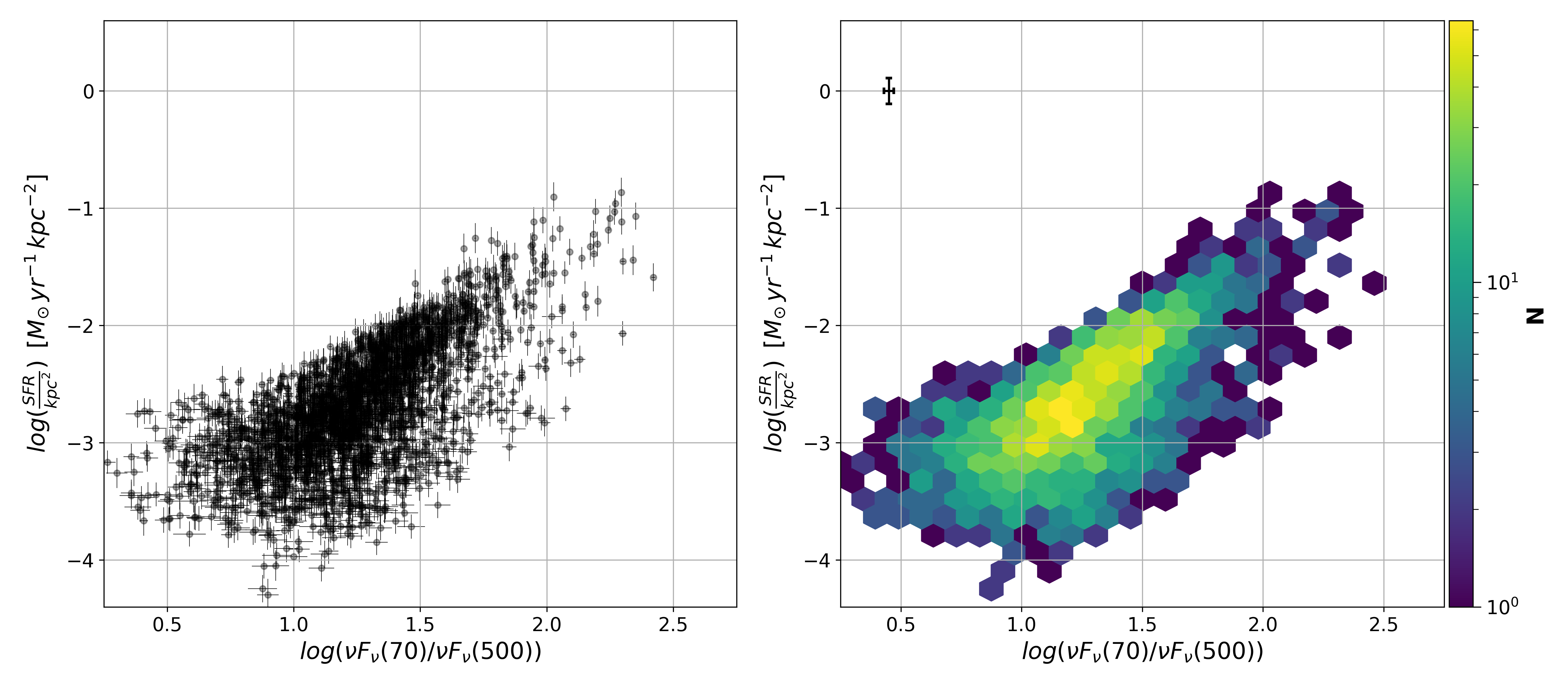}
\caption{ \textit{Regime 2}. Left: The black points correspond to the 36$''\times$36$''$ regions within the different Regime 2 galaxies. Right: 2D hexagonal binning showing the number density of the points in the left plots. The black brackets show the mean error in all regions. Top panel: the $f_{70}/f_{500}$ versus $f_{8}/f_{24}$ color-color relation. Middle panel: $\Sigma_{SFR}$ versus $f_{8}/f_{24}$ color. Plotted in black dashes is the log($x$) versus log($1/x$) relation. Bottom panel: $\Sigma_{SFR}$  versus $f_{70}/f_{500}$ color. The axes are matched to Figure \ref{fig:f5} to allow for a direct visual comparison with Regime 1. Note that even for the numerous galaxies/environments in this regime, there is a clear overall relation between $\Sigma_{SFR}$ and the $f_{70}/f_{500}$ color. }
\label{fig:f6}
\end{figure*}

With our sample divided into two regimes in this way, we compare the differences in the observed correlations. Figure \ref{fig:f5} shows the $f_{70}/f_{500}$ versus $f_{8}/f_{24}$ color-color relation, the $\Sigma_{SFR}$ versus $f_{8}/f_{24}$ relation and the $\Sigma_{SFR}$ versus $f_{70}/f_{500}$ relation for all galaxies in Regime 1. Figure \ref{fig:f6} shows the same relations but for all galaxies within Regime 2. The stark difference between the two regimes can clearly be seen in Figures \ref{fig:f5} and \ref{fig:f6}. 

In the top panel of Figure \ref{fig:f5}, we see a clear overall relation between the colors for Regime 1 galaxies that begins to ``turn over" at lower $f_{70}/f_{500}$ colors (log($\nu f_{\nu}(70)/ \nu f_{\nu}(500)$)$\lesssim 1.5$). Due to the strong positive correlation between $\Sigma_{SFR}$ and $f_{70}/f_{500}$ shown in the bottom panel of Figure \ref{fig:f5}, this suggests a slight ``turnover" in the color-color relation for lower $\Sigma_{SFR}$; seen in the middle panel of Figure \ref{fig:f5} at log($\Sigma_{SFR}$)$\lesssim -2.0$. The log($x$) versus log($1/x$) relation in Figure \ref{fig:f5} demonstrates that the $\Sigma_{SFR}$ versus $f_{8}/f_{24}$ relation is more than just simply the direct relation between the SFR and the 24 $\mu$m flux used to calculate it. 

In contrast to Regime 1, there is no net color-color relation in Regime 2 galaxies, shown by the overall vertical trend in the top panel of Figure \ref{fig:f6}. Yet for Regime 2, there is still a strong overall correlation between $\Sigma_{SFR}$ and $f_{70}/f_{500}$ (bottom panel, Figure \ref{fig:f6}). This suggests that the $f_{70}/f_{500}$ color is universally a strong indicator of the SFR, across our whole sample of galaxies. On the contrary, the $f_{8}/f_{24}$ color is only a good SFR indicator in a select few cases.

\begin{figure}
\centering
\includegraphics[width=7.4cm]{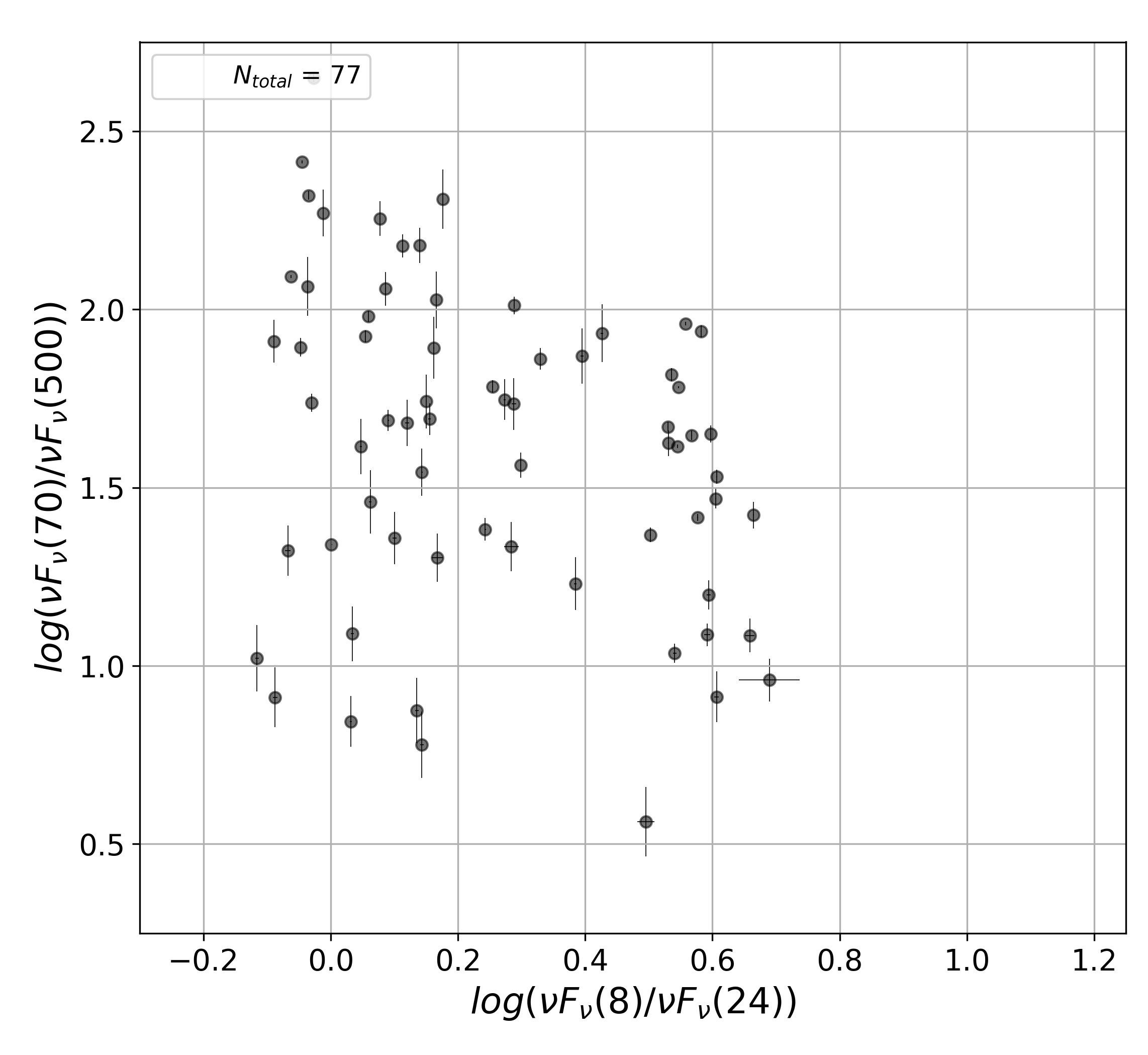}
\includegraphics[width=7.4cm]{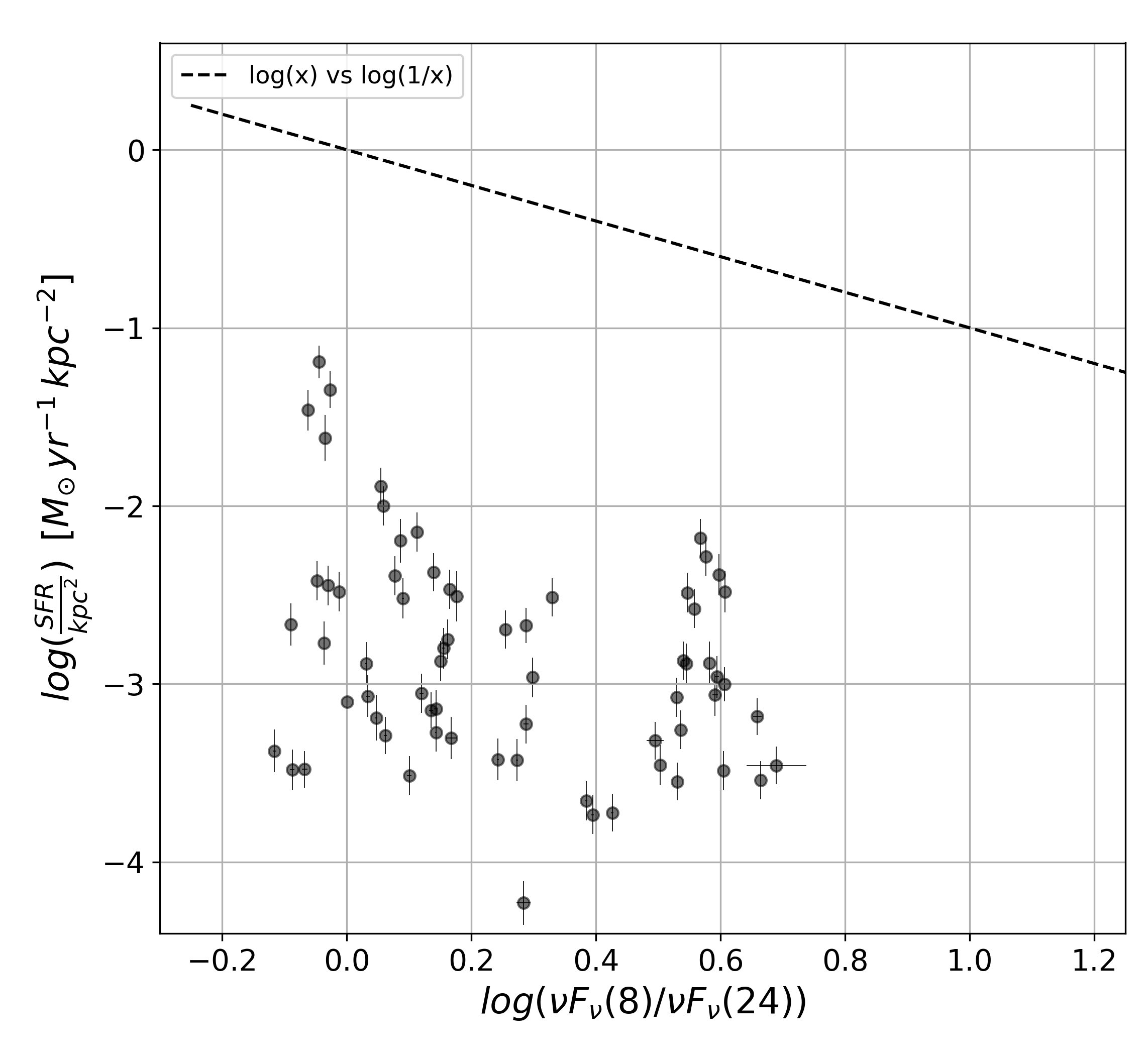}
\includegraphics[width=7.4cm]{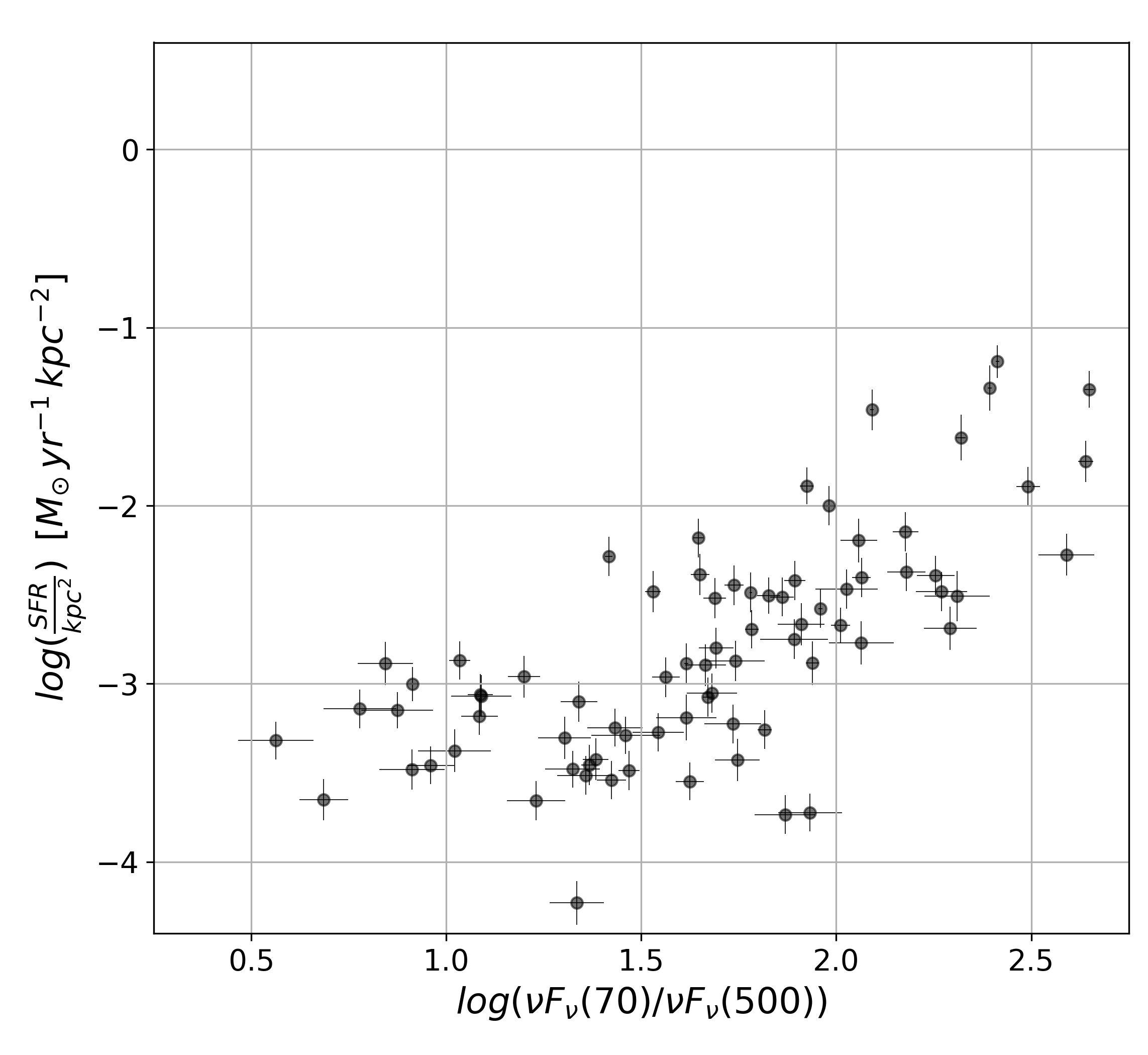}
\caption{ \textit{Unclassified Sources}. The points correspond to the 36$''\times$36$''$ regions within the Unclassified sources listed in Table \hyperlink{t3}{3}, for a relaxed SNR $\geq5.0$ cutoff in the four IR bands.  See Figures \ref{fig:f5} and \ref{fig:f6} for a description of the scatter plots. The axes are matched to Figure \ref{fig:f5}.
}
\label{fig:fig_other}
\end{figure}

We plot the trends for the Unclassified sources in Figure \ref{fig:fig_other}. For these sources, the initial strict SNR cutoffs are relaxed to maximize the number of regions. We require that SNR $\geq$ 5.0 for 8, 24, 70, and 500 $\mu m$. There are only a small number of regions within these galaxies that meet the relaxed criteria (N = 77). Similar to Regime 2 galaxies, this subset of sources exhibits no clear overall relation between the MIR and FIR colors and between $\Sigma_{SFR}$ and $f_{8}/f_{24}$, yet demonstrates a relatively strong correlation between $\Sigma_{SFR}$ and $f_{70}/f_{500}$. Although, given the small number of points and lower thresholds, this remains somewhat uncertain and we are unable to individually characterize these sources.

\begin{table}
\centering
\caption{\hypertarget{t3}{ Regimes}} 
\begin{center}
\hspace{-9mm} 
\begin{tabular}{ c  r  c  r  c }
\hline
\hline
\rule{0pt}{3ex}
Galaxy & $\rho_{1}$ $^{\mbox{\textit{a}}}$ & $P_{1}$ $^{\mbox{\textit{b}}}$ & $\rho_{2}$ $^{\mbox{\textit{c}}}$ & $P_{2}$ $^{\mbox{\textit{d}}}$ \\
\hline
\rule{0pt}{3ex}
& & Regime 1 $^{\mbox{\textit{e}}}$& & \\
& & [N=7] & & \\
\hline
\rule{0pt}{3ex}
NGC5236 & -0.784 & 1.69e-38 & -0.800 & 3.32e-41\\
NGC1316 & -0.808 & 8.39e-04 & -0.753 & 2.98e-03\\
NGC4579 & -0.502 & 1.32e-03 & -0.551 & 3.33e-04\\
NGC5194 & -0.664 & 1.45e-15 & -0.606 & 1.40e-12\\
NGC4449 & -0.630 & 2.49e-07 & -0.508 & 7.54e-05\\
NGC4536 & -0.650 & 2.45e-07 & -0.665 & 1.02e-07\\
NGC3190 & -0.720 & 8.24e-03 & -0.832 & 7.85e-04\\
\hline
\rule{0pt}{3ex}
& & Regime 2 $^{\mbox{\textit{f}}}$ & &\\
& & [N=34] & &\\
\hline
\rule{0pt}{3ex}
IC2574 & -0.608 & 1.63e-03 & -0.434 & 3.41e-02\\
NGC4236 & -0.502 & 2.36e-05 & -0.451 & 1.84e-04\\
NGC5713 & -0.091 & 7.29e-01 & 0.147 & 5.73e-01\\
NGC5457 & 0.054 & 2.42e-01 & 0.149 & 1.12e-03\\
NGC3938 & 0.282 & 5.45e-02 & -0.002 & 9.90e-01\\
NGC4254 & -0.407 & 1.02e-03 & -0.649 & 1.19e-08\\
NGC3351 & -0.255 & 4.92e-02 & -0.269 & 3.73e-02\\
NGC0628 & 0.194 & 2.13e-02 & 0.407 & 5.89e-07\\
NGC5474 & 0.463 & 7.62e-03 & 0.723 & 2.92e-06\\
NGC7793 & 0.255 & 7.65e-03 & 0.248 & 9.57e-03\\
NGC4736 & -0.355 & 9.14e-03 & -0.335 & 1.41e-02\\
NGC4321 & -0.360 & 1.10e-03 & -0.658 & 4.53e-11\\
NGC1291 & 0.272 & 2.10e-01 & 0.558 & 5.63e-03\\
NGC2403 & 0.108 & 2.14e-01 & 0.043 & 6.21e-01\\
NGC1482 & 0.491 & 1.50e-01 & 0.152 & 6.76e-01\\
NGC5055 & -0.185 & 3.68e-02 & -0.137 & 1.22e-01\\
NGC4826 & 0.391 & 2.98e-02 & 0.209 & 2.59e-01\\
NGC0337 & 0.025 & 9.30e-01 & 0.089 & 7.52e-01\\
NGC4594 & 0.094 & 5.64e-01 & 0.045 & 7.81e-01\\
NGC3627 & -0.462 & 1.58e-04 & -0.191 & 1.37e-01\\
NGC1097 & -0.368 & 1.66e-04 & -0.154 & 1.27e-01\\
NGC4725 & -0.450 & 2.55e-05 & -0.707 & 1.67e-13\\
NGC3521 & -0.111 & 3.26e-01 & -0.186 & 9.77e-02\\
NGC3621 & -0.227 & 3.05e-02 & 0.257 & 1.41e-02\\
NGC7331 & -0.129 & 2.40e-01 & 0.470 & 5.55e-06\\
NGC4559 & 0.012 & 9.26e-01 & 0.214 & 8.93e-02\\
NGC1512 & -0.152 & 3.69e-01 & -0.054 & 7.52e-01\\
NGC2841 & -0.081 & 5.63e-01 & -0.153 & 2.73e-01\\
NGC3049 & 0.266 & 4.04e-01 & 0.371 & 2.36e-01\\
NGC0925 & 0.459 & 1.02e-05 & 0.470 & 5.56e-06\\
NGC4569 & -0.285 & 1.77e-01 & -0.503 & 1.23e-02\\
NGC2976 & 0.217 & 2.17e-01 & 0.223 & 2.06e-01\\
NGC3198 & -0.047 & 7.39e-01 & 0.233 & 9.28e-02\\
NGC4631 & 0.393 & 2.19e-05 & 0.296 & 1.67e-03 \\

\hline
\hline
\rule{0pt}{3ex}
& & Unclassified $^{\mbox{\textit{g}}}$& &\\
& & [N=15] & &\\
\hline
\rule{0pt}{3ex}
NGC3773 & &
NGC4625 & &
NGC1266 \\
NGC3265 & &
NGC2798 & &
NGC0855 \\
NGC5866 & &
NGC1404 & &
HoII \\
DDO053 & &
NGC2915 & &
HoI \\
M81DwB & &
DDO154 & &
DDO165 \\
\hline
\end{tabular}
\end{center}
\begin{flushleft} 
\currtabletypesize{\sc Note}--- \\
\rule{0pt}{3ex}
$^{\mbox{\textit{a}}}$ Spearman correlation coefficient ($\rho_{1}$) for the $f_{70}/f_{500}$ versus $f_{8}/f_{24}$ color-color relation.\\
$^{\mbox{\textit{b}}}$ The p-value corresponding to $\rho_{1}$ for a ``two-sided" alternative hypothesis. Roughly, the probability of an uncorrelated system producing datasets with a correlation at least as strong as the one measured.\\
$^{\mbox{\textit{c}}}$  $\rho_{2}$ for the $\Sigma_{SFR}$ versus $f_{8}/f_{24}$ relation.\\
$^{\mbox{\textit{d}}}$ The p-value corresponding to $\rho_{2}$.\\
$^{\mbox{\textit{e}}}$ Regime 1. Require $|\rho| \geq$ 0.5 for both the $f_{70}/f_{500}$ versus $f_{8}/f_{24}$ relation and the $\Sigma_{SFR}$ versus $f_{8}/f_{24}$ relation.\\
$^{\mbox{\textit{f}}}$ Regime 2. Require $|\rho| < $ 0.5 for either of the relations.\\
$^{\mbox{\textit{g}}}$ Galaxies unable to be put in either Regime due to an insufficient number of regions (N$\leq10$) meeting the initial SNR cutoffs. \\
\vspace{1mm}
\end{flushleft}
\end{table}

\begin{figure*}
\centering
\includegraphics[width=18cm]{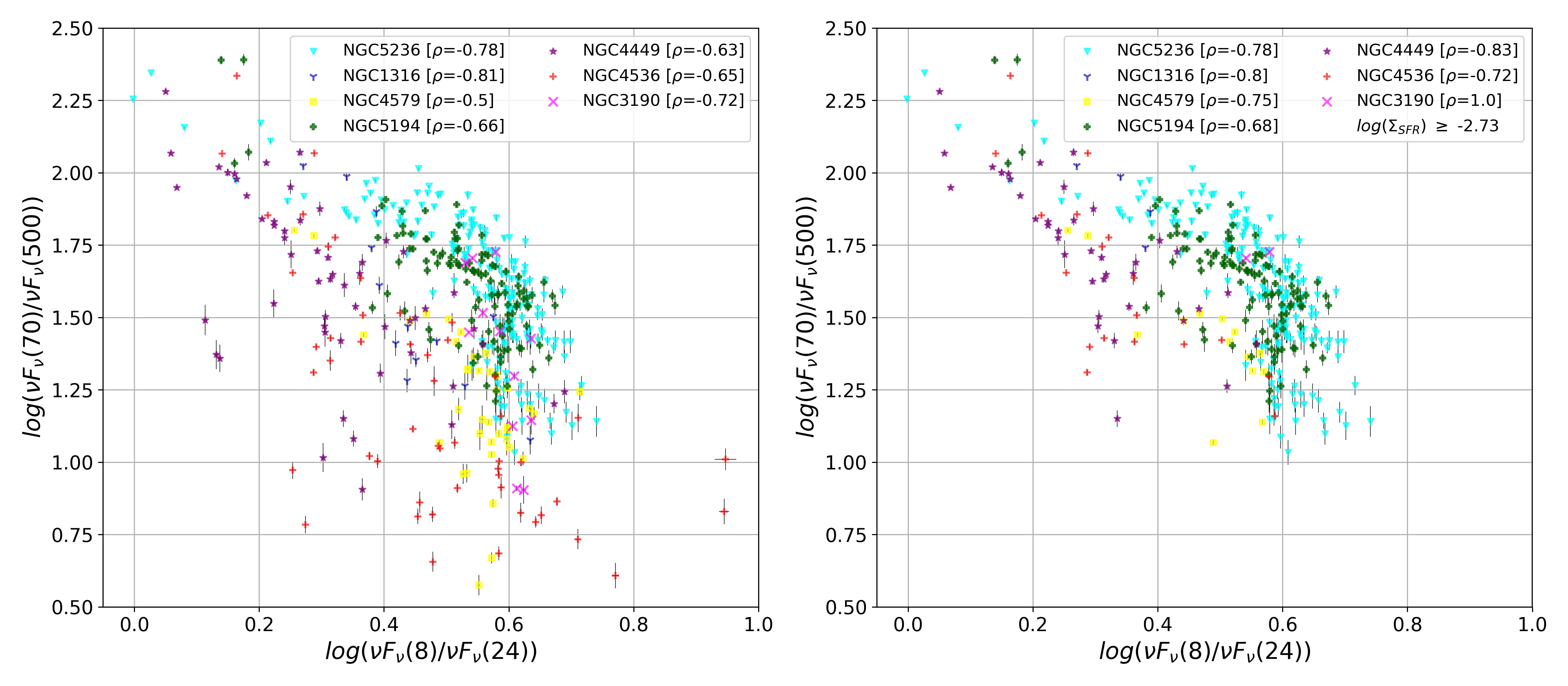}
\caption{SFR surface density cutoff. Shows a comparison of the $f_{70}/f_{500}$ versus $f_{8}/f_{24}$ color-color relation for regions in Regime 1 galaxies before a log($\Sigma_{SFR}$)$\geq -2.73 \,\, M_{\odot} yr^{-1} kpc^{-2}$ cutoff (Left panel) and after the cutoff (Right panel). This cutoff corresponds to the minimum $\Sigma_{SFR}$ measured in NGC5236. The updated correlation coefficients $\rho$ are shown in the legend of the right panel. Note the generally stronger correlation between the IR colors after the cutoff. }
\label{fig:f7}
\end{figure*}

 To investigate the apparent ``turnover" in the color-color trend at low SFR surface density for Regime 1 galaxies, we apply a minimum log($\Sigma_{SFR}$)$\geq -2.73 \,\, M_{\odot} yr^{-1} kpc^{-2}$ cutoff to all regions, which corresponds to the minimum value of $\Sigma_{SFR}$ for NGC5236. NGC5236 shows strong IR color-color correlation throughout its regions and thus should provide a reasonable minimum $\Sigma_{SFR}$ for correlation in the other galaxies. Figure \ref{fig:f7} shows the color-color relations for Regime 1 before and after the $\Sigma_{SFR}$ cutoff, along with the updated correlation coefficients. 
 It can clearly be seen in Figure \ref{fig:f7} that the limitation to high $\Sigma_{SFR}$ regions noticeably strengthens the color-color relations within these galaxies. Specifically, note the large change in the correlation coefficients $\rho$ for NGC4449 (-0.63 to -0.83) and for NGC4579 (-0.5 to -0.75). These results suggest that galaxies that are uniformly dominated by high levels of star formation (like NGC5236) will tend to demonstrate a strong $f_{70}/f_{500}$ versus $f_{8}/f_{24}$ color-color relation over a relatively large range in IR colors. On the contrary, galaxies with non-uniform or lower levels of star formation (NGC5457) will tend to demonstrate weaker color-color trends, as differences in ISM composition and dust heating by the underlying diffuse stellar population become important.

\hypertarget{5.2.1}{\subsubsection{Separating the Regimes}}
As previously discussed, the two regimes in our sample of galaxies are separated based on correlation coefficients. We require for Regime 1 galaxies that $|\rho| \geq$ 0.5 for both the $f_{70}/f_{500}$ versus $f_{8}/f_{24}$ relation and the $\Sigma_{SFR}$ versus $f_{8}/f_{24}$ relation. For Regime 2, we require that $|\rho| < $ 0.5 for either relation. The adopted cutoff between the regimes at $|\rho| =$ 0.5 is a particularly important assumption and will have a large impact on our regime classifications. 

There are few important reasons that we have settled on the cutoff at $|\rho| =$ 0.5. Most crucial, we make the assumption that NGC4449 marks the transition between the two regimes. This assumption is based off Figures \ref{fig:f3} and \ref{fig:f4}, which show that the $f_{8}/f_{24}$ color in NGC4449 remains tightly connected to the SFR surface density only in the central regions of the galaxy. This suggests that in NGC4449, the IR colors are only strongly correlated in the high SFR surface density, central regions; in contrast to NGC5236, which exhibits a strong relation across the vast majority of its regions, and NGC5457, which shows little to no relation in any regions. From Table \hyperlink{t3}{3}, NGC4449 has measured correlation coefficients of $\rho_{1}=0.63$ and  $\rho_{2}=0.51$. A correlation coefficient cutoff higher than $|\rho| =$ 0.5 would place NGC4449 in Regime 2. A lower cutoff (e.g. $|\rho| =$ 0.4) could be reasonable, but would add more confusion into the regime separation, specifically when comparing with expectations from visual classification. The cutoff at $|\rho| =$ 0.5 is found to agree well with visual classification; all galaxies classified as Regime 1 by the cutoff were also first visually classified as Regime 1. In this sense, the adopted correlation coefficient cutoff is strict for Regime 1, ensuring that all galaxies placed in this regime demonstrate strong IR color-color correlation across a significant fraction of their star-forming regions. 

Yet, the correlation coefficient cutoff applied to define the regimes is imperfect. There are a few galaxies in our sample that may be misclassified using this method. Most notably, the Regime 1 galaxy NGC3190 has high correlation coefficients ($\rho_{1}=-0.72$ and $\rho_{2}=-0.83$; Table \hyperlink{t3}{3}), but has a very limited range in the $f_{8}/f_{24}$ color ($\sim$0.1 dex; Figure \ref{fig:f5}). Compared with the other Regime 1 galaxies in Figure \ref{fig:f5}, the color-color trend of NGC3190 is near vertical, more like a typical Regime 2 galaxy. NGC3190 highlights a potential limitation of the classifications based on correlation coefficients and may be a misclassified Regime 2 galaxy. There are two galaxies, NGC4236 and NGC4725, that are classified as Regime 2 by the correlation coefficients but are borderline Regime 1, demonstrating moderate IR color-color correlation across a large number of regions. Table \hyperlink{t3}{3} shows both galaxies have correlation coefficients near the regime cutoff ($\rho_{1}$ and $\rho_{2} \leq -0.45$) at high significance levels (low p-values $\lesssim 10^{-4}$). These galaxies may better represent ``in-between" cases, more like NGC4449, where there are a mix of regions that do and do not exhibit the strong IR color-color correlation. In addition, the observed turnover in the color-color trends of Regime 1 galaxies somewhat complicates the interpretation of the correlation coefficients as the discriminant between the regimes, generally lowering the measured correlations.

\hypertarget{5.3}{\subsection{Distance Effects}}

Given that our galaxies are located over a range of distances ($\sim$3--$25 \, Mpc$), we may expect an associated effect on the observed IR color-color correlations. Galaxies that lie at larger distances will have $36''$ regions that correspond to larger physical scales. Over the distance range of our sample, this is a factor of $\sim$5 variation in physical scale (Table \hyperlink{t1}{1}). The larger physical scales in more distant galaxies will tend to average over features, such as strong spiral arm and interarm regions, leading to derived SFRs that are more uniform across the disk and thus weaker correlation between their IR colors.

Previously, we observed in Figure \ref{fig:f3} that the Regime 1 galaxies NGC5236 and NGC4449, which lie at similar distances, display very different trends and correlation strength between their IR colors. In addition, Figure \ref{fig:f5} demonstrates that NGC5194 and NGC4536, which lie at 8.6 and $15.3\, Mpc$ respectively (a factor of 1.2 and 2.1 times further than the quintessential Regime 2 galaxy NGC5457), both exhibit strong correlation between their IR colors ($|\rho| \geq 0.65$). This hints that distance effects are likely not dominant in our sample.

\begin{figure*}
\centering
\includegraphics[width=\textwidth]{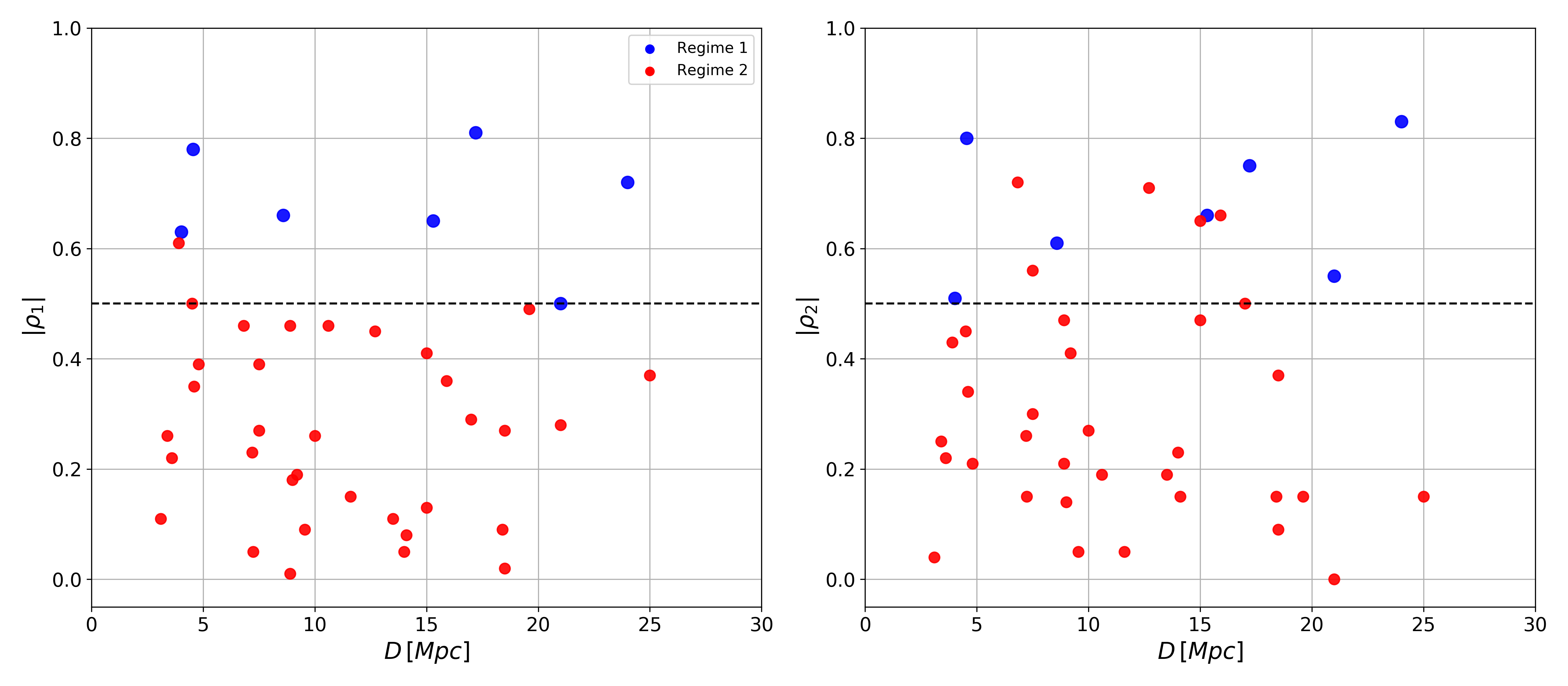}
\caption{ Left: The absolute value of the correlation coefficient of the IR color-color relation ($|\rho_{1}|$) versus distance to each galaxy in Regime 1 (blue) and Regime 2 (red).  Right: the same for the correlation coefficient of the $\Sigma_{SFR}$ versus $f_{8}/f_{24}$ relation ($|\rho_{2}|$). The black dashed line shows the value of the correlation coefficient cutoff between the Regimes (0.5). For Regime 1, we require $|\rho_{1}|$ and $|\rho_{2}| \geq 0.5$. For Regime 2, $|\rho_{1}|$ or $|\rho_{2}| < 0.5$. Note Regime 2 galaxies are not simply the most distant ones.   }
\label{fig:f11}
\end{figure*}

To further test whether variations in distance are important, we plot the relation between the distance to each galaxy and the correlation coefficients $\rho_{1}$ and $\rho_{2}$ in Figure \ref{fig:f11}. It is clear that both Regimes span a similar range in distance from $\sim$3--$25 \, Mpc$. Regime 2 galaxies are not simply those that lie at the largest distances.  There are numerous galaxies at low distances that demonstrate weakly correlated IR colors and several at high distances that demonstrate strong correlation (Figure \ref{fig:f11}). Furthermore, we find no evidence for decreasing correlation coefficients ($|\rho|$) with increasing distance within either Regime (Figure \ref{fig:f11}). These results provide good evidence that the large discrepancies between the IR color-color correlations observed in our sample are not due to the different spatial scales sampled in each galaxy by our 36$''\times$36$''$ bins. 

\hypertarget{5.4}{\subsection{Metallicity Effects}}

In Section \hyperlink{5.2}{5.2}, we discussed the effect of the relative strength of star formation on the observed strength of the $f_{70}/f_{500}$ versus $f_{8}/f_{24}$ color-color correlation. This may not be the only important effect we are seeing across our sample. For example, NGC5457 has an extremely large range in metallicity from its central regions to its outskirts \citep{2003PASP..115..928K}, while NGC5236 is more uniform \citep{Hernandez_2017}. The IRAC 8 $\mu$m flux is well known to have a significant dependence on metallicity or the composition of the local ISM. A change in the 8 $\mu$m flux due to a change in metallicity throughout a galaxy could spread out the  $f_{8}/f_{24}$ color distribution for different sub-galactic regions. This could potentially ``wash out" the IR color-color relation for galaxies with strong metallicity gradients. So in addition to the strength of SF, the composition of the ISM may also have a significant effect on the observed IR color-color trends. 

To investigate how differences in metallicity affect the observed IR color-color trends, we retrieve metallicity gradients from the literature for a sub-sample of our galaxies. This sub-sample consists of both Regime 1 and Regime 2 galaxies that have low inclination angles and have published metallicity gradients spanning a large portion of the galaxies. We define metallicity as the abundance ratio of oxygen to hydrogen or  12$+log(O/H)$. Our gradients are obtained from a variety of different studies, shown in Table \hyperlink{t4}{4}. These studies use observations of collisionally excited nebular emission lines from HII regions to derive metallicity distributions/gradients throughout the galaxies. Specifically, the metallicities of the HII regions are derived from so-called \textit{strong-line} empirical calibrations of the various optical forbidden line ratios ([OIII], [OII], [NII], etc). It is important that there is consistency between the studies in the calibrations used to derive the metallicities. There exists poorly understood systematic differences in the oxygen abundances found by empirical relations compared with theoretical calibrations; empirical calibrations tend to yield a factor of 1.5 to 5 times lower oxygen abundances compared to the theoretical ones  \citep{2003PASP..115..928K,Garnett_2004,2004ApJ...615..228B,2008ApJ...681.1183K}.  

\begin{figure*}
\centering
\includegraphics[width=17.5cm]{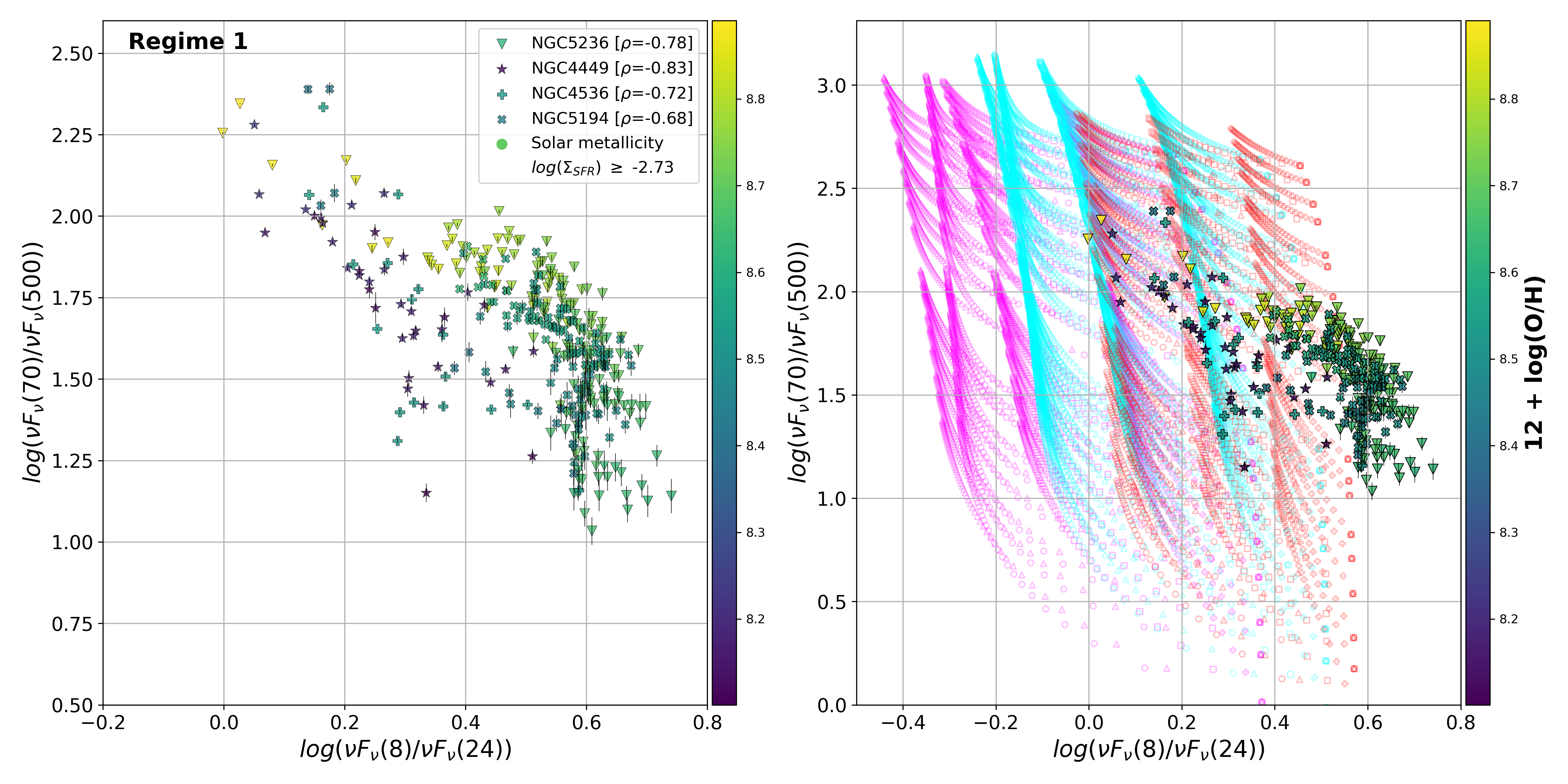}

\vspace{5mm}

\includegraphics[width=17.5cm]{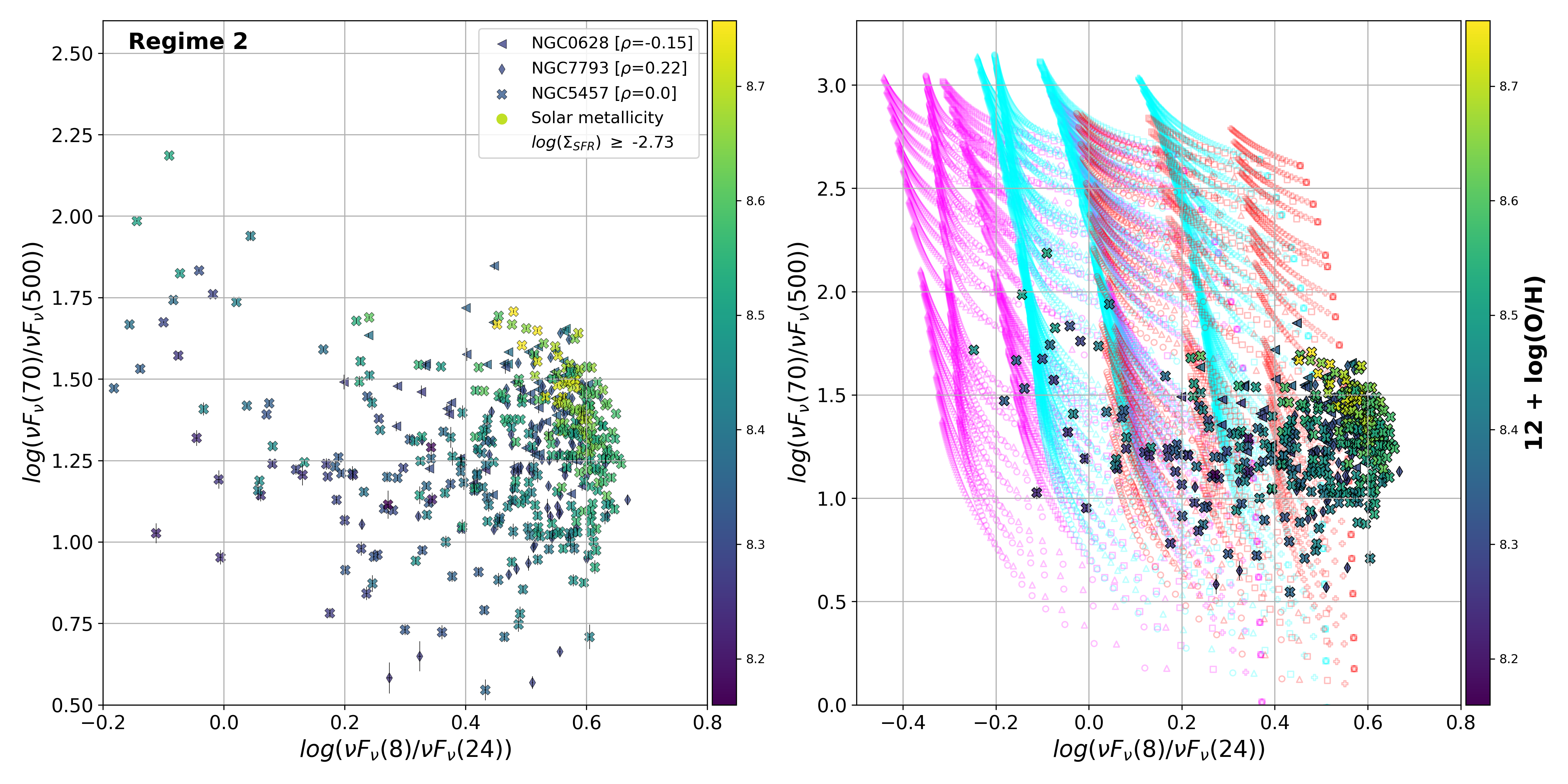}
\caption{Color-color metallicity relations. Top: Sub-sample of Regime 1 galaxies with derived metallicity distributions. Bottom: Regime 2 galaxies. Left panels: The IR color-color relations for the sub-samples, where now the points/regions are color-coded by metallicity.  The color of solar metallicity ($12+log(O/H)=8.7$) is shown in the legend, along with the correlation coefficents $\rho$ and the $\Sigma_{SFR}$ cutoff. Right panels: Same as the left panels with the \cite{2007ApJ...657..810D} model expectations plotted on top (colored tracks) for the range of parameters; $U_{min}$ = 0.1, ..., 25 (bottom to top), $q_{PAH}$ = $1.77\%$ (magenta), $2.50\%$ (cyan), and $4.56\%$ (red), $U_{max}$ = $10^3$ (open plus), $10^4$ (open square), $10^5$ (open triangle), and $10^6$ (open circle), and $\gamma$ = 0.0, ..., 1.0 (right to left).  Note the limit for log($\nu f_{\nu}(8)/ \nu f_{\nu}(24)$)$\lesssim 0.5$ is built into the \cite{2007ApJ...657..810D} models and does not represent a physical limitation. }
\label{fig:f8}
\end{figure*}

We use the central metallicities and gradients given in Table \hyperlink{t4}{4}, along with distances between the galactic centers and our $36''$ regions, to calculate the full radial distribution of metallicities spanning each galaxy. The left panels of Figure \ref{fig:f8} show the $f_{70}/f_{500}$ versus $f_{8}/f_{24}$ color-color relation for this sub-sample of galaxies, now with data points color-coded by their derived metallicity. For Regime 1 galaxies (top left panel of Figure \ref{fig:f8}), we observe a slight split in the ``overall" color-color relation between the galaxies, roughly traced by the metallicity. The two galaxies with higher metallicity, NGC5236 and NGC5194, inhabit an upper branch in the relation (towards higher $f_{8}/f_{24}$ colors). The metal-poor, less actively star-forming NGC4449 dominates the lower branch, along with NGC4536. Within the Regime 2 galaxy NGC5457 (lower left panel of Figure \ref{fig:f8}), we also observe the trend that the lowest metallicity regions lie at the lowest $f_{8}/f_{24}$ colors. These results suggest that the differences in metallicity are contributing to the spreading out of the observed color-color relation both within NGC5457 and between the different galaxies in our sample. 

We compare these results to dust model expectations from \cite{2007ApJ...657..810D} \citep[also see][]{2014ApJ...780..172D}. These dust models assume a mixture of dust grains, combining carbonaceous grains (which include PAHs) and amorphous silicate grains. The grain size distribution is targeted at reproducing the local Milky Way, LMC and SMC extinction curves. A range of PAH-to-dust mass fractions ($q_{PAH}$) is considered for each grain distribution, with the maximum at around the Milky Way dust value of $\sim$4.6\%. In these models, the dust mixture is assumed to be heated by a combination of two starlight intensity components. One component is the diffuse starlight component that permeates the ISM, defined as the energy density $U_{min}$. The other component is a power law distribution of intensities between $U_{min}$ and $U_{max}$, given by $dM_{dust}/dU \, \propto \, U^{-\alpha}$, where $\alpha$ is the value of the powerlaw slope. \cite{2007ApJ...657..810D} found that most IR SEDs are insensitive to $U_{max}$ and as a result, it is typically fixed at a large value. The two intensity components are subsequently added together in proportion to (1-$\gamma$) and $\gamma$, where $0\leq \gamma \leq 1$. This parameter $\gamma$ is thus related to the fraction of starlight intensity due to current SF. Additionally, the total emission in each band of the IR SED is proportional to the total dust mass ($M_{dust}$). Thus these dust models depend on 5 main parameters: $U_{min}$, $\alpha$, $q_{PAH}$, $\gamma$, and $M_{dust}$.

The right hand panels of Figure \ref{fig:f8} show the \cite{2007ApJ...657..810D} dust model expectations, plotted as the various colored tracks. For fixed $U_{max}$, the low metallicity region of the plot (left) corresponds to low $q_{PAH}$. This is consistent with \cite{2020ApJ...889..150A} who derive a linear relation between metallicity and $q_{PAH}$ for the sample of KINGFISH galaxies. These dust emission models clearly predict uncorrelated IR colors, contrary to what we observe for Regime 1 galaxies.  

\begin{figure*}
\centering
\includegraphics[width=13.7cm]{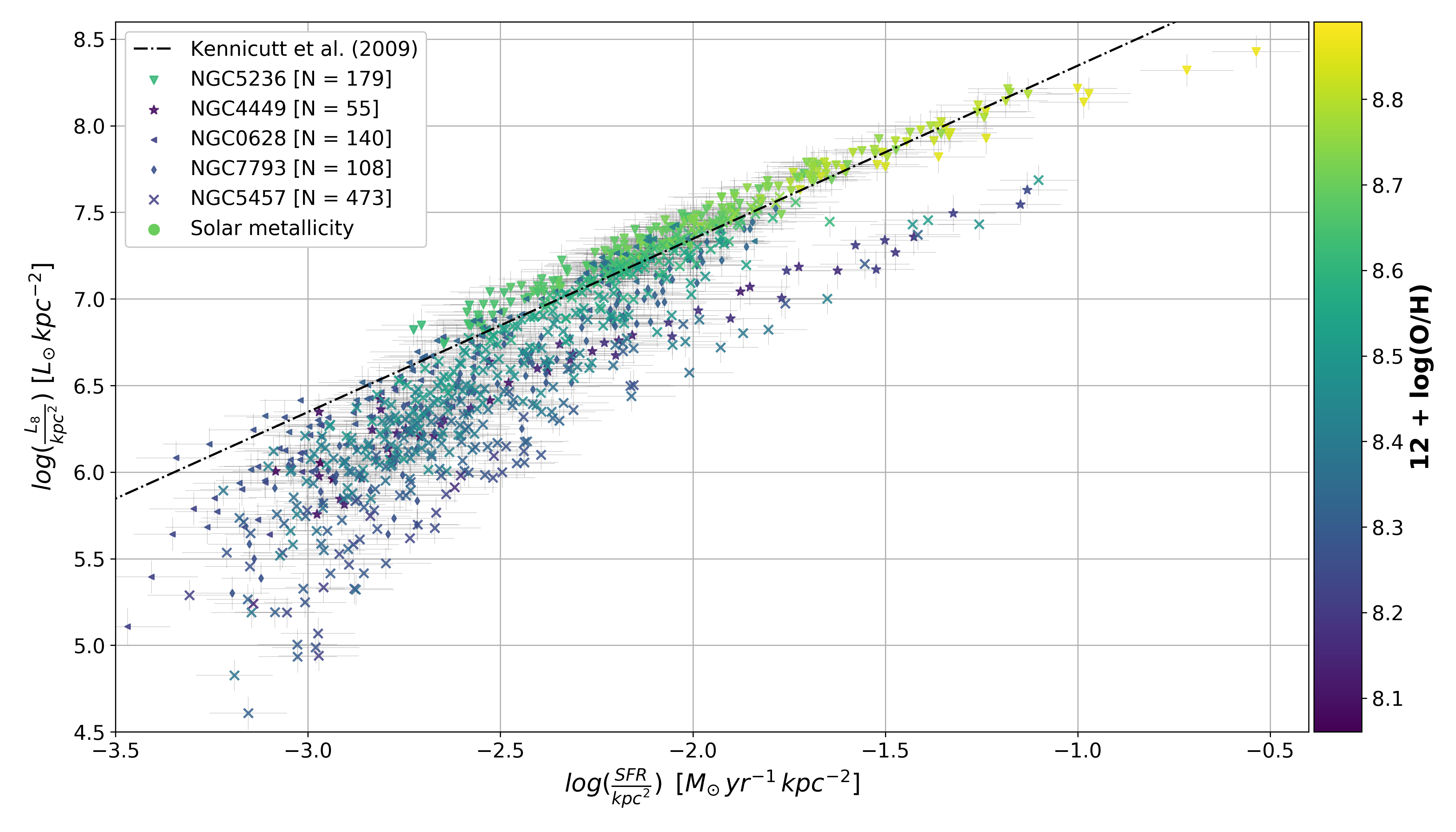}
\caption{ The 8 $\mu$m luminosity surface density versus $\Sigma_{SFR}$ for a sub-sample of galaxies with derived metallicity distributions. The individual points represent the 36$''\times$36$''$ sub-galactic regions and are color-coded on an absolute galactic metallicity scale, where we define metallicity by the abundance ratio, 12$+log(O/H)$.  The color of solar metallicity ($8.7$) is shown in the legend, along with the number of star-forming regions for each galaxy. The black line shows the \cite{2009ApJ...703.1672K} relation. Note the large overall spread in the $\Sigma_{L_{8}}$ versus $\Sigma_{SFR}$ relation, especially within the galaxy NGC5457.}
\label{fig:f9}
\end{figure*}

Figure \ref{fig:f9} shows the 8 $\mu$m luminosity surface density ($\Sigma_{L_{8}}$ or $L_{8}/kpc^{2}$) versus $\Sigma_{SFR}$ relation, with points color-coded by metallicity. Again, we observe a split or ``branching" by metallicity at high $\Sigma_{SFR}$. NGC5236, with the highest metallicity, inhabits the upper ``branch" (high $\Sigma_{SFR}$, high $\Sigma_{L_{8}}$) and is consistent with the relation from \cite{2009ApJ...703.1672K}, while the metal-poor NGC4449 inhabits the lower branch. Interestingly, NGC5457 spans both branches at intermediate $\Sigma_{SFR}$, likely due to its extremely large metallicity range (more than a factor of 2 greater than the rest of the sub-sample). We suggest that the large range of environments/metallicities in NGC5457 adds dispersion to the 8 $\mu$m luminosity, which leads to the large observed spread in the relation between $\Sigma_{SFR}$ and $\Sigma_{L_{8}}$ and contributes to its observed uncorrelated IR colors. More generally, both Figures \ref{fig:f8} and \ref{fig:f9} provide strong evidence that variations in metallicity across our galaxy sample contribute to the observed differences in the IR color-color relations.

\begin{center}
\begin{table}
\centering
\caption{\hypertarget{t4}{ Metallicity Gradients}}  
$[12+log(O/H)]$ \\
\begin{tabular}{ c c c c c}
\hline
\hline
\rule{0pt}{3ex}
Galaxy & Regime & Central &  Gradient$^{\mbox{\textit{*}}}$ & Range$^{\mbox{\textit{+}}}$ \\
& & Metallicity & $[dex/kpc]$ &  \\
\hline
\rule{0pt}{3ex} NGC0628$^{\mbox{\textit{a}}}$ & 2 & 8.45 & -0.0193 & 8.21--8.45 \\
\rule{0pt}{3ex} NGC5194$^{\mbox{\textit{a}}}$ & 1 & 8.67 & -0.0221 & 8.40--8.67\\
\rule{0pt}{3ex} NGC7793$^{\mbox{\textit{a}}}$ & 2 & 8.35 & -0.0217 & 8.25--8.35\\
\rule{0pt}{3ex} NGC5236$^{\mbox{\textit{b}}}$ & 1 & 8.90 & -0.0434 & 8.61--8.90\\
\rule{0pt}{3ex} NGC5457$^{\mbox{\textit{c}}}$ & 2 & 8.78 & -0.0290 & 8.13--8.78\\
\rule{0pt}{3ex} NGC4449$^{\mbox{\textit{d}}}$ & 1 & 8.26 & -0.0551 & 8.06--8.26\\
\rule{0pt}{3ex} NGC4536$^{\mbox{\textit{e}}}$ & 1 & 8.55 & -0.0048 & 8.47--8.55\\
\hline
\end{tabular}
\begin{flushleft} 
\rule{0pt}{3ex}
\currtabletypesize{\sc Note}--- \\
\rule{0pt}{4ex}
$^{\mbox{\textit{*}}}$ Gradients assumed constant throughout the galaxies. \\
\rule{0pt}{3ex}
$^{\mbox{\textit{+}}}$ Full metallicity range across our grids for each galaxy (center to outskirts).  \\
\rule{0pt}{3ex}
$^{\mbox{\textit{a}}}$ Central metallicities and gradients obtained from \cite{Moustakas_2010} using \cite{2005ApJ...631..231P} empirically calibrated values for consistency with the other studies. \\
\rule{0pt}{1ex}
$^{\mbox{\textit{b}}}$ \cite{Hernandez_2017}\\
\rule{0pt}{1ex}
$^{\mbox{\textit{c}}}$ \cite{2003PASP..115..928K}\\
\rule{0pt}{1ex}
$^{\mbox{\textit{d}}}$ \cite{2015MNRAS.450.3254P}\\
\rule{0pt}{1ex}
$^{\mbox{\textit{e}}}$ \cite{2016MNRAS.462.1281M}\\
\rule{0pt}{3ex}
\rule{0pt}{3ex}
\end{flushleft}
\end{table}
\end{center}

\begin{figure*}
\centering
\hspace{-4mm}
\includegraphics[width=13.7cm]{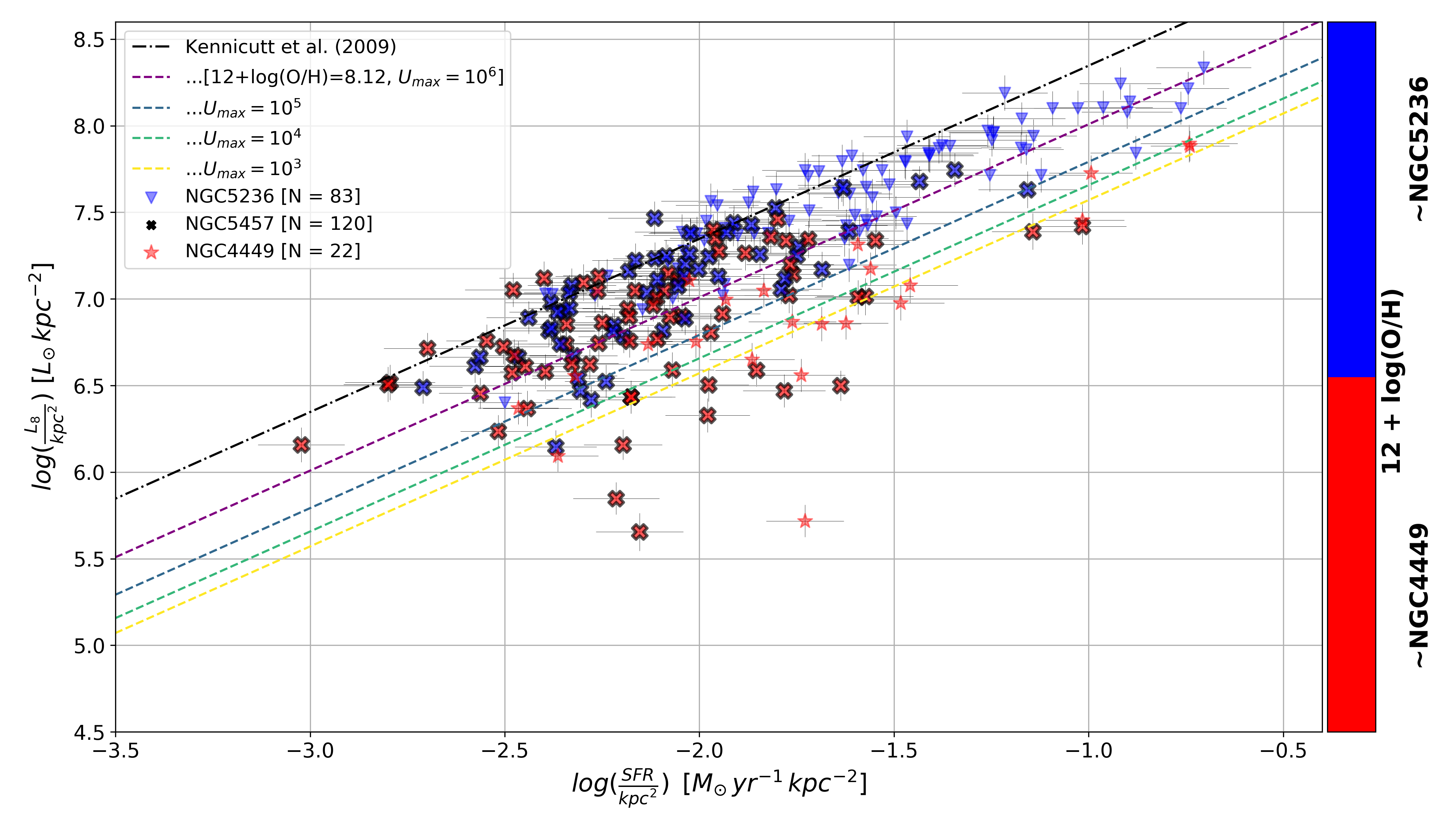}
\caption{The 8 $\mu$m luminosity surface density versus $\Sigma_{SFR}$ in the case of subtracting the diffuse local background. The individual points represent circular apertures $10''$ in radius centered on a sample of bright, isolated star-forming clumps in the three test galaxies at a common spatial resolution of ${\sim}225 \, pc$. The local background is calculated in encompassing annuli of equal area and is removed. NGC5236 is denoted by blue triangles, NGC4449 by red stars, and NGC5457 by x's. The regions in NGC5457 are colored blue when their derived metallicities are closer to the central metallicity of NGC5236 and red when closer to NGC4449. The black dashed line shows the relation from \cite{2009ApJ...703.1672K}, consistent with solar metallicities. The colored lines show the same relation but shifted to the lowest metallicity region (12+$log(O/H)$=8.12) by deriving the corresponding decrease in $L_{8}$ from the relation between oxygen abundance and $q_{PAH}$ from \cite{2020ApJ...889..150A} and the \cite{2007ApJ...657..810D} dust models; for $U_{min}$=(1.0), $q_{PAH}$=(3.19 to 0.47), and $U_{max}$=$(10^{6},10^{5},10^{4},10^{3})$ (purple, blue, green, yellow respectively).} 
\label{fig:f10}
\end{figure*}

\vspace{-4mm}
\hypertarget{5.4.1}{\subsubsection{Accounting for the local background}}

A significant component of the 8 $\mu$m luminosity has been associated with the diffuse, cold ISM \citep{2008MNRAS.389..629B,2014ApJ...784..130C,2014ApJ...797..129L}, which suggests that the local diffuse background emission may play an important role in the derived $\Sigma_{L_{8}}$ versus $\Sigma_{SFR}$ relation shown in Figure \ref{fig:f9}. As a result, we attempt to remove the contribution of the diffuse component by removing the local background from our measurements. We apply a common physical resolution, local background subtracted analysis to the bright star-forming knots in the three test galaxies, NGC5236, NGC5457, and NGC4449. 

For this analysis, various bright and isolated star-forming clumps are first visually identified in the three galaxies using the MIPS 24 $\mu$m image. All data (IRAC 3.6, 8.0 $\mu$m, MIPS 24 $\mu$m, and FUV) are then convolved to the physical resolution of MIPS 24 $\mu$m in the most distant galaxy (NGC5457). This is accomplished by using the \cite{2011PASP..123.1218A} kernels (resampled/renormalized) to convolve all images in NGC5236 to $10''$ angular resolution, in NGC4449 to $12''$ and in NGC5457 to MIPS 24 $\mu$m resolution directly. The resulting common physical resolution is ${\sim}225 \, pc$ for all images in each galaxy. We then measure the flux in circular apertures with radii of $10''$, centered on the identified star-forming clumps.  We use an encapsulating annulus around the apertures of equal area to remove the local background from the apertures in each image. The background within one annuli pixel is found by iteratively calculating the $3\sigma$ clipped mode of the annulus. The total background within the aperture is estimated by multiplying the single pixel mode by the pixel area of the aperture, which is then subtracted from the flux measurements. The known stellar contribution to the 8.0 and 24 $\mu$m fluxes is removed using the 3.6 $\mu$m in the same method as presented in Section \hyperlink{4.1}{4.1}. We calculate the image uncertainties as the standard deviation of the iteratively $3\sigma$ clipped, convolved images and apply minimum cutoffs, requiring all measurements in each image have SNR $\geq 5.0$. The $\Sigma_{L_{8}}$, $\Sigma_{SFR}$ and the uncertainties are calculated in the same method outlined in Section \hyperlink{4.2}{4.2}.

Figure \ref{fig:f10} shows the $\Sigma_{L_{8}}$ versus $\Sigma_{SFR}$ relation for this case of subtracting the local diffuse background, at a common physical resolution of ${\sim}225 \, pc$. It is clear that the large scatter in the $\Sigma_{L_{8}}$ versus $\Sigma_{SFR}$ relation first demonstrated in Figure \ref{fig:f9} is still present after accounting for the local background emission and relative physical resolution. The lower metallicity (red) regions still cluster towards lower $\Sigma_{L_{8}}$ at fixed $\Sigma_{SFR}$, although the clear bifurcation in the relation at high $\Sigma_{SFR}$ is much less pronounced in this case. The \cite{2009ApJ...703.1672K} relation marks the solar metallicity regime, consistent with the slightly super-solar metallicity galaxy NGC5236. The sub-solar galaxy NGC4449 is consistent with the relation when scaled down to the its lowest metallicity ($12+log(O/H)=8.12$) and most quiescent regions ($U_{max}=10^{4},10^{3}$). We conclude from Figure \ref{fig:f10} that the high scatter observed in the $\Sigma_{L_{8}}$ versus $\Sigma_{SFR}$ relation is consistent with large variations in metallicity and the varying intensity of radiation heating the dust under different star-forming conditions, with little contribution from the heating by the local diffuse stellar populations.

\hypertarget{6}{\section{Discussion}}

In the above sections, we have identified two distinct regimes within our sample of galaxies: 1) Regime 1 galaxies that demonstrate a strong ``red" FIR versus ``blue" MIR color-color correlation within their $kpc$--scale star-forming regions and 2) Regime 2 galaxies that demonstrate uncorrelated IR colors. This bifurcation within our sample appears to be driven by two main effects: 1) the local strength of the star formation and 2) the metal content of the ISM. 

We have found evidence that as the surface density of SFR decreases within our Regime 1 galaxies, the IR color-color correlations begin to ``turn over", shown in Figure \ref{fig:f5}. This observed turnover suggests something fundamental about the nature of the IR color-color correlations. That is, the turnover in the relation for a specific galaxy may mark the transition point between the Regimes within that individual galaxy, driven by the decreasing SFR surface density towards the outer regions. For example, Figure \ref{fig:f3} shows that the color-color relation goes nearly vertical (${\rho}\,{\sim}\,{0}$) in the outer, low SFR surface density regions of NGC5236 (for log($\nu f_{\nu}(70)/ \nu f_{\nu}(500)$) $\lesssim 1.6$). This suggests that a strongly Regime 1 galaxy like NGC5236 can still have outer regions that are ``Regime 2-like", in which there is little to no IR color-color correlation. A galaxy with a more developed turnover (such that ${\rho}\,{\sim}\,{0}$ throughout) would just be a galaxy in which essentially no regions exhibit the strong color-color correlation and is therefore Regime 2. This idea implies that splitting our sample into the two Regimes is likely an oversimplification; yet, we still find it useful for exploring and developing an understanding of the variations in our sample. 

The exact SFR surface density where this transition between the Regimes takes place is likely to vary between galaxies, possibly depending on metallicity, the diffuse stellar populations, the interstellar radiation field, etc. Using the turnover in Regime 1 galaxies as a guide, a typical value of SFR surface density where this transition takes places is $\Sigma_{SFR}\,{\sim} \, $0.01 $M_{\odot} \, yr^{-1} \, kpc^{-2}$ (Figure \ref{fig:f5}). This transition SFR surface density is consistent with the transition $\Sigma_{SFR}$  found between normal star-forming disk galaxies and
compact starbursts and/or LIRGs (Luminous Infrared Galaxies) in the local Universe \citep[][their Figure 9]{2012ARA&A..50..531K}, suggesting that this $\Sigma_{SFR}$ is critical at separating more than one regime (e.g. IR colors and morphology) and may have an underlying physical significance.  

Performing a minimum cut on $\Sigma_{SFR}$ strengthens the observed IR color-color correlations in Regime 1 galaxies (Figure \ref{fig:f7}). This is interesting and somewhat to be expected given the observed turnover in the relations in the outer regions of the galaxies. We interpret this in the following way: regions or galaxies that exhibit high surface densities of SF will tend to have IR SEDs that are dominated by SF. The high levels of SF drown out any significant contribution to the IR SED from the heating by the underlying old stellar populations. The sides of the IR SED will be strongly correlated because both the MIR/FIR colors will be strong tracers of new SF, while receiving a minimal relative contribution from other sources. A $\Sigma_{SFR}\,\,{\geq} \,\, $0.01 $M_{\odot} \, yr^{-1} \, kpc^{-2}$ may represent the condition for which the above typically holds true.

 NGC5236 is well known to exhibit high levels of SF throughout its disk \citep{1998ApJ...498..541K,2010ApJ...714.1256C}. In NGC5236, we clearly see a strong relation between  $\Sigma_{SFR}$ and both the $f_{70}/f_{500}$ and $f_{8}/f_{24}$ colors, shown in Figures \ref{fig:f4} and \ref{fig:f5}. Additionally, there is a strong relation between $\Sigma_{L_{8}}$ and $\Sigma_{SFR}$, despite the fact that it hosts a relatively significant metallicity gradient (Figure \ref{fig:f9} and Table \hyperlink{t4}{4}). We propose that in NGC5236, the high levels of SF are dominating the IR SED, leading to the strong IR color-color correlation observed across the majority of its star-forming regions.

Our results for NGC4449 are in good agreement with the results of \cite{2018ApJ...852..106C}, but here we extend on their interpretation. \cite{2018ApJ...852..106C} found a MIR/FIR color-color trend for NGC4449 that remained strong across the full range of regions and colors probed. They derive the IR color-color trend for NGC4449 using a combination of the flux density at 8 and 24 $\mu$m for the MIR and the 70 and 1100 $\mu$m for the FIR, where the 1100 $\mu$m dust continuum emission is measured by the \textit{AzTEC} camera on the Large Millimeter Telescope (\textit{LMT}). As a result, \cite{2018ApJ...852..106C} achieve much higher spatial resolution, but only probe the central 5$\times$5 $kpc^2$ region of NGC4449; thus only the high $\Sigma_{SFR}$ areas in the central starburst. In our study, we instead utilize archival SPIRE 500 $\mu$m data to probe a much larger range in $\Sigma_{SFR}$ and more diverse environments, at the cost of spatial resolution. As a result, we probe NGC4449 down to the much lower $\Sigma_{SFR}$ in its outer regions.  Although NGC4449 exhibits an IR color-color correlation, it does not remain strong throughout the entire galaxy. An increase in the dispersion of the color-color trend is observed at lower values of $\Sigma_{SFR}$, shown in Figures \ref{fig:f3} and \ref{fig:f4}. The $\Sigma_{SFR}$ cutoff noticeably increases the overall strength in the trend for NGC4449, seen by a change in the correlation coefficient $\rho$ from -0.63 to -0.83 (Figure \ref{fig:f7}); more consistent with the tight correlation observed by \cite{2018ApJ...852..106C}. We suggest that these results arise from the fact that NGC4449, although demonstrating high levels of specific SFR, remains ``starbursting" only within the central regions \citep{2009ApJ...692.1305L}; a typical characteristic of starburst dwarf galaxies \citep{1999ApJ...522..183M,2005ApJS..156..345G}. In the less actively star-forming, outer regions of NGC4449, the IR SED is likely no longer dominated by strong SF and the effects due to the varying ISM heating and composition become important, specifically in the MIR.

In NGC5457, which exhibits a large variety of star-forming and passive environments \citep{Watkins_2017}, the IR SED throughout the galaxy likely consists of significant contributions from both new SF and the underlying stellar population. The dual sources of dust heating, along with the significant variance in ISM composition as a function of radius and the metallicity dependence of the 8 $\mu$m luminosity, provides a good explanation of the uncorrelated colors observed in NGC5457, shown in Figure \ref{fig:f3}---along with the majority of our sample of galaxies (Figure \ref{fig:f6}).

In addition to the relative strength of SF, there is a clear metallicity effect on the observed IR color-color correlations within our sample, shown in Figures \ref{fig:f8}, \ref{fig:f9} and \ref{fig:f10}. In the case of Regime 1 galaxies, we see a splitting of the overall MIR/FIR color-color correlations traced by the metallicity, shown in Figure \ref{fig:f8}. In the case of NGC5457, the extreme variations in metallicity directly correlate with the large spread in its $f_{8}/f_{24}$ colors (low metallicity, low $f_{8}/f_{24}$). This suggests that variations in metallicity play an important role in the dispersion of the observed color-color correlations.

Perhaps most informative, Figure \ref{fig:f9} shows the distinct ``splitting" or widening of the $\Sigma_{L_{8}}$ versus $\Sigma_{SFR}$ relation, traced by the metal abundance. This is shown by the significant gap at high $\Sigma_{SFR}$ between the metal-rich NGC5236 and sub-solar metallicity NGC4449 and the overall enormous dispersion in the relation for the highly diverse NGC5457. Figure \ref{fig:f9} suggests that the observed differences in the IR color-color trends arise as variation in ISM composition adds dispersion to the 8 $\mu$m luminosity, specifically at intermediate and low $\Sigma_{SFR}$. Additionally, accounting for the local diffuse background emission (Figure \ref{fig:f10}) does not significantly affect the large scatter observed in the $\Sigma_{L_{8}}$ versus $\Sigma_{SFR}$ relation, further suggesting differences in metallicity and star-forming conditions to be the dominant effects. Figures \ref{fig:f9} and \ref{fig:f10} provide a prime example of how the 8 $\mu$m luminosity can be a highly problematic indicator of the SFR when applied across diverse environments. We observe a factor of 3 variation in $\Sigma_{L_{8}}$ at the highest $\Sigma_{SFR}$ and over a factor of 10 variation at low  $\Sigma_{SFR}$, clearly correlating with changes in metal content. This is far from the one-to-one correlation of an ideal SFR indicator. These results are in good agreement with many previous studies highlighting the dependence of the 8 $\mu$m luminosity on the composition of the ISM \citep{2005ApJ...628L..29E,2007ApJ...666..870C,2007ApJ...656..770S,2014MNRAS.445..899C,Shivaei_2017,2020ApJ...889..150A}. Our results suggest that the 8 $\mu$m luminosity is not a robust monochromatic indicator of the SFR for general applications and cannot be applied to quantify active SF across diverse systems/samples. We join other researchers in cautioning those who plan to use MIR luminosities from MIRI on \textit{JWST} to study resolved SF in high redshift galaxies. 

For future directions of this work, it will be useful to include the emission at 12 $\mu$m from the Wide-field Infrared Survey Explorer (\textit{WISE}).  The \textit{WISE} 12 $\mu$m luminosity has been shown to be a relatively strong tracer of the SFR \citep{2012JApA...33..213S,Lee_2013,2017ApJ...850...68C}. It has been suggested to under-estimate the SFR in metal-poor galaxies, but on a comparatively less significant level than the IRAC 8 $\mu$m \citep{Lee_2013}. Conversely,  \cite{2017ApJ...850...68C} find that within the combined SINGS and KINGFISH sample of galaxies, the relation between SFR and the \textit{WISE} 12 $\mu$m luminosity remains insensitive to the range in metallicity probed by the sample. \cite{2012ApJ...748...80D} find that about 80\% of the 12 $\mu$m luminosity in star-forming galaxies is produced by stellar populations younger than $0.6 \, Gyr$. Although the picture still remains unclear, the 12 $\mu$m luminosity likely does not suffer from as strong of a dependence on the metal content as the 8 $\mu$m. As a result, it will be beneficial to test the $f_{70}/f_{500}$ versus $f_{12}/f_{24}$ colors and how the derived correlations compare to this study. We speculate that stronger and more wide-spread IR color-color correlations within our sample of galaxies are waiting to be uncovered if the metallicity-sensitive IRAC 8 $\mu$m emission is replaced with the \textit{WISE} 12 $\mu$m. The spatial resolution is already limited by the much lower resolution of SPIRE 500 $\mu$m, so the decrease in resolution from IRAC 8 to \textit{WISE} 12 $\mu$m will not be a concern. It will also be useful to include the 3.3 $\mu$m  PAH emission that will be observed with NIRCam on \textit{JWST} and test how it compares to its 8 $\mu$m counterpart. Additionally, it will be possible to add in a sample of extremely active star-forming galaxies (similar to NGC5236) to test whether or not they universally exhibit strong MIR/FIR color-color correlations within their star-forming regions, as suggested by our data.

\hypertarget{7}{\section{Conclusions}}

We have presented a novel analysis of a sample of local galaxies, combining archival multi-wavelength data from \textit{GALEX}, \textit{Spitzer Space Telescope} and \textit{Herschel Space Observatory} in order to investigate sub-galactic MIR and FIR color-color correlations within the infrared SED. Our main findings are the following:

\begin{itemize}
\item The ``blue side" MIR and ``red side" FIR color-color correlations within the IR SED found by \cite{2018ApJ...852..106C} do not hold in general for the resolved star-forming regions within our sample of galaxies. We identify two regimes within our sample: 1) Regime 1 galaxies that demonstrate a FIR versus MIR color-color correlation within their $kpc$--scale star-forming regions and 2) Regime 2 galaxies that demonstrate uncorrelated IR colors. Splitting the sample using correlation coefficients, we find that the majority of our galaxies do not exhibit strong $f_{70}/f_{500}$ versus $f_{8}/f_{24}$ color-color correlations. 

\item The division within our sample of galaxies appears to be driven by two main effects: 1) the local strength of the star formation and 2) the metal content of the ISM. The MIR/FIR color-color correlation strength is found to be tightly connected with the relative strength of the star formation, with an observed ``turnover" in the relations at low star formation rate surface densities. The metal content then becomes important at low $\Sigma_{SFR}$, when the IR SED receives significant contributions from various heating sources. This is highlighted by the SFRs, metallicity distributions, and observed color-color trends of the three test cases that  span the range of properties of our sample: NGC5236, NGC5457, and NGC4449. We find that galaxies uniformly dominated by high surface densities of star formation (e.g. NGC5236) show strong IR color-color correlations, while galaxies that exhibit lower levels of star formation and substantial variance in metallicity/composition (e.g. NGC5457) show weaker or uncorrelated colors.

\item  Our results for the galaxy NGC4449 indicate general consistency with the results from \cite{2018ApJ...852..106C}. However, probing down to much lower $\Sigma_{SFR}$ we find that the observed MIR/FIR color-color correlation significantly weakens in the outer, less actively star-forming regions, implying that the observed strong relation does not hold throughout the entirety of the galaxy.

\item Variations in metallicity play an important role in our sample. For galaxies with strong IR color-color correlations, we identify a ``splitting" or widening in the overall color-color relation, traced by differences in metallicity between galaxies. We identify a similar trend for regions within the highly non-uniform galaxy NGC5457. This suggests that differences in metal content of the ISM have a role in increasing the dispersion of the observed color-color trend, both between galaxies and within those that exhibit highly diverse environments. A distinct metallicity dependence is observed in the 8 $\mu$m luminosity surface density versus SFR surface density relation, which contributes to the substantial dispersion observed in the relation. This highlights the problematic use of the 8 $\mu$m luminosity as a general monochromatic indicator of the SFR, consistent with existing studies \citep{2005ApJ...628L..29E,2007ApJ...666..870C,2007ApJ...656..770S,2014MNRAS.445..899C,Shivaei_2017,2020ApJ...889..150A}.

\end{itemize}

Our results are largely dependent on both the SFR surface density and the well-known metal dependencies within the 8 $\mu$m luminosity. Initially focusing on the 8 $\mu$m luminosity for the MIR color is essential for informing upcoming studies that combine MIR and FIR data from \textit{JWST} and \textit{ALMA}. Yet, we identify the importance of testing other photometric bands. It will be useful to investigate the emission at 12 $\mu$m from \textit{WISE}, which has been suggested to be less metallicity-sensitive compared to the 8 $\mu$m \citep{2017ApJ...850...68C}, or the 3.3 $\mu$m  PAH emission that will be observed with \textit{JWST} at low and high redshift in hopes of uncovering stronger and more wide-spread sub-galactic color-color correlations within the IR SED.

\begin{acknowledgments}
The authors would like to thank the anonymous referee for many thoughtful and constructive comments that have helped improve this manuscript considerably. This research has made use of the NASA/IPAC Extragalactic Database (NED),
which is operated by the Jet Propulsion Laboratory, California Institute of Technology, under contract with the National Aeronautics and Space Administration. B.A.G. acknowledges support from NSF grant 1907791.
\end{acknowledgments}

\facilities{\textit{GALEX}, \textit{Spitzer} (IRAC, MIPS), \textit{Herschel} (PACS, SPIRE)}

\software{astropy \citep{astropy:2013,astropy:2018}, photutils \citep{larry_bradley_2019_3568287}, IRAF \citep{1986SPIE..627..733T,1993ASPC...52..173T}, SAOImageDS9 \citep{2003AJ....125..525J}}

\nocite{https://doi.org/10.26132/ned1}

\clearpage
\bibliographystyle{aasjournal}
\bibliography{papers}

\begin{thebibliography}{}
\expandafter\ifx\csname natexlab\endcsname\relax\def\natexlab#1{#1}\fi
\providecommand{\url}[1]{\href{#1}{#1}}
\providecommand{\dodoi}[1]{doi:~\href{http://doi.org/#1}{\nolinkurl{#1}}}
\providecommand{\doeprint}[1]{\href{http://ascl.net/#1}{\nolinkurl{http://ascl.net/#1}}}
\providecommand{\doarXiv}[1]{\href{https://arxiv.org/abs/#1}{\nolinkurl{https://arxiv.org/abs/#1}}}

\bibitem[{{Aniano} {et~al.}(2011){Aniano}, {Draine}, {Gordon}, \&
  {Sandstrom}}]{2011PASP..123.1218A}
{Aniano}, G., {Draine}, B.~T., {Gordon}, K.~D., \& {Sandstrom}, K. 2011, \pasp,
  123, 1218, \dodoi{10.1086/662219}

\bibitem[{Aniano {et~al.}(2011)Aniano, Draine, Gordon, \&
  Sandstrom}]{Aniano_2011}
Aniano, G., Draine, B.~T., Gordon, K.~D., \& Sandstrom, K. 2011, Publications
  of the Astronomical Society of the Pacific, 123, 1218, \dodoi{10.1086/662219}

\bibitem[{{Aniano} {et~al.}(2020){Aniano}, {Draine}, {Hunt}, {Sandstrom},
  {Calzetti}, {Kennicutt}, {Dale}, {Galametz}, {Gordon}, {Leroy}, {Smith},
  {Roussel}, {Sauvage}, {Walter}, {Armus}, {Bolatto}, {Boquien}, {Crocker}, {De
  Looze}, {Donovan Meyer}, {Helou}, {Hinz}, {Johnson}, {Koda}, {Miller},
  {Montiel}, {Murphy}, {Rela{\~n}o}, {Rix}, {Schinnerer}, {Skibba}, {Wolfire},
  \& {Engelbracht}}]{2020ApJ...889..150A}
{Aniano}, G., {Draine}, B.~T., {Hunt}, L.~K., {et~al.} 2020, \apj, 889, 150,
  \dodoi{10.3847/1538-4357/ab5fdb}

\bibitem[{{Astropy Collaboration} {et~al.}(2013){Astropy Collaboration},
  {Robitaille}, {Tollerud}, {Greenfield}, {Droettboom}, {Bray}, {Aldcroft},
  {Davis}, {Ginsburg}, {Price-Whelan}, {Kerzendorf}, {Conley}, {Crighton},
  {Barbary}, {Muna}, {Ferguson}, {Grollier}, {Parikh}, {Nair}, {Unther},
  {Deil}, {Woillez}, {Conseil}, {Kramer}, {Turner}, {Singer}, {Fox}, {Weaver},
  {Zabalza}, {Edwards}, {Azalee Bostroem}, {Burke}, {Casey}, {Crawford},
  {Dencheva}, {Ely}, {Jenness}, {Labrie}, {Lim}, {Pierfederici}, {Pontzen},
  {Ptak}, {Refsdal}, {Servillat}, \& {Streicher}}]{astropy:2013}
{Astropy Collaboration}, {Robitaille}, T.~P., {Tollerud}, E.~J., {et~al.} 2013,
  \aap, 558, A33, \dodoi{10.1051/0004-6361/201322068}

\bibitem[{{Astropy Collaboration} {et~al.}(2018){Astropy Collaboration},
  {Price-Whelan}, {Sip{\H{o}}cz}, {G{\"u}nther}, {Lim}, {Crawford}, {Conseil},
  {Shupe}, {Craig}, {Dencheva}, {Ginsburg}, {Vand erPlas}, {Bradley},
  {P{\'e}rez-Su{\'a}rez}, {de Val-Borro}, {Aldcroft}, {Cruz}, {Robitaille},
  {Tollerud}, {Ardelean}, {Babej}, {Bach}, {Bachetti}, {Bakanov}, {Bamford},
  {Barentsen}, {Barmby}, {Baumbach}, {Berry}, {Biscani}, {Boquien}, {Bostroem},
  {Bouma}, {Brammer}, {Bray}, {Breytenbach}, {Buddelmeijer}, {Burke},
  {Calderone}, {Cano Rodr{\'\i}guez}, {Cara}, {Cardoso}, {Cheedella}, {Copin},
  {Corrales}, {Crichton}, {D'Avella}, {Deil}, {Depagne}, {Dietrich}, {Donath},
  {Droettboom}, {Earl}, {Erben}, {Fabbro}, {Ferreira}, {Finethy}, {Fox},
  {Garrison}, {Gibbons}, {Goldstein}, {Gommers}, {Greco}, {Greenfield},
  {Groener}, {Grollier}, {Hagen}, {Hirst}, {Homeier}, {Horton}, {Hosseinzadeh},
  {Hu}, {Hunkeler}, {Ivezi{\'c}}, {Jain}, {Jenness}, {Kanarek}, {Kendrew},
  {Kern}, {Kerzendorf}, {Khvalko}, {King}, {Kirkby}, {Kulkarni}, {Kumar},
  {Lee}, {Lenz}, {Littlefair}, {Ma}, {Macleod}, {Mastropietro}, {McCully},
  {Montagnac}, {Morris}, {Mueller}, {Mumford}, {Muna}, {Murphy}, {Nelson},
  {Nguyen}, {Ninan}, {N{\"o}the}, {Ogaz}, {Oh}, {Parejko}, {Parley}, {Pascual},
  {Patil}, {Patil}, {Plunkett}, {Prochaska}, {Rastogi}, {Reddy Janga},
  {Sabater}, {Sakurikar}, {Seifert}, {Sherbert}, {Sherwood-Taylor}, {Shih},
  {Sick}, {Silbiger}, {Singanamalla}, {Singer}, {Sladen}, {Sooley},
  {Sornarajah}, {Streicher}, {Teuben}, {Thomas}, {Tremblay}, {Turner},
  {Terr{\'o}n}, {van Kerkwijk}, {de la Vega}, {Watkins}, {Weaver}, {Whitmore},
  {Woillez}, {Zabalza}, \& {Astropy Contributors}}]{astropy:2018}
{Astropy Collaboration}, {Price-Whelan}, A.~M., {Sip{\H{o}}cz}, B.~M., {et~al.}
  2018, \aj, 156, 123, \dodoi{10.3847/1538-3881/aabc4f}

\bibitem[{{Barnes} {et~al.}(2014){Barnes}, {van Zee}, {Dale}, {Staudaher},
  {Bullock}, {Calzetti}, {Chandar}, \& {Dalcanton}}]{2014ApJ...789..126B}
{Barnes}, K.~L., {van Zee}, L., {Dale}, D.~A., {et~al.} 2014, \apj, 789, 126,
  \dodoi{10.1088/0004-637X/789/2/126}

\bibitem[{{Bendo} {et~al.}(2008){Bendo}, {Draine}, {Engelbracht}, {Helou},
  {Thornley}, {Bot}, {Buckalew}, {Calzetti}, {Dale}, {Hollenbach}, {Li}, \&
  {Moustakas}}]{2008MNRAS.389..629B}
{Bendo}, G.~J., {Draine}, B.~T., {Engelbracht}, C.~W., {et~al.} 2008, \mnras,
  389, 629, \dodoi{10.1111/j.1365-2966.2008.13567.x}

\bibitem[{{Bendo} {et~al.}(2012){Bendo}, {Boselli}, {Dariush}, {Pohlen},
  {Roussel}, {Sauvage}, {Smith}, {Wilson}, {Baes}, {Cooray}, {Clements},
  {Cortese}, {Foyle}, {Galametz}, {Gomez}, {Lebouteiller}, {Lu}, {Madden},
  {Mentuch}, {O'Halloran}, {Page}, {Remy}, {Schulz}, \&
  {Spinoglio}}]{2012MNRAS.419.1833B}
{Bendo}, G.~J., {Boselli}, A., {Dariush}, A., {et~al.} 2012, \mnras, 419, 1833,
  \dodoi{10.1111/j.1365-2966.2011.19735.x}

\bibitem[{{Bigiel} {et~al.}(2008){Bigiel}, {Leroy}, {Walter}, {Brinks}, {de
  Blok}, {Madore}, \& {Thornley}}]{2008AJ....136.2846B}
{Bigiel}, F., {Leroy}, A., {Walter}, F., {et~al.} 2008, \aj, 136, 2846,
  \dodoi{10.1088/0004-6256/136/6/2846}

\bibitem[{{Binder} \& {Povich}(2018)}]{2018ApJ...864..136B}
{Binder}, B.~A., \& {Povich}, M.~S. 2018, \apj, 864, 136,
  \dodoi{10.3847/1538-4357/aad7b2}

\bibitem[{Bradley {et~al.}(2019)Bradley, Sipőcz, Robitaille, Tollerud,
  Vinícius, Deil, Barbary, Wilson, Busko, Günther, Cara, Conseil, Droettboom,
  Bostroem, Bray, Bratholm, Lim, Craig, Barentsen, Pascual, Donath, Greco,
  Perren, Kerzendorf, de~Val-Borro, Dencheva, de~Albernaz~Ferreira, Souchereau,
  D'Eugenio, \& Weaver}]{larry_bradley_2019_3568287}
Bradley, L., Sipőcz, B., Robitaille, T., {et~al.} 2019, astropy/photutils:
  v0.7.2, v0.7.2,  Zenodo, \dodoi{10.5281/zenodo.3568287}

\bibitem[{{Bresolin} {et~al.}(2004){Bresolin}, {Garnett}, \&
  {Kennicutt}}]{2004ApJ...615..228B}
{Bresolin}, F., {Garnett}, D.~R., \& {Kennicutt}, Robert~C., J. 2004, \apj,
  615, 228, \dodoi{10.1086/424377}

\bibitem[{{Calapa} {et~al.}(2014){Calapa}, {Calzetti}, {Draine}, {Boquien},
  {Kramer}, {Xilouris}, {Verley}, {Braine}, {Rela{\~n}o}, {van der Werf},
  {Israel}, {Hermelo}, \& {Albrecht}}]{2014ApJ...784..130C}
{Calapa}, M.~D., {Calzetti}, D., {Draine}, B.~T., {et~al.} 2014, \apj, 784,
  130, \dodoi{10.1088/0004-637X/784/2/130}

\bibitem[{{Calzetti} {et~al.}(2012){Calzetti}, {Liu}, \&
  {Koda}}]{2012ApJ...752...98C}
{Calzetti}, D., {Liu}, G., \& {Koda}, J. 2012, \apj, 752, 98,
  \dodoi{10.1088/0004-637X/752/2/98}

\bibitem[{{Calzetti} {et~al.}(2005){Calzetti}, {Kennicutt}, {Bianchi},
  {Thilker}, {Dale}, {Engelbracht}, {Leitherer}, {Meyer}, {Sosey}, {Mutchler},
  {Regan}, {Thornley}, {Armus}, {Bendo}, {Boissier}, {Boselli}, {Draine},
  {Gordon}, {Helou}, {Hollenbach}, {Kewley}, {Madore}, {Martin}, {Murphy},
  {Rieke}, {Rieke}, {Roussel}, {Sheth}, {Smith}, {Walter}, {White}, {Yi},
  {Scoville}, {Polletta}, \& {Lindler}}]{2005ApJ...633..871C}
{Calzetti}, D., {Kennicutt}, R.~C., J., {Bianchi}, L., {et~al.} 2005, \apj,
  633, 871, \dodoi{10.1086/466518}

\bibitem[{{Calzetti} {et~al.}(2007){Calzetti}, {Kennicutt}, {Engelbracht},
  {Leitherer}, {Draine}, {Kewley}, {Moustakas}, {Sosey}, {Dale}, {Gordon},
  {Helou}, {Hollenbach}, {Armus}, {Bendo}, {Bot}, {Buckalew}, {Jarrett}, {Li},
  {Meyer}, {Murphy}, {Prescott}, {Regan}, {Rieke}, {Roussel}, {Sheth}, {Smith},
  {Thornley}, \& {Walter}}]{2007ApJ...666..870C}
{Calzetti}, D., {Kennicutt}, R.~C., {Engelbracht}, C.~W., {et~al.} 2007, \apj,
  666, 870, \dodoi{10.1086/520082}

\bibitem[{{Calzetti} {et~al.}(2010){Calzetti}, {Wu}, {Hong}, {Kennicutt},
  {Lee}, {Dale}, {Engelbracht}, {van Zee}, {Draine}, {Hao}, {Gordon},
  {Moustakas}, {Murphy}, {Regan}, {Begum}, {Block}, {Dalcanton}, {Funes}, {Gil
  de Paz}, {Johnson}, {Sakai}, {Skillman}, {Walter}, {Weisz}, {Williams}, \&
  {Wu}}]{2010ApJ...714.1256C}
{Calzetti}, D., {Wu}, S.~Y., {Hong}, S., {et~al.} 2010, \apj, 714, 1256,
  \dodoi{10.1088/0004-637X/714/2/1256}

\bibitem[{{Calzetti} {et~al.}(2018){Calzetti}, {Wilson}, {Draine}, {Roussel},
  {Johnson}, {Heyer}, {Wall}, {Grasha}, {Battisti}, {Andrews}, {Kirkpatrick},
  {Rosa Gonz{\'a}lez}, {Vega}, {Puschnig}, {Yun}, {{\"O}stlin}, {Evans},
  {Tang}, {Lowenthal}, \& {S{\'a}nchez-Arguelles}}]{2018ApJ...852..106C}
{Calzetti}, D., {Wilson}, G.~W., {Draine}, B.~T., {et~al.} 2018, \apj, 852,
  106, \dodoi{10.3847/1538-4357/aaa1e2}

\bibitem[{{Chandar} {et~al.}(2005){Chandar}, {Leitherer}, {Tremonti},
  {Calzetti}, {Aloisi}, {Meurer}, \& {de Mello}}]{2005ApJ...628..210C}
{Chandar}, R., {Leitherer}, C., {Tremonti}, C.~A., {et~al.} 2005, \apj, 628,
  210, \dodoi{10.1086/430592}

\bibitem[{{Cluver} {et~al.}(2017){Cluver}, {Jarrett}, {Dale}, {Smith},
  {August}, \& {Brown}}]{2017ApJ...850...68C}
{Cluver}, M.~E., {Jarrett}, T.~H., {Dale}, D.~A., {et~al.} 2017, \apj, 850, 68,
  \dodoi{10.3847/1538-4357/aa92c7}

\bibitem[{{Cook} {et~al.}(2014){Cook}, {Dale}, {Johnson}, {Van Zee}, {Lee},
  {Kennicutt}, {Calzetti}, {Staudaher}, \& {Engelbracht}}]{2014MNRAS.445..899C}
{Cook}, D.~O., {Dale}, D.~A., {Johnson}, B.~D., {et~al.} 2014, \mnras, 445,
  899, \dodoi{10.1093/mnras/stu1787}

\bibitem[{{Crocker} {et~al.}(2015){Crocker}, {Chandar}, {Calzetti}, {Holwerda},
  {Leitherer}, {Popescu}, \& {Tuffs}}]{2015ApJ...808...76C}
{Crocker}, A.~F., {Chandar}, R., {Calzetti}, D., {et~al.} 2015, \apj, 808, 76,
  \dodoi{10.1088/0004-637X/808/1/76}

\bibitem[{Croxall {et~al.}(2016)Croxall, Pogge, Berg, Skillman, \&
  Moustakas}]{Croxall_2016}
Croxall, K.~V., Pogge, R.~W., Berg, D.~A., Skillman, E.~D., \& Moustakas, J.
  2016, The Astrophysical Journal, 830, 4, \dodoi{10.3847/0004-637x/830/1/4}

\bibitem[{{Daddi} {et~al.}(2010){Daddi}, {Elbaz}, {Walter}, {Bournaud},
  {Salmi}, {Carilli}, {Dannerbauer}, {Dickinson}, {Monaco}, \&
  {Riechers}}]{2010ApJ...714L.118D}
{Daddi}, E., {Elbaz}, D., {Walter}, F., {et~al.} 2010, \apjl, 714, L118,
  \dodoi{10.1088/2041-8205/714/1/L118}

\bibitem[{{Dale} {et~al.}(2009){Dale}, {Cohen}, {Johnson}, {Schuster},
  {Calzetti}, {Engelbracht}, {Gil de Paz}, {Kennicutt}, {Lee}, {Begum},
  {Block}, {Dalcanton}, {Funes}, {Gordon}, {Johnson}, {Marble}, {Sakai},
  {Skillman}, {van Zee}, {Walter}, {Weisz}, {Williams}, {Wu}, \&
  {Wu}}]{2009ApJ...703..517D}
{Dale}, D.~A., {Cohen}, S.~A., {Johnson}, L.~C., {et~al.} 2009, \apj, 703, 517,
  \dodoi{10.1088/0004-637X/703/1/517}

\bibitem[{{de Vaucouleurs} {et~al.}(1991){de Vaucouleurs}, {de Vaucouleurs},
  {Corwin}, {Buta}, {Paturel}, \& {Fouque}}]{1991rc3..book.....D}
{de Vaucouleurs}, G., {de Vaucouleurs}, A., {Corwin}, Herold~G., J., {et~al.}
  1991, {Third Reference Catalogue of Bright Galaxies}

\bibitem[{{Donoso} {et~al.}(2012){Donoso}, {Yan}, {Tsai}, {Eisenhardt},
  {Stern}, {Assef}, {Leisawitz}, {Jarrett}, \&
  {Stanford}}]{2012ApJ...748...80D}
{Donoso}, E., {Yan}, L., {Tsai}, C., {et~al.} 2012, \apj, 748, 80,
  \dodoi{10.1088/0004-637X/748/2/80}

\bibitem[{{Draine} \& {Li}(2007)}]{2007ApJ...657..810D}
{Draine}, B.~T., \& {Li}, A. 2007, \apj, 657, 810, \dodoi{10.1086/511055}

\bibitem[{{Draine} {et~al.}(2014){Draine}, {Aniano}, {Krause}, {Groves},
  {Sandstrom}, {Braun}, {Leroy}, {Klaas}, {Linz}, {Rix}, {Schinnerer},
  {Schmiedeke}, \& {Walter}}]{2014ApJ...780..172D}
{Draine}, B.~T., {Aniano}, G., {Krause}, O., {et~al.} 2014, \apj, 780, 172,
  \dodoi{10.1088/0004-637X/780/2/172}

\bibitem[{{Elbaz} {et~al.}(2011){Elbaz}, {Dickinson}, {Hwang},
  {D{\'\i}az-Santos}, {Magdis}, {Magnelli}, {Le Borgne}, {Galliano},
  {Pannella}, {Chanial}, {Armus}, {Charmandaris}, {Daddi}, {Aussel}, {Popesso},
  {Kartaltepe}, {Altieri}, {Valtchanov}, {Coia}, {Dannerbauer}, {Dasyra},
  {Leiton}, {Mazzarella}, {Alexander}, {Buat}, {Burgarella}, {Chary}, {Gilli},
  {Ivison}, {Juneau}, {Le Floc'h}, {Lutz}, {Morrison}, {Mullaney}, {Murphy},
  {Pope}, {Scott}, {Brodwin}, {Calzetti}, {Cesarsky}, {Charlot}, {Dole},
  {Eisenhardt}, {Ferguson}, {F{\"o}rster Schreiber}, {Frayer}, {Giavalisco},
  {Huynh}, {Koekemoer}, {Papovich}, {Reddy}, {Surace}, {Teplitz}, {Yun}, \&
  {Wilson}}]{2011A&A...533A.119E}
{Elbaz}, D., {Dickinson}, M., {Hwang}, H.~S., {et~al.} 2011, \aap, 533, A119,
  \dodoi{10.1051/0004-6361/201117239}

\bibitem[{{Engelbracht} {et~al.}(2005){Engelbracht}, {Gordon}, {Rieke},
  {Werner}, {Dale}, \& {Latter}}]{2005ApJ...628L..29E}
{Engelbracht}, C.~W., {Gordon}, K.~D., {Rieke}, G.~H., {et~al.} 2005, \apjl,
  628, L29, \dodoi{10.1086/432613}

\bibitem[{{Eskew} {et~al.}(2012){Eskew}, {Zaritsky}, \&
  {Meidt}}]{2012AJ....143..139E}
{Eskew}, M., {Zaritsky}, D., \& {Meidt}, S. 2012, \aj, 143, 139,
  \dodoi{10.1088/0004-6256/143/6/139}

\bibitem[{Freedman {et~al.}(2001)Freedman, Madore, Gibson, Ferrarese, Kelson,
  Sakai, Mould, Robert C.~Kennicutt, Ford, Graham, Huchra, Hughes, Illingworth,
  Macri, \& Stetson}]{Freedman_2001}
Freedman, W.~L., Madore, B.~F., Gibson, B.~K., {et~al.} 2001, The Astrophysical
  Journal, 553, 47, \dodoi{10.1086/320638}

\bibitem[{Garnett {et~al.}(2004)Garnett, Robert C.~Kennicutt, \&
  Bresolin}]{Garnett_2004}
Garnett, D.~R., Robert C.~Kennicutt, J., \& Bresolin, F. 2004, The
  Astrophysical Journal, 607, L21, \dodoi{10.1086/421489}

\bibitem[{{Genzel} {et~al.}(2010){Genzel}, {Tacconi}, {Gracia-Carpio},
  {Sternberg}, {Cooper}, {Shapiro}, {Bolatto}, {Bouch{\'e}}, {Bournaud},
  {Burkert}, {Combes}, {Comerford}, {Cox}, {Davis}, {Schreiber},
  {Garcia-Burillo}, {Lutz}, {Naab}, {Neri}, {Omont}, {Shapley}, \&
  {Weiner}}]{2010MNRAS.407.2091G}
{Genzel}, R., {Tacconi}, L.~J., {Gracia-Carpio}, J., {et~al.} 2010, \mnras,
  407, 2091, \dodoi{10.1111/j.1365-2966.2010.16969.x}

\bibitem[{{Gil de Paz} \& {Madore}(2005)}]{2005ApJS..156..345G}
{Gil de Paz}, A., \& {Madore}, B.~F. 2005, \apjs, 156, 345,
  \dodoi{10.1086/427068}

\bibitem[{{Gil de Paz} {et~al.}(2007){Gil de Paz}, {Boissier}, {Madore},
  {Seibert}, {Joe}, {Boselli}, {Wyder}, {Thilker}, {Bianchi}, {Rey}, {Rich},
  {Barlow}, {Conrow}, {Forster}, {Friedman}, {Martin}, {Morrissey}, {Neff},
  {Schiminovich}, {Small}, {Donas}, {Heckman}, {Lee}, {Milliard}, {Szalay}, \&
  {Yi}}]{2007ApJS..173..185G}
{Gil de Paz}, A., {Boissier}, S., {Madore}, B.~F., {et~al.} 2007, \apjs, 173,
  185, \dodoi{10.1086/516636}

\bibitem[{Gordon {et~al.}(2008)Gordon, Engelbracht, Rieke, Misselt, Smith, \&
  Robert C.~Kennicutt}]{Gordon_2008}
Gordon, K.~D., Engelbracht, C.~W., Rieke, G.~H., {et~al.} 2008, The
  Astrophysical Journal, 682, 336, \dodoi{10.1086/589567}

\bibitem[{{Helou} {et~al.}(2004){Helou}, {Roussel}, {Appleton}, {Frayer},
  {Stolovy}, {Storrie-Lombardi}, {Hurt}, {Lowrance}, {Makovoz}, {Masci},
  {Surace}, {Gordon}, {Alonso-Herrero}, {Engelbracht}, {Misselt}, {Rieke},
  {Rieke}, {Willner}, {Pahre}, {Ashby}, {Fazio}, \&
  {Smith}}]{2004ApJS..154..253H}
{Helou}, G., {Roussel}, H., {Appleton}, P., {et~al.} 2004, \apjs, 154, 253,
  \dodoi{10.1086/422640}

\bibitem[{Hernandez {et~al.}(2017)Hernandez, Larsen, Trager, Kaper, \&
  Groot}]{Hernandez_2017}
Hernandez, S., Larsen, S., Trager, S., Kaper, L., \& Groot, P. 2017, Monthly
  Notices of the Royal Astronomical Society, 473, 826–837,
  \dodoi{10.1093/mnras/stx2397}

\bibitem[{{Hony} {et~al.}(2015){Hony}, {Gouliermis}, {Galliano}, {Galametz},
  {Cormier}, {Chen}, {Dib}, {Hughes}, {Klessen}, {Roman-Duval}, {Smith},
  {Bernard}, {Bot}, {Carlson}, {Gordon}, {Indebetouw}, {Lebouteiller}, {Lee},
  {Madden}, {Meixner}, {Oliveira}, {Rubio}, {Sauvage}, \&
  {Wu}}]{2015MNRAS.448.1847H}
{Hony}, S., {Gouliermis}, D.~A., {Galliano}, F., {et~al.} 2015, \mnras, 448,
  1847, \dodoi{10.1093/mnras/stv107}

\bibitem[{{Jameson} {et~al.}(2016){Jameson}, {Bolatto}, {Leroy}, {Meixner},
  {Roman-Duval}, {Gordon}, {Hughes}, {Israel}, {Rubio}, {Indebetouw}, {Madden},
  {Bot}, {Hony}, {Cormier}, {Pellegrini}, {Galametz}, \&
  {Sonneborn}}]{2016ApJ...825...12J}
{Jameson}, K.~E., {Bolatto}, A.~D., {Leroy}, A.~K., {et~al.} 2016, \apj, 825,
  12, \dodoi{10.3847/0004-637X/825/1/12}

\bibitem[{{Jarrett} {et~al.}(2003){Jarrett}, {Chester}, {Cutri}, {Schneider},
  \& {Huchra}}]{2003AJ....125..525J}
{Jarrett}, T.~H., {Chester}, T., {Cutri}, R., {Schneider}, S.~E., \& {Huchra},
  J.~P. 2003, \aj, 125, 525, \dodoi{10.1086/345794}

\bibitem[{{Karachentsev} {et~al.}(2003){Karachentsev}, {Sharina}, {Dolphin},
  {Grebel}, {Geisler}, {Guhathakurta}, {Hodge}, {Karachentseva}, {Sarajedini},
  \& {Seitzer}}]{2003A&A...398..467K}
{Karachentsev}, I.~D., {Sharina}, M.~E., {Dolphin}, A.~E., {et~al.} 2003, \aap,
  398, 467, \dodoi{10.1051/0004-6361:20021598}

\bibitem[{{Kawamura} {et~al.}(2009){Kawamura}, {Mizuno}, {Minamidani},
  {Filipovi{\'c}}, {Staveley-Smith}, {Kim}, {Mizuno}, {Onishi}, {Mizuno}, \&
  {Fukui}}]{2009ApJS..184....1K}
{Kawamura}, A., {Mizuno}, Y., {Minamidani}, T., {et~al.} 2009, \apjs, 184, 1,
  \dodoi{10.1088/0067-0049/184/1/1}

\bibitem[{{Kennicutt}(1998)}]{1998ApJ...498..541K}
{Kennicutt}, Robert~C., J. 1998, \apj, 498, 541, \dodoi{10.1086/305588}

\bibitem[{{Kennicutt} {et~al.}(2003){Kennicutt}, {Armus}, {Bendo}, {Calzetti},
  {Dale}, {Draine}, {Engelbracht}, {Gordon}, {Grauer}, {Helou}, {Hollenbach},
  {Jarrett}, {Kewley}, {Leitherer}, {Li}, {Malhotra}, {Regan}, {Rieke},
  {Rieke}, {Roussel}, {Smith}, {Thornley}, \& {Walter}}]{2003PASP..115..928K}
{Kennicutt}, Robert~C., J., {Armus}, L., {Bendo}, G., {et~al.} 2003, \pasp,
  115, 928, \dodoi{10.1086/376941}

\bibitem[{{Kennicutt} {et~al.}(2007){Kennicutt}, {Calzetti}, {Walter}, {Helou},
  {Hollenbach}, {Armus}, {Bendo}, {Dale}, {Draine}, {Engelbracht}, {Gordon},
  {Prescott}, {Regan}, {Thornley}, {Bot}, {Brinks}, {de Blok}, {de Mello},
  {Meyer}, {Moustakas}, {Murphy}, {Sheth}, \& {Smith}}]{2007ApJ...671..333K}
{Kennicutt}, Robert~C., J., {Calzetti}, D., {Walter}, F., {et~al.} 2007, \apj,
  671, 333, \dodoi{10.1086/522300}

\bibitem[{{Kennicutt} {et~al.}(2009){Kennicutt}, {Hao}, {Calzetti},
  {Moustakas}, {Dale}, {Bendo}, {Engelbracht}, {Johnson}, \&
  {Lee}}]{2009ApJ...703.1672K}
{Kennicutt}, Robert~C., J., {Hao}, C.-N., {Calzetti}, D., {et~al.} 2009, \apj,
  703, 1672, \dodoi{10.1088/0004-637X/703/2/1672}

\bibitem[{{Kennicutt} \& {Evans}(2012)}]{2012ARA&A..50..531K}
{Kennicutt}, R.~C., \& {Evans}, N.~J. 2012, \araa, 50, 531,
  \dodoi{10.1146/annurev-astro-081811-125610}

\bibitem[{Kennicutt \& Evans(2012)}]{Kennicutt_2012}
Kennicutt, R.~C., \& Evans, N.~J. 2012, Annual Review of Astronomy and
  Astrophysics, 50, 531–608, \dodoi{10.1146/annurev-astro-081811-125610}

\bibitem[{{Kennicutt} {et~al.}(2011){Kennicutt}, {Calzetti}, {Aniano},
  {Appleton}, {Armus}, {Beir{\~a}o}, {Bolatto}, {Brandl}, {Crocker}, {Croxall},
  {Dale}, {Donovan Meyer}, {Draine}, {Engelbracht}, {Galametz}, {Gordon},
  {Groves}, {Hao}, {Helou}, {Hinz}, {Hunt}, {Johnson}, {Koda}, {Krause},
  {Leroy}, {Li}, {Meidt}, {Montiel}, {Murphy}, {Rahman}, {Rix}, {Roussel},
  {Sandstrom}, {Sauvage}, {Schinnerer}, {Skibba}, {Smith}, {Srinivasan},
  {Vigroux}, {Walter}, {Wilson}, {Wolfire}, \& {Zibetti}}]{2011PASP..123.1347K}
{Kennicutt}, R.~C., {Calzetti}, D., {Aniano}, G., {et~al.} 2011, \pasp, 123,
  1347, \dodoi{10.1086/663818}

\bibitem[{{Kewley} \& {Ellison}(2008)}]{2008ApJ...681.1183K}
{Kewley}, L.~J., \& {Ellison}, S.~L. 2008, \apj, 681, 1183,
  \dodoi{10.1086/587500}

\bibitem[{{Knapen} {et~al.}(2010){Knapen}, {Sharp}, {Ryder},
  {Falc{\'o}n-Barroso}, {Fathi}, \& {Guti{\'e}rrez}}]{2010MNRAS.408..797K}
{Knapen}, J.~H., {Sharp}, R.~G., {Ryder}, S.~D., {et~al.} 2010, \mnras, 408,
  797, \dodoi{10.1111/j.1365-2966.2010.17180.x}

\bibitem[{{Kroupa}(2001)}]{2001MNRAS.322..231K}
{Kroupa}, P. 2001, \mnras, 322, 231, \dodoi{10.1046/j.1365-8711.2001.04022.x}

\bibitem[{{Kruijssen} \& {Longmore}(2014)}]{2014MNRAS.439.3239K}
{Kruijssen}, J.~M.~D., \& {Longmore}, S.~N. 2014, \mnras, 439, 3239,
  \dodoi{10.1093/mnras/stu098}

\bibitem[{{Lada} {et~al.}(2013){Lada}, {Lombardi}, {Roman-Zuniga}, {Forbrich},
  \& {Alves}}]{2013ApJ...778..133L}
{Lada}, C.~J., {Lombardi}, M., {Roman-Zuniga}, C., {Forbrich}, J., \& {Alves},
  J.~F. 2013, \apj, 778, 133, \dodoi{10.1088/0004-637X/778/2/133}

\bibitem[{{Lebouteiller} {et~al.}(2011){Lebouteiller}, {Bernard-Salas},
  {Whelan}, {Brandl}, {Galliano}, {Charmandaris}, {Madden}, \&
  {Kunth}}]{2011ApJ...728...45L}
{Lebouteiller}, V., {Bernard-Salas}, J., {Whelan}, D.~G., {et~al.} 2011, \apj,
  728, 45, \dodoi{10.1088/0004-637X/728/1/45}

\bibitem[{Lee {et~al.}(2013)Lee, Hwang, \& Ko}]{Lee_2013}
Lee, J.~C., Hwang, H.~S., \& Ko, J. 2013, The Astrophysical Journal, 774, 62,
  \dodoi{10.1088/0004-637x/774/1/62}

\bibitem[{{Lee} {et~al.}(2009){Lee}, {Kennicutt}, {Funes}, {Sakai}, \&
  {Akiyama}}]{2009ApJ...692.1305L}
{Lee}, J.~C., {Kennicutt}, Robert~C., J., {Funes}, S.~J. J.~G., {Sakai}, S., \&
  {Akiyama}, S. 2009, \apj, 692, 1305, \dodoi{10.1088/0004-637X/692/2/1305}

\bibitem[{{Leroy} {et~al.}(2008){Leroy}, {Walter}, {Brinks}, {Bigiel}, {de
  Blok}, {Madore}, \& {Thornley}}]{2008AJ....136.2782L}
{Leroy}, A.~K., {Walter}, F., {Brinks}, E., {et~al.} 2008, \aj, 136, 2782,
  \dodoi{10.1088/0004-6256/136/6/2782}

\bibitem[{{Leroy} {et~al.}(2009){Leroy}, {Walter}, {Bigiel}, {Usero}, {Weiss},
  {Brinks}, {de Blok}, {Kennicutt}, {Schuster}, {Kramer}, {Wiesemeyer}, \&
  {Roussel}}]{2009AJ....137.4670L}
{Leroy}, A.~K., {Walter}, F., {Bigiel}, F., {et~al.} 2009, \aj, 137, 4670,
  \dodoi{10.1088/0004-6256/137/6/4670}

\bibitem[{{Leroy} {et~al.}(2013){Leroy}, {Walter}, {Sandstrom}, {Schruba},
  {Munoz-Mateos}, {Bigiel}, {Bolatto}, {Brinks}, {de Blok}, {Meidt}, {Rix},
  {Rosolowsky}, {Schinnerer}, {Schuster}, \& {Usero}}]{2013AJ....146...19L}
{Leroy}, A.~K., {Walter}, F., {Sandstrom}, K., {et~al.} 2013, \aj, 146, 19,
  \dodoi{10.1088/0004-6256/146/2/19}

\bibitem[{{Li} {et~al.}(2013){Li}, {Crocker}, {Calzetti}, {Wilson},
  {Kennicutt}, {Murphy}, {Brandl}, {Draine}, {Galametz}, {Johnson}, {Armus},
  {Gordon}, {Croxall}, {Dale}, {Engelbracht}, {Groves}, {Hao}, {Helou}, {Hinz},
  {Hunt}, {Krause}, {Roussel}, {Sauvage}, \& {Smith}}]{2013ApJ...768..180L}
{Li}, Y., {Crocker}, A.~F., {Calzetti}, D., {et~al.} 2013, \apj, 768, 180,
  \dodoi{10.1088/0004-637X/768/2/180}

\bibitem[{Lin {et~al.}(2020)Lin, Calzetti, Kong, Adamo, Cignoni, Cook, Dale,
  Grasha, Grebel, Messa, \& et~al.}]{Lin_2020}
Lin, Z., Calzetti, D., Kong, X., {et~al.} 2020, The Astrophysical Journal, 896,
  16, \dodoi{10.3847/1538-4357/ab9106}

\bibitem[{{Liu} {et~al.}(2011){Liu}, {Koda}, {Calzetti}, {Fukuhara}, \&
  {Momose}}]{2011ApJ...735...63L}
{Liu}, G., {Koda}, J., {Calzetti}, D., {Fukuhara}, M., \& {Momose}, R. 2011,
  \apj, 735, 63, \dodoi{10.1088/0004-637X/735/1/63}

\bibitem[{{Lu} {et~al.}(2014){Lu}, {Bendo}, {Boselli}, {Baes}, {Wu}, {Madden},
  {De Looze}, {R{\'e}my-Ruyer}, {Boquien}, {Wilson}, {Galametz}, {Lam},
  {Cooray}, {Spinoglio}, \& {Zhao}}]{2014ApJ...797..129L}
{Lu}, N., {Bendo}, G.~J., {Boselli}, A., {et~al.} 2014, \apj, 797, 129,
  \dodoi{10.1088/0004-637X/797/2/129}

\bibitem[{{Lundgren} {et~al.}(2004){Lundgren}, {Wiklind}, {Olofsson}, \&
  {Rydbeck}}]{2004A&A...413..505L}
{Lundgren}, A.~A., {Wiklind}, T., {Olofsson}, H., \& {Rydbeck}, G. 2004, \aap,
  413, 505, \dodoi{10.1051/0004-6361:20031507}

\bibitem[{{Madau} \& {Dickinson}(2014)}]{2014ARA&A..52..415M}
{Madau}, P., \& {Dickinson}, M. 2014, \araa, 52, 415,
  \dodoi{10.1146/annurev-astro-081811-125615}

\bibitem[{{Madden} {et~al.}(2006){Madden}, {Galliano}, {Jones}, \&
  {Sauvage}}]{2006A&A...446..877M}
{Madden}, S.~C., {Galliano}, F., {Jones}, A.~P., \& {Sauvage}, M. 2006, \aap,
  446, 877, \dodoi{10.1051/0004-6361:20053890}

\bibitem[{{Madden} {et~al.}(2013){Madden}, {R{\'e}my-Ruyer}, {Galametz},
  {Cormier}, {Lebouteiller}, {Galliano}, {Hony}, {Bendo}, {Smith}, {Pohlen},
  {Roussel}, {Sauvage}, {Wu}, {Sturm}, {Poglitsch}, {Contursi}, {Doublier},
  {Baes}, {Barlow}, {Boselli}, {Boquien}, {Carlson}, {Ciesla}, {Cooray},
  {Cortese}, {de Looze}, {Irwin}, {Isaak}, {Kamenetzky}, {Karczewski}, {Lu},
  {MacHattie}, {O'Halloran}, {Parkin}, {Rangwala}, {Schirm}, {Schulz},
  {Spinoglio}, {Vaccari}, {Wilson}, \& {Wozniak}}]{2013PASP..125..600M}
{Madden}, S.~C., {R{\'e}my-Ruyer}, A., {Galametz}, M., {et~al.} 2013, \pasp,
  125, 600, \dodoi{10.1086/671138}

\bibitem[{{Makarov} {et~al.}(2014){Makarov}, {Prugniel}, {Terekhova},
  {Courtois}, \& {Vauglin}}]{2014A&A...570A..13M}
{Makarov}, D., {Prugniel}, P., {Terekhova}, N., {Courtois}, H., \& {Vauglin},
  I. 2014, \aap, 570, A13, \dodoi{10.1051/0004-6361/201423496}

\bibitem[{{Marlowe} {et~al.}(1999){Marlowe}, {Meurer}, \&
  {Heckman}}]{1999ApJ...522..183M}
{Marlowe}, A.~T., {Meurer}, G.~R., \& {Heckman}, T.~M. 1999, \apj, 522, 183,
  \dodoi{10.1086/307603}

\bibitem[{McGaugh \& Schombert(2014)}]{McGaugh_2014}
McGaugh, S.~S., \& Schombert, J.~M. 2014, The Astronomical Journal, 148, 77,
  \dodoi{10.1088/0004-6256/148/5/77}

\bibitem[{{Meidt} {et~al.}(2012){Meidt}, {Schinnerer}, {Knapen}, {Bosma},
  {Athanassoula}, {Sheth}, {Buta}, {Zaritsky}, {Laurikainen}, {Elmegreen},
  {Elmegreen}, {Gadotti}, {Salo}, {Regan}, {Ho}, {Madore}, {Hinz}, {Skibba},
  {Gil de Paz}, {Mu{\~n}oz-Mateos}, {Men{\'e}ndez-Delmestre}, {Seibert}, {Kim},
  {Mizusawa}, {Laine}, \& {Comer{\'o}n}}]{2012ApJ...744...17M}
{Meidt}, S.~E., {Schinnerer}, E., {Knapen}, J.~H., {et~al.} 2012, \apj, 744,
  17, \dodoi{10.1088/0004-637X/744/1/17}

\bibitem[{Meidt {et~al.}(2014)Meidt, Schinnerer, van~de Ven, Zaritsky,
  Peletier, Knapen, Sheth, Regan, Querejeta, Muñoz-Mateos, \&
  et~al.}]{Meidt_2014}
Meidt, S.~E., Schinnerer, E., van~de Ven, G., {et~al.} 2014, The Astrophysical
  Journal, 788, 144, \dodoi{10.1088/0004-637x/788/2/144}

\bibitem[{{Meurer} {et~al.}(1995){Meurer}, {Heckman}, {Leitherer}, {Kinney},
  {Robert}, \& {Garnett}}]{1995AJ....110.2665M}
{Meurer}, G.~R., {Heckman}, T.~M., {Leitherer}, C., {et~al.} 1995, \aj, 110,
  2665, \dodoi{10.1086/117721}

\bibitem[{{Momose} {et~al.}(2010){Momose}, {Okumura}, {Koda}, \&
  {Sawada}}]{2010ApJ...721..383M}
{Momose}, R., {Okumura}, S.~K., {Koda}, J., \& {Sawada}, T. 2010, \apj, 721,
  383, \dodoi{10.1088/0004-637X/721/1/383}

\bibitem[{{Moreno-Raya} {et~al.}(2016){Moreno-Raya}, {L{\'o}pez-S{\'a}nchez},
  {Moll{\'a}}, {Galbany}, {V{\'\i}lchez}, \& {Carnero}}]{2016MNRAS.462.1281M}
{Moreno-Raya}, M.~E., {L{\'o}pez-S{\'a}nchez}, {\'A}.~R., {Moll{\'a}}, M.,
  {et~al.} 2016, \mnras, 462, 1281, \dodoi{10.1093/mnras/stw1706}

\bibitem[{{Morrissey} {et~al.}(2007){Morrissey}, {Conrow}, {Barlow}, {Small},
  {Seibert}, {Wyder}, {Budav{\'a}ri}, {Arnouts}, {Friedman}, {Forster},
  {Martin}, {Neff}, {Schiminovich}, {Bianchi}, {Donas}, {Heckman}, {Lee},
  {Madore}, {Milliard}, {Rich}, {Szalay}, {Welsh}, \&
  {Yi}}]{2007ApJS..173..682M}
{Morrissey}, P., {Conrow}, T., {Barlow}, T.~A., {et~al.} 2007, \apjs, 173, 682,
  \dodoi{10.1086/520512}

\bibitem[{Moustakas {et~al.}(2010)Moustakas, Kennicutt, Tremonti, Dale, Smith,
  \& Calzetti}]{Moustakas_2010}
Moustakas, J., Kennicutt, R.~C., Tremonti, C.~A., {et~al.} 2010, The
  Astrophysical Journal Supplement Series, 190, 233,
  \dodoi{10.1088/0067-0049/190/2/233}

\bibitem[{{NASA/IPAC Extragalactic Database
  (NED)}(2019)}]{https://doi.org/10.26132/ned1}
{NASA/IPAC Extragalactic Database (NED)}. 2019, NASA/IPAC Extragalactic
  Database (NED),  IPAC, \dodoi{10.26132/NED1}

\bibitem[{Oh {et~al.}(2008)Oh, de~Blok, Walter, Brinks, \& Kennicutt}]{Oh_2008}
Oh, S.-H., de~Blok, W. J.~G., Walter, F., Brinks, E., \& Kennicutt, R.~C. 2008,
  The Astronomical Journal, 136, 2761, \dodoi{10.1088/0004-6256/136/6/2761}

\bibitem[{{Onodera} {et~al.}(2010){Onodera}, {Kuno}, {Tosaki}, {Kohno},
  {Nakanishi}, {Sawada}, {Muraoka}, {Komugi}, {Miura}, {Kaneko}, {Hirota}, \&
  {Kawabe}}]{2010ApJ...722L.127O}
{Onodera}, S., {Kuno}, N., {Tosaki}, T., {et~al.} 2010, \apjl, 722, L127,
  \dodoi{10.1088/2041-8205/722/2/L127}

\bibitem[{{Pilyugin} {et~al.}(2015){Pilyugin}, {Grebel}, \&
  {Zinchenko}}]{2015MNRAS.450.3254P}
{Pilyugin}, L.~S., {Grebel}, E.~K., \& {Zinchenko}, I.~A. 2015, \mnras, 450,
  3254, \dodoi{10.1093/mnras/stv932}

\bibitem[{{Pilyugin} \& {Thuan}(2005)}]{2005ApJ...631..231P}
{Pilyugin}, L.~S., \& {Thuan}, T.~X. 2005, \apj, 631, 231,
  \dodoi{10.1086/432408}

\bibitem[{{Povich} {et~al.}(2007){Povich}, {Stone}, {Churchwell}, {Zweibel},
  {Wolfire}, {Babler}, {Indebetouw}, {Meade}, \&
  {Whitney}}]{2007ApJ...660..346P}
{Povich}, M.~S., {Stone}, J.~M., {Churchwell}, E., {et~al.} 2007, \apj, 660,
  346, \dodoi{10.1086/513073}

\bibitem[{{Querejeta} {et~al.}(2015){Querejeta}, {Meidt}, {Schinnerer},
  {Cisternas}, {Mu{\~n}oz-Mateos}, {Sheth}, {Knapen}, {van de Ven}, {Norris},
  {Peletier}, {Laurikainen}, {Salo}, {Holwerda}, {Athanassoula}, {Bosma},
  {Groves}, {Ho}, {Gadotti}, {Zaritsky}, {Regan}, {Hinz}, {Gil de Paz},
  {Menendez-Delmestre}, {Seibert}, {Mizusawa}, {Kim}, {Erroz-Ferrer}, {Laine},
  \& {Comer{\'o}n}}]{2015ApJS..219....5Q}
{Querejeta}, M., {Meidt}, S.~E., {Schinnerer}, E., {et~al.} 2015, \apjs, 219,
  5, \dodoi{10.1088/0067-0049/219/1/5}

\bibitem[{{Rahman} {et~al.}(2011){Rahman}, {Bolatto}, {Wong}, {Leroy},
  {Walter}, {Rosolowsky}, {West}, {Bigiel}, {Ott}, {Xue}, {Herrera-Camus},
  {Jameson}, {Blitz}, \& {Vogel}}]{2011ApJ...730...72R}
{Rahman}, N., {Bolatto}, A.~D., {Wong}, T., {et~al.} 2011, \apj, 730, 72,
  \dodoi{10.1088/0004-637X/730/2/72}

\bibitem[{{Rahman} {et~al.}(2012){Rahman}, {Bolatto}, {Xue}, {Wong}, {Leroy},
  {Walter}, {Bigiel}, {Rosolowsky}, {Fisher}, {Vogel}, {Blitz}, {West}, \&
  {Ott}}]{2012ApJ...745..183R}
{Rahman}, N., {Bolatto}, A.~D., {Xue}, R., {et~al.} 2012, \apj, 745, 183,
  \dodoi{10.1088/0004-637X/745/2/183}

\bibitem[{Relaño \& Kennicutt(2009)}]{Rela_o_2009}
Relaño, M., \& Kennicutt, R.~C. 2009, The Astrophysical Journal, 699,
  1125–1143, \dodoi{10.1088/0004-637x/699/2/1125}

\bibitem[{{Roussel}(2013)}]{2013PASP..125.1126R}
{Roussel}, H. 2013, \pasp, 125, 1126, \dodoi{10.1086/673310}

\bibitem[{{Sandstrom} {et~al.}(2012){Sandstrom}, {Bolatto}, {Bot}, {Draine},
  {Ingalls}, {Israel}, {Jackson}, {Leroy}, {Li}, {Rubio}, {Simon}, {Smith},
  {Stanimirovi{\'c}}, {Tielens}, \& {van Loon}}]{2012ApJ...744...20S}
{Sandstrom}, K.~M., {Bolatto}, A.~D., {Bot}, C., {et~al.} 2012, \apj, 744, 20,
  \dodoi{10.1088/0004-637X/744/1/20}

\bibitem[{{Schruba} {et~al.}(2010){Schruba}, {Leroy}, {Walter}, {Sand strom},
  \& {Rosolowsky}}]{2010ApJ...722.1699S}
{Schruba}, A., {Leroy}, A.~K., {Walter}, F., {Sand strom}, K., \& {Rosolowsky},
  E. 2010, \apj, 722, 1699, \dodoi{10.1088/0004-637X/722/2/1699}

\bibitem[{{Shi} {et~al.}(2012){Shi}, {Kong}, {Wicker}, {Chen}, {Gong}, \&
  {Fan}}]{2012JApA...33..213S}
{Shi}, F., {Kong}, X., {Wicker}, J., {et~al.} 2012, Journal of Astrophysics and
  Astronomy, 33, 213, \dodoi{10.1007/s12036-012-9145-5}

\bibitem[{Shivaei {et~al.}(2017)Shivaei, Reddy, Shapley, Siana, Kriek,
  Mobasher, Coil, Freeman, Sanders, Price, \& et~al.}]{Shivaei_2017}
Shivaei, I., Reddy, N.~A., Shapley, A.~E., {et~al.} 2017, The Astrophysical
  Journal, 837, 157, \dodoi{10.3847/1538-4357/aa619c}

\bibitem[{{Skibba} {et~al.}(2011){Skibba}, {Engelbracht}, {Dale}, {Hinz},
  {Zibetti}, {Crocker}, {Groves}, {Hunt}, {Johnson}, {Meidt}, {Murphy},
  {Appleton}, {Armus}, {Bolatto}, {Brandl}, {Calzetti}, {Croxall}, {Galametz},
  {Gordon}, {Kennicutt}, {Koda}, {Krause}, {Montiel}, {Rix}, {Roussel},
  {Sandstrom}, {Sauvage}, {Schinnerer}, {Smith}, {Walter}, {Wilson}, \&
  {Wolfire}}]{2011ApJ...738...89S}
{Skibba}, R.~A., {Engelbracht}, C.~W., {Dale}, D., {et~al.} 2011, \apj, 738,
  89, \dodoi{10.1088/0004-637X/738/1/89}

\bibitem[{{Smith} {et~al.}(2007){Smith}, {Draine}, {Dale}, {Moustakas},
  {Kennicutt}, {Helou}, {Armus}, {Roussel}, {Sheth}, {Bendo}, {Buckalew},
  {Calzetti}, {Engelbracht}, {Gordon}, {Hollenbach}, {Li}, {Malhotra},
  {Murphy}, \& {Walter}}]{2007ApJ...656..770S}
{Smith}, J.~D.~T., {Draine}, B.~T., {Dale}, D.~A., {et~al.} 2007, \apj, 656,
  770, \dodoi{10.1086/510549}

\bibitem[{{Sofue} {et~al.}(1999){Sofue}, {Tutui}, {Honma}, {Tomita},
  {Takamiya}, {Koda}, \& {Takeda}}]{1999ApJ...523..136S}
{Sofue}, Y., {Tutui}, Y., {Honma}, M., {et~al.} 1999, \apj, 523, 136,
  \dodoi{10.1086/307731}

\bibitem[{{Thim} {et~al.}(2003){Thim}, {Tammann}, {Saha}, {Dolphin}, {Sandage},
  {Tolstoy}, \& {Labhardt}}]{2003ApJ...590..256T}
{Thim}, F., {Tammann}, G.~A., {Saha}, A., {et~al.} 2003, \apj, 590, 256,
  \dodoi{10.1086/374888}

\bibitem[{{Tody}(1986)}]{1986SPIE..627..733T}
{Tody}, D. 1986, in Society of Photo-Optical Instrumentation Engineers (SPIE)
  Conference Series, Vol. 627, Instrumentation in astronomy VI, ed. D.~L.
  {Crawford}, 733, \dodoi{10.1117/12.968154}

\bibitem[{{Tody}(1993)}]{1993ASPC...52..173T}
{Tody}, D. 1993, in Astronomical Society of the Pacific Conference Series,
  Vol.~52, Astronomical Data Analysis Software and Systems II, ed. R.~J.
  {Hanisch}, R.~J.~V. {Brissenden}, \& J.~{Barnes}, 173

\bibitem[{{Tremonti} {et~al.}(2001){Tremonti}, {Calzetti}, {Leitherer}, \&
  {Heckman}}]{2001ApJ...555..322T}
{Tremonti}, C.~A., {Calzetti}, D., {Leitherer}, C., \& {Heckman}, T.~M. 2001,
  \apj, 555, 322, \dodoi{10.1086/321436}

\bibitem[{{Wall} {et~al.}(2016){Wall}, {Puerari}, {Tilanus}, {Israel},
  {Austermann}, {Aretxaga}, {Wilson}, {Yun}, {Scott}, {Perera}, {Roberts}, \&
  {Hughes}}]{2016MNRAS.459.1440W}
{Wall}, W.~F., {Puerari}, I., {Tilanus}, R., {et~al.} 2016, \mnras, 459, 1440,
  \dodoi{10.1093/mnras/stw687}

\bibitem[{Watkins {et~al.}(2017)Watkins, Mihos, \& Harding}]{Watkins_2017}
Watkins, A.~E., Mihos, J.~C., \& Harding, P. 2017, The Astrophysical Journal,
  851, 51, \dodoi{10.3847/1538-4357/aa8fcd}

\bibitem[{{Whitmore} {et~al.}(2014){Whitmore}, {Brogan}, {Chandar}, {Evans},
  {Hibbard}, {Johnson}, {Leroy}, {Privon}, {Remijan}, \&
  {Sheth}}]{2014ApJ...795..156W}
{Whitmore}, B.~C., {Brogan}, C., {Chandar}, R., {et~al.} 2014, \apj, 795, 156,
  \dodoi{10.1088/0004-637X/795/2/156}

\bibitem[{{Zibetti} {et~al.}(2009){Zibetti}, {Charlot}, \&
  {Rix}}]{2009MNRAS.400.1181Z}
{Zibetti}, S., {Charlot}, S., \& {Rix}, H.-W. 2009, \mnras, 400, 1181,
  \dodoi{10.1111/j.1365-2966.2009.15528.x}

\end{thebibliography}

\end{document}